\documentclass[12pt]{article}%
\usepackage{amsmath}
\usepackage{graphicx}
\usepackage{amsfonts}
\usepackage{amssymb}%
\numberwithin{equation}{section}
\begin{document}

\title{About dual two-dimensional oscillator and Coulomb-like theories on
pseudosphere}
\author{G.V. Grigoryan\thanks{%
Yerevan Physics Institute. Yerevan. Armenia; e-mail:gagri@yerphi.am}, R.P.
Grigoryan\thanks{%
Yerevan Physics Institute. Yerevan. Armenia; e-mail:rogri@yerphi.am} I.V.
Tyutin\thanks{%
Lebedev Physical Institute. Moscow. Russia; e-mail:tyutin@lpi.ru}}
\date{ }
\maketitle

\begin{abstract}
We present a mathematically rigorous quantum-mechanical treatment of a
two-dimensional nonrelativistic quantum dual theories (with oscillator and
Coulomb like potentials) on pseudosphere and compare their spectra and the
sets of eigenfunctions. We construct all self-adjoint Schrodinger operators
for these theories and represent rigorous solutions of the corresponding
spectral problems. Solving the first part of the problem, we use a method of
specifying s.a. extensions by (asymptotic) s.a. boundary conditions. Solving
spectral problems, we follow the Krein's method of guiding functionals. We
show, that there is one to one correspondence between the spectral points of
dual theories in the planes energy-coupling constants not only for discrete,
but also for continuous spectra.
\end{abstract}

\section{Introduction}

It is well known \cite{Ter-Ant}, that if one introduces in a radial part of
the $D$ dimensional oscillator ($D > 2$)

\begin{equation}  \label{eq1}
{\frac{{d^{2}R}}{{du^{2}}}} + {\frac{{D - 1}}{{u}}}{\frac{{dR}}{{du}}} - {%
\frac{{L\left( {L + D - 2} \right)}}{{u^{2}}}}R + {\frac{{2\mu} }{{\hbar ^{2}%
}}}\left( {E - {\frac{{\mu \omega ^{2}u^{2}}}{{2}}}} \right)R = 0
\end{equation}
(here $R$ is the radial part of the wave function for the $D$ dimensional
oscillator ($D>2$) and $L=0,\mathrm{1},\mathrm{2},...$ are the eigenvalues
of the global angular momentum ) $r=u^{2}$ then equation (\ref{eq1})
transforms into

\begin{equation}  \label{eq2}
{\frac{{d^{2}R}}{{dr^{2}}}} + {\frac{{d - 1}}{{r}}}{\frac{{dR}}{{dr}}} - {%
\frac{{l\left( {l + d - 2} \right)}}{{r^{2}}}}R + {\frac{{2\mu} }{{\hbar ^{2}%
}}}\left( {\mathcal{E} + {\frac{{\alpha} }{{r}}}} \right)R = 0
\end{equation}

\noindent where $d=D/2+1\quad l=L/2\quad \mathcal{E}=-{\frac{{\mu \omega ^{2}%
}}{{8}}}\quad \alpha =E/4$, which formally is identical to the radial
equation for $d$-dimensional hydrogen atom.

Equations (\ref{eq1}) and (\ref{eq2}) are dual to each other and the duality
transformation is $r=u^{2}$. For discreet spectrum of these equations (and
wave fuctions regular at the origin) it was proved, that to each state of
equation (\ref{eq1}) corresponds a state in (\ref{eq2}), and visa versa \cite%
{{Ners-Ter-Ant},{Hak-Ter-Ant}}. However the correspondence of the states in
general (for discrete, as well as continuous spectra and for all values of
the parameters of the theory) the problems was not considered.

In \cite{TGG} we constructed all self-adjoint Schrodinger operators for
nonrelativistic one-dimen\-sional quantum dual theories and represented
rigorous solutions of the corresponding spectral problems. We have shown
that there is one to one correspondence between the spectra of dual theories
for discreet , as well as continuous spectra.

In this paper will solve the quantum problem of two dimensional quantum dual
theories (with oscillator and Coulomb like potentials ) on pseudosphere and
compare their spectra and the sets of eigenfunctions. As it was in one
dimensional case, we again have a correspondence of the states for all
values of the parameters ${E}_{O}$, ${\lambda }$, $E_{C}$, and $g$, except
when the angular momentum $m=1$, when the duality is one-to-one only in the
case of parameter of s.a. extension $\zeta=\pi/2$ (see below in coulomb
case). The interest to these models was stimulated also by the fact that
among the theorists dealing with similar problems exists a notion, that the
"Hamiltonian isn't self adjoit at high energies" \cite{Ners}.  In section 2
we will consider the quantum problem for the oscillator, will find solutions
of the equation for all values of the variable and parameters. In Section 3
we will consider the quantum problem for Coulomb-like system. The results
will be compared in section 4, where we will show the one-to one
correspondense of the spectra and proper functions of the Hamiltonians of
both problems.

\section{Quantum two-dimentional oscillator-like interaction on pseudosphere}

\subsection{Preliminaries}

Here, we consider the QM of a paticle moving on pseudosphere (two-sheet
hyperboloid) in an ``oscillator'' potential. We describe the pseudosphere by
the coordinates $\mathbf{u}=\{u^{1},u^{2}\}$ of its stereographic projection
on the plane, such that the coordinate $\mathbf{s}=\{s^{1},s^{2},s^{3}\}$ of
the ambient space, $(s^{3})^{2}-(s^{1})^{2}-(s^{2})^{2}=R^{2}$ are 
\begin{equation*}
s^{1}=\frac{2R^{2}u^{1}}{R^{2}-r^{2}},\;s^{2}=\frac{2R^{2}u^{2}}{R^{2}-r^{2}}%
,\;s^{3}=R\frac{R^{2}+r^{2}}{R^{2}-r^{2}},\;r=\sqrt{(u^{1})^{2}+(u^{2})^{2}}%
, 
\end{equation*}
where $R$ is a radius of the pseudosphere.The disc $r<R$ describes the upper
sheet $s^{3}>R$, the exterior of the disc $r>R$ describes the lower sheet $%
s^{3}<-R$. There is an useful one-to-one map $P$, $P^{2}=1$, of the interior
of disc onto the exterior of disc and inversely:%
\begin{equation*}
P\mathbf{u}=\mathbf{u}_{P}=\{u_{P}^{1},u_{P}^{2}\}=\left\{ \frac{R^{2}}{r^{2}%
}u^{1},\frac{R^{2}}{r^{2}}u^{2}\right\} ,\;r_{P}=\frac{R^{2}}{r},\;P^{2}%
\mathbf{u}=\mathbf{u}, 
\end{equation*}
We will call this map by the parity transformation and the operator $P$ by
the parity operator. Note that $P$-transformation commutes with the rotation
of $u$-plane around origin.

The metric $g_{jk}(\mathbf{u})$, the invariant volume element $d\Lambda (%
\mathbf{u})$, and the Beltrami-Laplace operator $\Delta_{BL}(\mathbf{u})$
have the following form in the coordinates $\mathbf{u}$:%
\begin{align*}
g_{jk}(\mathbf{u}) & =\frac{R^{4}}{(R^{2}-r^{2})^{2}}\delta_{jk},\;g^{jk}(%
\mathbf{u})=\frac{(R^{2}-r^{2})^{2}}{R^{4}}\delta^{jk},\;\sqrt{g(\mathbf{u})}%
=\frac{R^{4}}{(R^{2}-r^{2})^{2}}, \\
d\Lambda(\mathbf{u}) & =\frac{R^{4}}{(R^{2}-r^{2})^{2}}d^{2}u,\;\Delta _{BL}(%
\mathbf{u})=\frac{(R^{2}-r^{2})^{2}}{R^{4}}\Delta(\mathbf{u}),\;\Delta(%
\mathbf{u})=\partial_{u^{k}}\partial_{u^{k}},\;j,k=1,2.
\end{align*}
The wave functions $\Psi(\mathbf{u)}$ of the QM-problem under consideration
belong to the Hilbert space $\mathfrak{H}=L_{\Lambda}^{2}(\mathbb{R}^{2})$
with the scalar product%
\begin{equation*}
\left( \Psi_{1},\Psi_{2}\right) =\int_{\mathbb{R}^{2}}\overline{\Psi _{1}(%
\mathbf{u)}}\Psi_{2}(\mathbf{u)}d\Lambda(\mathbf{u}),\;\forall\Psi
_{1},\Psi_{2}\in\mathfrak{H,}
\end{equation*}
and the s.a. Hamiltonians are associated with a differential operation $%
\check{H}$,%
\begin{equation*}
\check{H}=\check{H}(\mathbf{u})=-\Delta_{BL}(\mathbf{u})+V(\mathbf{u}),%
\mathbf{\;}V(\mathbf{u})=\frac{4(q-1)r^{2}}{(R^{2}+r^{2})^{2}},\;q=\frac{%
R^{4}}{4}\lambda, 
\end{equation*}
where $V(\mathbf{u)}$ is an ``oscillator'' potential and $q$ is a coupling
constant ($\lambda$ is a coupling constant in the plane limit $R\rightarrow
\infty$) . We note that the introduced invariant volume element, the
Beltrami-Laplace operator, ``oscillator'' potential, differential operation
and scalar product are invariant under parity transformation:%
\begin{align*}
& Pd\Lambda(\mathbf{u})=d\Lambda(\mathbf{u}_{P})=d\Lambda(\mathbf{u}%
),\;P\Delta_{BL}(\mathbf{u})P=\Delta_{BL}(\mathbf{u}_{P})=\Delta _{BL}(%
\mathbf{u}), \\
& PV(\mathbf{u})=V(\mathbf{u}_{P})=V(\mathbf{u)},\;P\check{H}(\mathbf{u})P=%
\check{H}(\mathbf{u}_{P})=\check{H}(\mathbf{u}),\;\left(
\Psi_{1P},\Psi_{2P}\right) =\left( \Psi_{1},\Psi_{2}\right) ,
\end{align*}
where $\Psi_{P}(\mathbf{u})=P\Psi(\mathbf{u})=\Psi(\mathbf{u}_{P})$.
Furthermore, a relation holds:%
\begin{equation*}
\int_{r_{<}<R}\overline{\Psi_{1}(\mathbf{u}_{<}\mathbf{)}}\Psi_{2}(\mathbf{u}%
_{<})d\Lambda(\mathbf{u}_{<})=\int_{r_{>}>R}\overline{\tilde{\Psi }_{1}(%
\mathbf{u}_{>}\mathbf{)}}\tilde{\Psi}_{2}(\mathbf{u}_{>})d\Lambda (\mathbf{u}%
_{>}), 
\end{equation*}
where $\mathbf{u}_{<}=P\mathbf{u}_{>}$, $\tilde{\Psi}_{k}(\mathbf{u}%
_{<})=\Psi_{kP}(\mathbf{u}_{>})=\Psi_{k}(P\mathbf{u}_{>})$.

\subsubsection{Polar coordinates $r$, $\protect\varphi$}

Rewrite the above introduced quantities in the polar coordinates $r$, $%
\varphi$, $u^{1}=r\cos\varphi$, $u^{2}=r\sin\varphi$. We have:%
\begin{align*}
& d\Lambda(\mathbf{u})=d\omega(r)d\varphi,\;d\omega(r)=\frac{R^{4}r}{%
(R^{2}-r^{2})^{2}}dr, \\
& \check{H}=-\Delta_{BLr}-\Delta_{BL\varphi}+V(r),\;\Delta_{BL\varphi }=%
\frac{(R^{2}-r^{2})^{2}}{R^{4}r^{2}}\partial_{\varphi}^{2},\;V(r)=\frac{%
4(q-1)r^{2}}{(R^{2}+r^{2})^{2}}, \\
& \Delta_{BLr}=\frac{(R^{2}-r^{2})^{2}}{R^{4}}\Delta_{r},\;\Delta _{r}=\frac{%
1}{r}\partial_{r}r\partial_{r},
\end{align*}%
\begin{align*}
& \Psi(\mathbf{u})=\sum_{m\in Z}\Lambda_{m}(\mathbf{u}),\;\Lambda _{m}(%
\mathbf{u})=\frac{1}{\sqrt{2\pi}}e^{im\varphi}\Psi_{m}(r), \\
& \left( \Lambda_{1m_{1}},\Lambda_{2m_{2}}\right) =\delta_{m_{1}m_{2}}\left(
\Lambda_{1m_{1}}\Lambda_{2m_{1}}\right) ,\;\left( \Lambda
_{1m},\Lambda_{2m}\right) =\int_{0}^{\infty}\overline{\Psi_{1m}(r)}\Psi
_{2m}(r)d\omega(r) \\
& \check{H}\Psi(\mathbf{u})=\sum_{m\in Z}\frac{1}{\sqrt{2\pi}}e^{im\varphi }%
\check{H}_{m}\Psi_{m}(r),\;\check{H}_{m}=-\Delta_{BLr}+\frac{%
m^{2}(R^{2}-r^{2})^{2}}{R^{4}r^{2}}+V(r).
\end{align*}

Represent $\Psi_{m}(r)$ in the form%
\begin{equation*}
\Psi_{m}(r)=\frac{|R^{2}-r^{2}|}{R^{2}\sqrt{r}}\psi_{m}(r). 
\end{equation*}
Then we have%
\begin{align}
& \left( \Lambda_{1m},\Lambda_{2m}\right) =\langle\psi_{1m},\psi
_{2m}\rangle=\int_{0}^{\infty}\overline{\psi_{1m}(r)}\psi_{2m}(r)dr,  \notag
\\
& \check{H}_{m}\Psi_{m}(r)=\frac{|R^{2}-r^{2}|}{R^{2}\sqrt{r}}\check{h}%
_{m}\psi_{m}(r),  \notag \\
& \check{h}_{m}=\frac{\sqrt{r}}{R^{2}-r^{2}}\check{H}_{m}\frac{R^{2}-r^{2}}{%
\sqrt{r}}=  \notag \\
& =-\partial_{r}p_{0}(r)\partial_{r}+\frac{%
10R^{2}r^{2}-9r^{4}-R^{4}+4m^{2}(R^{2}-r^{2})^{2}}{4R^{4}r^{2}}%
+V(r),\;p_{0}(r)=\frac{(R^{2}-r^{2})^{2}}{R^{4}}.   \label{Opso2.1.1.1}
\end{align}

\subsubsection{$P$-transformation}

In the polar coordinates, we have%
\begin{equation*}
Pr=r_{P}=\frac{R^{2}}{r},\;P\varphi=\varphi_{P}=\varphi. 
\end{equation*}%
\begin{align*}
& Pr\partial_{r}=r_{P}\partial_{r_{P}}=-\frac{R^{2}}{r}\frac{%
r^{2}\partial_{r}}{R^{2}}=-r\partial_{r},\;P\frac{dr}{r}=\frac{dr_{P}}{r_{P}}%
=-\frac{r}{R^{2}}\frac{R^{2}dr}{r^{2}}=-\frac{dr}{r}, \\
& I_{\pm}(r)=\frac{(R^{2}\pm r^{2})^{2}}{R^{4}r^{2}},\;PI_{\pm}(r)=I_{\pm
}(r_{P})=I_{\pm}(r), \\
& d\omega(r)=\frac{1}{I_{-}(r)}\frac{dr}{r},\;Pd\omega(r)=d\omega
(r),\;d\Lambda(\mathbf{u})=\frac{1}{I_{-}(r)}\frac{dr}{r}d\varphi
,\;Pd\Lambda(\mathbf{u})=d\Lambda(\mathbf{u}), \\
& \Delta_{BLr}=\frac{1}{4}I_{-}(r)(r\partial_{r})^{2},\;P\Delta
_{BLr}(r)=\Delta_{BLr}(r_{P})=\Delta_{BLr}(r), \\
& \Delta_{BL\varphi}=I_{-}(r)\partial_{\varphi}^{2},\;P\Delta_{BL\varphi }(%
\mathbf{u})=\Delta_{BL\varphi}(\mathbf{u}_{P})=\Delta_{BL\varphi}(\mathbf{u}%
), \\
& V(r)=\frac{4(q-1)}{R^{4}I_{+}(r)},\;PV(r)=V(r_{P})=V(r), \\
& P\check{H}(\mathbf{u})P=\check{H}(\mathbf{u}_{P})=\check{H}(\mathbf{u}),\;P%
\check{H}_{m}(r)P=\check{H}_{m}(r_{P})=\check{H}_{m}(r).
\end{align*}
Operator $P$ acts on the radial wave finctions by the following way
(subscript ``$m$'' is omitted),%
\begin{equation*}
P\Psi(r)=\Psi(r_{P}(r))=\Psi(R^{2}/r). 
\end{equation*}

It is convenient to introduce some notation:%
\begin{align*}
& r=\{\rho,\xi\},\;0\leq\xi\leq R,\;R\leq\rho<\infty, \\
& P_{\rho\xi}\xi=\rho_{P}=\frac{R^{2}}{\rho},\;P_{\xi\rho}\rho=\xi _{P}=%
\frac{R^{2}}{\xi},\;\Psi(r)=\left( 
\begin{array}{c}
\Psi^{>}(\rho) \\ 
\Psi^{<}(\xi)%
\end{array}
\right) , \\
& P_{\rho\xi}P_{\xi\rho}\rho=P_{\rho\xi}\xi_{P}=P_{\rho\xi}\frac{R^{2}}{\xi }%
=\frac{R^{2}}{\rho_{P}}=\rho,\;P_{\xi\rho}P_{\rho\xi}\xi=\xi,
\end{align*}%
\begin{align*}
&
P_{\rho\xi}\Psi^{<}(\xi)=\Psi^{<}(R^{2}/\rho),\;P_{\xi\rho}\Psi^{>}(\rho)=%
\Psi^{>}(R^{2}/\xi),\;\check{H}(r)=\left( 
\begin{array}{cc}
\check{H}^{>}(\rho) & 0 \\ 
0 & \check{H}^{<}(\xi)%
\end{array}
\right) , \\
& P\Psi(r)=\Psi_{P}(r)=\left( 
\begin{array}{c}
\Psi_{P}^{|>}(\rho) \\ 
\Psi_{P}^{|<}(\xi)%
\end{array}
\right)
\Longrightarrow\Psi_{P}^{|>}(\rho)=\Psi^{<}(R^{2}/\rho)=\Psi^{<}(\rho_{P}),
\\
& \Psi_{P}^{|<}(\xi)=\Psi^{>}(R^{2}/\xi)=\Psi^{>}(\xi_{P})\Longrightarrow
P=\left( 
\begin{array}{cc}
0 & P_{\rho\xi} \\ 
P_{\xi\rho} & 0%
\end{array}
\right) ,
\end{align*}
where $\Psi_{P}^{|>}(\rho)\equiv(\Psi_{P})^{>}(\rho)$ and so forth.

Checking:%
\begin{align*}
& \check{H}(r)=P\check{H}(r)P=\left( 
\begin{array}{cc}
P_{\rho\xi}\check{H}^{<}(\xi)P_{\xi\rho} & 0 \\ 
0 & P_{\xi\rho}\check{H}^{>}(\rho)P_{\rho\xi}%
\end{array}
\right) = \\
& \,=\left( 
\begin{array}{cc}
\check{H}^{<}(\rho_{P}) & 0 \\ 
0 & \check{H}^{>}(\xi_{P})%
\end{array}
\right) =\check{H}(r_{P})=\check{H}(r).
\end{align*}

Turn out to the ``$h$-space''. We have%
\begin{align*}
& \Psi(r)=\frac{|R^{2}-r^{2}|}{R^{2}\sqrt{r}}\psi(r),\;P\Psi(r)=\frac{%
|R^{2}-r_{P}^{2}|}{R^{2}\sqrt{r_{P}}}\psi(r_{P})=\frac{|R^{2}-r^{2}|}{%
Rr^{3/2}}\psi(r_{P})= \\
& \,=\frac{|R^{2}-r^{2}|}{R^{2}\sqrt{r}}\tilde{P}\psi(r)\Longrightarrow 
\tilde{P}\psi(r)=\frac{R}{r}\psi(r_{P})=P\frac{r}{R}\psi(r)\Longrightarrow \\
& \,\Longrightarrow\tilde{P}=\frac{R}{r}P=\frac{r_{P}}{R}P=P\frac{r}{R}%
=.\left( 
\begin{array}{cc}
0 & \frac{R}{\rho}P_{\rho\xi} \\ 
\frac{R}{\xi}P_{\xi\rho} & 0%
\end{array}
\right) = \\
& \,=\left( 
\begin{array}{cc}
0 & \frac{\rho_{P}}{R}P_{\rho\xi} \\ 
\frac{\xi_{P}}{R}P_{\xi\rho} & 0%
\end{array}
\right) =\left( 
\begin{array}{cc}
0 & P_{\rho\xi}\frac{\xi}{R} \\ 
P_{\xi\rho}\frac{\rho}{R} & 0%
\end{array}
\right) .
\end{align*}%
\begin{equation*}
\tilde{P}^{2}\psi(r)=\left( 
\begin{array}{cc}
\frac{R}{\rho}\frac{R}{\rho_{P}}P_{\rho\xi}P_{\xi\rho} & 0 \\ 
0 & \frac{R}{\xi}\frac{R}{\xi_{P}}P_{\xi\rho}P_{\rho\xi}%
\end{array}
\right) \left( 
\begin{array}{c}
\psi^{>}(\rho) \\ 
\psi^{<}(\xi)%
\end{array}
\right) =\left( 
\begin{array}{c}
\psi^{>}(\rho) \\ 
\psi^{<}(\xi)%
\end{array}
\right) =\psi(r), 
\end{equation*}%
\begin{align}
& \tilde{P}\check{h}(r)\tilde{P}=\frac{R}{r}\check{h}(r_{P})\frac{r}{R}=%
\frac{R}{r}P\frac{\sqrt{r}}{R^{2}-r^{2}}\check{H}_{m}(r)\frac{R^{2}-r^{2}}{%
\sqrt{r}}P\frac{r}{R}=  \notag \\
& \,=\frac{R}{r}\frac{r^{3/2}}{R(R^{2}-r^{2})}\check{H}_{m}(r)\frac{%
R(R^{2}-r^{2})}{r^{3/2}}\frac{r}{R}=\frac{\sqrt{r}}{R^{2}-r^{2}}\check{H}%
_{m}(r)\frac{R^{2}-r^{2}}{\sqrt{r}}=\check{h}(r),   \label{Opso2.1.1.2.2}
\end{align}%
\begin{align*}
& \psi(r)=\left( 
\begin{array}{c}
\psi^{>}(\rho) \\ 
\psi^{<}(\xi)%
\end{array}
\right) ,\;\tilde{P}\psi(r)=P\frac{r}{R}\psi(r)=\psi_{\tilde{P}}(r)=\left( 
\begin{array}{c}
\psi_{\tilde{P}}^{|>}(\rho) \\ 
\psi_{\tilde{P}}^{|<}(\xi)%
\end{array}
\right) \Longrightarrow \\
& \,\Longrightarrow\psi_{\tilde{P}}^{|>}(\rho)=P_{\rho\xi}\frac{\xi}{R}%
\psi^{<}(\xi)=\frac{\rho_{P}}{R}\psi^{<}(\rho_{P}), \\
& \psi_{\tilde{P}}^{|<}(\xi)=P_{\xi\rho}\frac{\rho}{R}\psi^{>}(\rho )=\frac{%
\xi_{P}}{R}\psi^{>}(\xi_{P}).
\end{align*}%
\begin{equation}
\check{h}(r)=.\left( 
\begin{array}{cc}
\check{h}^{>}(\rho) & 0 \\ 
0 & \check{h}^{<}(\xi)%
\end{array}
\right) ,\;\check{h}^{>}(\rho)=\frac{R}{\rho}\check{h}^{<}(\rho _{P})\frac{%
\rho}{R},\;\check{h}^{<}(\xi)=\frac{R}{\xi}\check{h}^{>}(\xi _{P})\frac{\xi}{%
R},   \label{Opso2.1.1.2.3}
\end{equation}
two last equalities follow from eq. (\ref{Opso2.1.1.2.2}),%
\begin{align}
& \check{h}^{>}(\rho)\psi_{\tilde{P}}^{|>}(\rho)=\frac{R}{\rho}\check{h}%
^{<}(\rho_{P})\frac{\rho}{R}\frac{\rho_{P}}{R}\psi^{<}(\rho_{P})=\frac{R}{%
\rho}\check{h}^{<}(\rho_{P})\psi^{<}(\rho_{P})=\frac{R}{\rho}P_{\rho\xi }%
\check{h}^{<}(\xi)\psi^{<}(\xi),  \label{Opso2.1.1.2.4a} \\
& \,[\check{h}^{<}(\xi)-W]\psi^{<}(\xi)=0\Longrightarrow\lbrack\check{h}%
^{>}(\rho)-W]\psi_{\tilde{P}}^{|>}(\rho)=0   \label{Opso2.1.1.2.4b}
\end{align}

\subsubsection{Some relations}

Any functions $\Lambda_{m}(\mathbf{u})=\frac{1}{\sqrt{2\pi}}\mathrm{e}%
^{im\varphi_{u}}\Psi_{m}(r)$ can be represented in the form%
\begin{align*}
& \Lambda_{m}(\mathbf{u})=\frac{1}{\sqrt{2\pi}}\mathrm{e}^{im\varphi_{u}}%
\Psi_{m,+}(r)+\frac{1}{\sqrt{2\pi}}\mathrm{e}^{im\varphi_{u}}\Psi_{m.-}(r),
\\
& \Psi_{m,\zeta}(r)=\frac{1+\zeta P}{2}\Psi_{m}(r),\;P\Psi_{m,\zeta}(r)=%
\zeta\Psi_{m,\zeta}(r),\;\zeta=\pm, \\
& \Psi_{m,\zeta}(r)=\left( 
\begin{array}{c}
\Psi_{m,\zeta}^{>}(\rho) \\ 
\Psi_{m,\zeta}^{<}(\xi)%
\end{array}
\right) ,\;\Psi_{m,\zeta}^{<}(\xi)=\frac{1}{2}\left[ \Psi_{m}^{<}(\xi
)+\zeta\Psi_{m}^{>}(\xi_{P})\right] , \\
& \Psi_{m,\zeta}^{>}(\rho)=\frac{1}{2}\left[ \Psi_{m}^{>}(\rho)+\zeta
\Psi_{m}^{<}(\rho_{P})\right] =\zeta P_{\rho\xi}\Psi_{m,\zeta}^{<}(\xi
)=\zeta\Psi_{m,\zeta}^{<}(\rho_{P}).
\end{align*}

As a consequence, we have%
\begin{align*}
& \Lambda_{m}(\mathbf{u})=\frac{1}{\sqrt{2\pi}}\mathrm{e}^{im\varphi_{u}}%
\frac{|R^{2}-r^{2}|}{R^{2}\sqrt{r}}\psi_{m}(r)=\frac{1}{\sqrt{2\pi}}\mathrm{e%
}^{im\varphi_{u}}\frac{|R^{2}-r^{2}|}{R^{2}\sqrt{r}}\psi_{m,+}(r)+ \\
& \,+\frac{1}{\sqrt{2\pi}}\mathrm{e}^{im\varphi_{u}}\frac{|R^{2}-r^{2}|}{%
R^{2}\sqrt{r}}\psi_{m,-}(r),\;\Psi_{m,\zeta}(r)=\frac{|R^{2}-r^{2}|}{R^{2}%
\sqrt{r}}\psi_{m,\zeta}(r), \\
& \psi_{m,\zeta}(r)=\left( 
\begin{array}{c}
\psi_{m,\zeta}^{>}(\rho) \\ 
\psi_{m,\zeta}^{<}(\xi)%
\end{array}
\right) ,\;\psi_{m,\zeta}^{<}(\xi)=\frac{1}{2}\left[ \psi_{m}^{<}(\xi )+\zeta%
\frac{\xi_{P}}{R}\psi_{m}^{>}(\xi_{P})\right] , \\
& \psi_{m,\zeta}^{>}(\rho)=\frac{1}{2}\left[ \psi_{m}^{>}(\rho )+\zeta\frac{%
\rho_{P}}{R}\psi_{m}^{<}(\rho_{P})\right] =\zeta P_{\rho\xi }\frac{\xi}{R}%
\psi_{m,\zeta}^{<}(\xi)= \\
& =\zeta\frac{\rho_{P}}{R}\psi_{m,\zeta}^{<}(\rho_{P})=\zeta\psi_{\tilde {P}%
m,\zeta}^{|>}(\rho),\;\tilde{P}\psi_{m,\zeta}(r)=\zeta\psi_{m,\zeta}(r).
\end{align*}

Below we omit the subscript ``$\zeta$'' for the lower component of functions 
$\Psi_{m,\zeta}(r)$ and $\psi_{m,\zeta}(r)$, that is,%
\begin{align}
& \Psi_{m,\zeta}(r)=\left( 
\begin{array}{c}
\Psi_{m,\zeta}^{>}(\rho) \\ 
\Psi_{m}^{<}(\xi)%
\end{array}
\right) ,\;\Psi_{m,\zeta}^{>}(\rho)=\zeta\Psi_{m}^{<}(\rho_{P}),  \notag \\
& \psi_{m,\zeta}(r)=\left( 
\begin{array}{c}
\psi_{m,\zeta}^{>}(\rho) \\ 
\psi_{m}^{<}(\xi)%
\end{array}
\right) ,\;\psi_{m,\zeta}^{>}(\rho)=\zeta\frac{\rho_{P}}{R}%
\psi_{m,\zeta}^{<}(\rho_{P}),  \label{Opso2.1.2.2} \\
& P\Psi_{m,\zeta}(r)=\zeta\Psi_{m,\zeta}(r),\;\tilde{P}\psi_{m,\zeta
}(r)=\zeta\psi_{m,\zeta}(r).  \notag
\end{align}

Let $\check{h}(r)\psi_{m,\zeta}(r)\equiv\psi_{h,m,\zeta}(r)$. Then we have%
\begin{equation*}
\tilde{P}\psi_{h,m,\zeta}(r)=\left( \tilde{P}\check{h}(r)\tilde{P}\right)
\left( \tilde{P}\psi_{m,\zeta}(r)\right) =\zeta\check{h}(r)\psi_{m,\zeta
}(r)=\zeta\psi_{h,m,\zeta}(r) 
\end{equation*}
Note the relation ($\zeta$ is fixed)%
\begin{align*}
&
\int_{0}^{\infty}|\psi_{m,\zeta}(r)|^{2}dr=2\int_{0}^{R}|%
\psi_{m}^{<}(r)|^{2}dr, \\
& \int_{0}^{\infty}\overline{\psi_{1m,\zeta}(r)}\check{h}(r)\psi_{2m,\zeta
}(r)dr=2\int_{0}^{R}\overline{\psi_{1m}^{<}(r)}\check{h}(r)%
\psi_{2m}^{<}(r)dr.
\end{align*}

Let 
\begin{equation*}
\psi^{(>)}(r)=\left( 
\begin{array}{c}
\psi^{>}(\rho) \\ 
0%
\end{array}
\right) ,\;\psi^{(<)}(r)=\left( 
\begin{array}{c}
0 \\ 
\psi^{<}(\xi)%
\end{array}
\right) . 
\end{equation*}

Then:%
\begin{align*}
& \tilde{P}\psi^{(>)}(r)=\left( 
\begin{array}{c}
0 \\ 
\psi_{\tilde{P}}^{|<}(\xi)%
\end{array}
\right) ,\;\tilde{P}\psi^{(<)}(r)=\left( 
\begin{array}{c}
\psi_{\tilde{P}}^{|>}(\rho) \\ 
0%
\end{array}
\right) , \\
& \int_{R}^{\infty}\overline{\psi_{1\tilde{P}}^{|>}(\rho)}\psi_{2\tilde{P}%
}^{|>}(\rho)d\rho=\int_{R}^{\infty}\overline{\psi_{1}^{<}(\rho_{P})}\psi
_{2}^{<}(\rho_{P})\frac{R^{2}d\rho}{\rho^{2}}= \\
& \,=-\int_{R}^{\infty}\overline{\psi_{1}^{<}(\rho_{P})}\psi_{2}^{<}(\rho
_{P})d\frac{R^{2}}{\rho}=\int_{0}^{R}\overline{\psi_{1}^{<}(\rho_{P})}\psi
_{2}^{<}(\rho_{P})d\rho_{P}=\int_{0}^{R}\overline{\psi_{1}^{<}(\xi)}\psi
_{2}^{<}(\xi)d\xi, \\
& \int_{0}^{R}\overline{\psi_{1\tilde{P}}^{|<}(\xi)}\psi_{2\tilde{P}%
}^{|<}(\xi)d\xi=\int_{R}^{\infty}\overline{\psi_{1}^{>}(\rho)}%
\psi_{2}^{>}(\rho)d\rho, \\
& \int_{0}^{\infty}\overline{\psi_{1\tilde{P}}(r)}\psi_{2\tilde{P}%
}(r)dr=\int_{0}^{\infty}\overline{\psi_{1}(r)}\psi_{2}(r)dr.
\end{align*}

\subsection{Reduction to radial problem}

In the case under consideration, the Hilbert space $\mathfrak{H}$ is a
direct orthogonal sum of subspaces $\mathfrak{H}_{m,\zeta}$, that are the
eigenspaces of the rotation operator $\hat{U}_{S}$ and the parity operator $P
$, 
\begin{align*}
& \mathfrak{H}=\sideset{}{^{\,\lower1mm\hbox{$\oplus$}}} \sum_{m\in \mathbb{%
Z,\zeta=\pm}}\mathfrak{H}_{m,\zeta},\;\hat{U}_{S}\mathfrak{H}%
_{m,\zeta}=e^{-im\theta}\mathfrak{H}_{m,\zeta},\;P\mathfrak{H}%
_{m,\zeta}=\zeta\mathfrak{H}_{m,\zeta}, \\
& \mathfrak{H}_{m,\zeta}=\hat{P}_{m,\zeta}\mathfrak{H,}
\end{align*}
where $\theta$ is the rotation angle corresponding to $S$, and $\hat {P}%
_{m,\zeta}$ is an orthohonal projector on subspace $\mathfrak{H}_{m,\zeta}$. 
$\mathfrak{H}_{m,\zeta}$ consists of eigenfunctions $\Psi_{m,\zeta }(\mathbf{%
u})$ of angular momentum operator $\hat{L}_{z}=-i\hbar
\partial/\partial\varphi_{u}$ and parity operator $P$, $\Psi_{m,\zeta }(%
\mathbf{u})=\frac{1}{\sqrt{2\pi}}\mathrm{e}^{im\varphi_{u}}\frac{%
|R^{2}-r^{2}|}{2R^{2}\sqrt{r}}\psi_{m,\zeta}(r)$, where%
\begin{equation*}
\tilde{P}\psi_{m,\zeta}(r)=\zeta\psi_{m,\zeta}(r), 
\end{equation*}
and $\psi_{m,\zeta}(r)\in\mathfrak{h}_{m,\zeta}=L^{2}(\mathbb{R}_{+})$, $%
\mathfrak{h}_{m,\zeta}$ is the Hilbert space of s.-integrable functions on
the semi-axis $\mathbb{R}_{+}$ with scalar product 
\begin{equation*}
\left\langle f_{1},f_{2}\right\rangle =\int_{\mathbb{R}_{+}}\overline {%
f_{1}\left( r\right) }f_{2}\left( r\right) dr. 
\end{equation*}

We define an initial symmetric operator $\hat{H}$ associated with $\check{H}$%
\ as follows:%
\begin{equation*}
\hat{H}:\left\{ 
\begin{array}{l}
D_{H}=\{\psi(\mathbf{u}):\ \psi\in\mathcal{D}(\mathbb{R}^{2}\backslash
\{r=R\})\} \\ 
\hat{H}\psi=\check{H}\psi,\ \forall\psi\in D_{H}\ 
\end{array}
\right. , 
\end{equation*}
where $\mathcal{D}(\mathbb{R}^{2}\backslash\{r=R\})$ is the space of smooth
and compactly supported functions vanishing in a neighborhood of origin and
of the circle $r=R$. The domain $D_{H}$ is dense in $\mathfrak{H}$ and the
symmetricity of $\hat{H}$ is obvious. It is also obvious that the operator $%
\hat{H}$ commutes\footnote{%
We remind the reader of the notion of commutativity in this case (where one
of the operators, $U_{S}$ or $P$, is bounded and defined everywhere): we say
that the operators $\hat{H}$ and $U_{S}$ commute if $U_{S}\hat{H}$ $\subseteq%
\hat{H}U_{S}$, i.e., if $\psi\in D_{H}$, then also $U_{S}\psi\in D_{H}$ and $%
U_{S}\hat{H}\psi=\hat{H}U_{S}\psi $, and analogously for $P$.} with the
unitary operators $\hat{U}_{S}$ and the s.a. operator $P$,%
\begin{equation*}
\hat{H}=\sideset{}{^{\,\lower1mm\hbox{$\oplus$}}} \sum_{m\in\mathbb{Z,,\zeta
=\pm}}\hat{H}_{m,\zeta},\;\hat{H}_{m,\zeta}=\hat{P}_{m,\zeta}\hat{H}%
,\;D_{H_{m,\zeta}}=\hat{P}_{m,\zeta}D_{H}, 
\end{equation*}

\begin{align*}
& \hat{H}\Psi_{m,\zeta}(\mathbf{u})=\hat{H}_{m,\zeta}\Psi_{m,\zeta }(\mathbf{%
u})=\check{H}_{m}\Psi_{m,\zeta}(\mathbf{u})= \\
& \,=\frac{1}{\sqrt{2\pi}}\mathrm{e}^{im\varphi_{u}}\frac{|R^{2}-r^{2}|}{%
R^{2}\sqrt{r}}\hat{h}_{m,\zeta}\psi_{m,\zeta}\left( r\right) ,
\end{align*}
where $\hat{h}_{m,\zeta}$ is a symmetric operator defined in the Hilbert
space $\mathfrak{h}_{m,\zeta}=L^{2.\zeta}(\mathbb{R}_{+})$, $L^{2.\zeta}(%
\mathbb{R}_{+})=\hat{P}_{\zeta}L^{2}(\mathbb{R}_{+})$, $\hat{P}_{\zeta
}=(1+\zeta\tilde{P})/2$:%
\begin{equation}
\hat{h}_{m,\zeta}:\left\{ 
\begin{array}{l}
D_{h_{m,\zeta}}=\mathcal{D}_{\zeta}(\mathbb{R}_{+}\backslash\{R\}) \\ 
\hat{h}_{m,\zeta}\psi_{m,\zeta}=\check{h}_{m}\psi_{m,\zeta},\;\forall
\psi_{m,\zeta}\in D_{h_{m,\zeta}},%
\end{array}
\right. ,   \label{Opso2.2.0}
\end{equation}
the differential operation $\check{h}_{m}$ is given by eq. (\ref{Opso2.1.1.1}%
) and $\mathcal{D}_{\zeta}(\mathbb{R}_{+}\backslash\{R\})=\hat{P}_{\zeta }%
\mathcal{D}(\mathbb{R}_{+}\backslash\{R\})$.

In what follows, the s.a. operators $\hat{f}$ which commute with the
operators $\hat{U}_{S}$ and $P$, we will call rotationally- and
parity-invariant. Such operators can be represented in the form 
\begin{equation*}
\hat{f}=\sideset{}{^{\,\lower1mm\hbox{$\oplus$}}} \sum_{m\in\mathbb{Z,,\zeta
=\pm}}\hat{f}_{m,\zeta},\;\hat{f}_{m,\zeta}=\hat{P}_{m,\zeta}\hat{f}, 
\end{equation*}
and $\hat{f}_{m,\zeta}$ are s.a. operators in sufspaces $\mathfrak{H}%
_{m,\zeta}$.

Let $\hat{h}_{\mathfrak{e}m,\zeta}$ is a s.a$.$operator associated with the
differential operation $\check{h}_{m}$ in the Hilbert space $\mathfrak{h}%
_{m,\zeta}$. Then the operator $\hat{H}_{\mathfrak{e}m,\zeta}$,%
\begin{equation}
\hat{H}_{\mathfrak{e}m,\zeta}\Psi_{m,\zeta}(\mathbf{u})=\frac{1}{\sqrt{2\pi}}%
\mathrm{e}^{im\varphi_{u}}\frac{|R^{2}-r^{2}|}{R^{2}\sqrt{r}}\hat{h}_{%
\mathfrak{e}m,\zeta}\psi_{m,\zeta}\left( r\right) ,\;\psi_{m,\zeta }\left(
r\right) \in D_{h_{\mathfrak{e}m,\zeta}},   \label{Opso2.2.1}
\end{equation}
is a s.a. operator associated with $\check{H}_{m}$ in the Hilbert space $%
\mathfrak{H}_{m,\zeta}$ and operator $\hat{H}_{\mathfrak{e}}$,%
\begin{equation}
\hat{H}_{\mathfrak{e}}=\sideset{}{^{\,\lower1mm\hbox{$\oplus$}}} \sum _{m\in%
\mathbb{Z,\zeta=\pm}}\hat{H}_{\mathfrak{e}m,\zeta},   \label{Opso2.2.2}
\end{equation}
is a s.a. operator in the Hilbert space $\mathfrak{H}$.

Conversely, let $\hat{H}_{\mathfrak{e}}$ be a rotationally- and
parity-invariant s.a. extension of $\hat{H}$. Then it has the form~(\ref%
{Opso2.2.2}), where $\hat{H}_{\mathfrak{e}m,\zeta}$ are s.a. operators in $%
\mathfrak{H}_{m,\zeta}$. The operator $\hat{H}_{\mathfrak{e}m,\zeta}$ acts
in subspace $\mathfrak{H}_{m,\zeta}$ by the rule (\ref{Opso2.2.1}) with some
operator $\hat{h}_{\mathfrak{e}m,\zeta}$ which is obviously a s.a. operator
associated with the symmetric operator $\hat{h}_{m,\zeta}$ in the Hilbert
space $\mathfrak{h}_{m,\zeta}$.

In what follows, we restrict ourselves to the consideration of the s.a.
operators $\hat{H}_{\mathfrak{e}}$ which are the rotationally- and
parity-invariant s.a. extension of $\hat{H}$. As it was explained above,
this means that $\hat{H}_{\mathfrak{e}}$ has a structure of eq. (\ref%
{Opso2.2.2}), acts by the rule (\ref{Opso2.2.1}) and $\hat{h}_{\mathfrak{e}%
m,\zeta}$ is an s.a. extension of the symmetric operator $\hat{h}_{m}$.

Thus, the problem of constructing a rotationally-invariant s.a. Hamiltonian $%
\hat{H}_{\mathfrak{e}}$ is reduced to constructing s.a. radial Hamiltonians $%
\hat{h}_{\mathfrak{e}m,\zeta}$.

\subsection{$|m|\geq1$}

\subsubsection{Useful solutions, $r<R$}

We need solutions of an equation%
\begin{equation}
(\check{h}_{Om}-W_{O})\psi_{Om}^{<}(r)=0,   \label{Opso2.3.1.1}
\end{equation}
where $\check{h}_{m}$ is given by eq. (\ref{Opso2.1.1.1}) and $W_{O}=W$ is
complex energy,%
\begin{equation*}
W=|W|e^{i\varphi_{W}},\;0\leq\varphi_{W}\leq\pi,\;\operatorname{Im}W\geq0. 
\end{equation*}

It is convenient for our aims first to consider solutions more general
equation%
\begin{align}
& \,(\check{h}_{Om,\delta}-W_{O})\psi_{m,\delta}(r)=0,\;\check{h}_{Om,\delta
}=\check{h}_{m,\delta}=\left. \check{h}_{m}\right| _{m\rightarrow m+\delta }=
\label{Opso2.3.1.2} \\
& \,=\frac{\sqrt{r}}{R^{2}-r^{2}}\check{H}_{m,\delta}\frac{R^{2}-r^{2}}{%
\sqrt{r}},\;\check{H}_{m,\delta}=\left. \check{H}_{m}\right| _{m\rightarrow
m+\delta},\;|\delta|<1;  \notag \\
& \,(\check{H}_{m,\delta}-W)\Psi_{m,\delta}=0.\;\Psi_{m,\delta}(r)=\frac{%
R^{2}-r^{2}}{R^{2}\sqrt{r}}\psi_{m,\delta}(r).   \label{Opso2.3.1.2a}
\end{align}

Introduce a new variable $x$,%
\begin{align*}
& x=\frac{4R^{2}r^{2}}{(R^{2}+r^{2})^{2}},\;r=R\frac{1-\sqrt{1-x}}{\sqrt{x}}%
,\;r\partial_{r}=2x\sqrt{1-x}\partial_{x}, \\
& R^{2}+r^{2}=2R^{2}\frac{1-\sqrt{1-x}}{x},\;R^{2}-r^{2}=2R^{2}\frac{\sqrt{%
1-x}(1-\sqrt{1-x})}{x}, \\
& V(r)=\frac{4(q-1)r^{2}}{(R^{2}+r^{2})^{2}}=\frac{(q-1)x}{R^{2}},\;\frac{%
(R^{2}-r^{2})^{2}}{R^{4}r^{2}}=\frac{4(1-x)}{R^{2}x}, \\
& \Delta_{BLr}=\frac{(R^{2}-r^{2})^{2}}{R^{4}}\frac{1}{r}\partial
_{r}r\partial_{r}=\frac{16(1-x)^{3/2}}{R^{2}}\partial_{x}x\sqrt{1-x}%
\partial_{x},
\end{align*}
and new function $\phi_{\xi_{\mu}\xi_{\nu}\delta}(x)$, 
\begin{align*}
& \psi_{m,\delta}(r)=\frac{R^{2}\sqrt{r}}{R^{2}-r^{2}}x^{\xi_{\mu}\mu
_{\delta}}(1-x)^{1/4+\xi_{\nu}\nu}\phi_{\xi_{\mu}\xi_{\nu}\delta}(x), \\
&
\Psi_{m,\delta}(r)=x^{\xi_{\mu}\mu_{\delta}}(1-x)^{1/4+\xi_{\nu}\nu}\phi_{%
\xi_{\mu}\xi_{\nu}\delta}(x), \\
& \mu_{\delta}=\frac{1}{2}|m+\delta|,\;\nu=\frac{1}{4}\sqrt{q-w}%
,\;w=R^{2}W,\;\xi_{\mu},\xi_{\nu}=\pm1,
\end{align*}
and $\sqrt{q-w}$ is understood as the principal value, real $q$ is
understood as the limit $\operatorname{Im}q\rightarrow-0$. Note that $\operatorname{Re}\nu>0$%
. $\operatorname{Im}\nu<0$ for $\operatorname{Im}W>0$.

Then we obtain%
\begin{align}
&
[x(1-x)\partial_{x}^{2}+(\gamma_{\xi_{\mu}\delta}-(1+\alpha_{\xi_{\mu}\xi_{%
\nu}\delta}+\beta_{\xi_{\mu}\xi_{\nu}\delta})x)\partial_{x}-\alpha
_{\xi_{\mu}\xi_{\nu}\delta}\beta_{\xi_{\mu}\xi_{\nu}\delta}]\phi_{\xi_{\mu}%
\xi_{\nu}\delta}(x)=0,  \label{Opso2.3.1.3} \\
&
\alpha_{\xi_{\mu}\xi_{\nu}\delta}=1/2+\xi_{\mu}\mu_{\delta}+\xi_{\nu}\nu+%
\sigma,\;\beta_{\xi_{\mu}\xi_{\nu}\delta}=1/2+\xi_{\mu}\mu_{\delta}+\xi_{%
\nu}\nu-\sigma,  \notag \\
& \gamma_{\xi_{\mu}\delta}=1+2\xi_{\mu}\mu_{\delta},\;\sigma=\left\{ 
\begin{array}{l}
\frac{1}{4}\sqrt{q},\;q\geq0 \\ 
i\varkappa/4,\;\varkappa=\sqrt{|q|},\;q<0%
\end{array}
\right. .  \notag
\end{align}

Eq. (\ref{Opso2.3.1.3}) is the equation for hypergeometric functions, in the
terms of which we can express solutions of eq. (\ref{Opso2.3.1.2}). We will
use the following solutions:%
\begin{align*}
& O_{1,m,\delta}^{<}(r;W)=\frac{R^{2}\sqrt{r}}{R^{2}-r^{2}}%
x^{\mu_{\delta}}(1-x)^{1/4+\nu}\mathcal{F}(\alpha_{1\delta},\beta_{1\delta};%
\gamma_{1\delta };x)= \\
& \,=Q_{1,m,\delta}(W)O_{3,m,\delta}^{<}(r;W)+Q_{2,m,\delta}(W)v_{m,\delta
}^{<}(r;W)= \\
& \,=\left. O_{1,m,\delta}^{<}(r;W)\right| _{\sigma\rightarrow-\sigma
}=\left. O_{1,m,\delta}^{<}(r;W)\right| _{\nu\rightarrow-\nu}, \\
& Q_{1,m,\delta}(W)=\frac{\Gamma(\gamma_{1\delta})\Gamma(-2\nu)}{%
\Gamma(\alpha_{4\delta})\Gamma(\beta_{4\delta})},\;Q_{2,m,\delta }(W)=\frac{%
\Gamma(\gamma_{1\delta})\Gamma(2\nu)}{\Gamma(\alpha_{1\delta
})\Gamma(\beta_{1\delta})}=\left. Q_{1,m,\delta}(W)\right| _{\nu
\rightarrow-\nu}, \\
& O_{2,m,\delta}^{<}(r;W)=\frac{1}{\Gamma(\gamma_{2\delta})}\frac{R^{2}\sqrt{%
r}}{R^{2}-r^{2}}x^{-\mu_{\delta}}(1-x)^{1/4+\nu}\mathcal{F}%
(\alpha_{2\delta},\beta_{2\delta};\gamma_{2\delta};x)= \\
& \,=\frac{1}{\Gamma(\gamma_{2\delta})}\left. O_{1,m,\delta}^{<}(r;W)\right|
_{\mu_{\delta}\rightarrow-\mu_{\delta}}=\left. O_{2,m,\delta
}^{<}(r;W)\right| _{\sigma\rightarrow-\sigma}=\left.
O_{2,m,\delta}^{<}(r;W)\right| _{\nu\rightarrow-\nu},
\end{align*}%
\begin{align*}
& (r;W)=\frac{R^{2}\sqrt{r}}{R^{2}-r^{2}}x^{\mu_{\delta}}(1-x)^{1/4+\nu }%
\mathcal{F}(\alpha_{1\delta},\beta_{1\delta};\gamma_{3};1-x), \\
& v_{m,\delta}^{<}(r;W)=\frac{R^{2}\sqrt{r}}{R^{2}-r^{2}}x^{\mu_{%
\delta}}(1-x)^{1/4-\nu}\mathcal{F}(\alpha_{4\delta},\beta_{4\delta};\gamma
_{4};1-x)=\left. O_{3,m,\delta}^{<}(r;W)\right| _{\nu\rightarrow-\nu},
\end{align*}%
\begin{align*}
& \alpha_{1\delta,2\delta}=1/2\pm\mu_{\delta}+\nu+\sigma,\;\beta
_{1\delta,2\delta}=1/2\pm\mu_{\delta}+\nu-\sigma, \\
&
\alpha_{4\delta}=1/2+\mu_{\delta}-\nu+\sigma,\;\beta_{4\delta}=1/2+\mu_{%
\delta}-\nu-\sigma, \\
& \gamma_{1\delta,2\delta}=1\pm2\mu_{\delta},\;\gamma_{3,4}=1\pm2\nu.
\end{align*}
Note that $O_{1,m,\delta}^{<}(r;W)$ and $O_{2,m,\delta}^{<}(r;W)$ are
real-entire in $W$ solutions of eq. (\ref{Opso2.3.1.2}).

Represent $O_{3,m,\delta}^{<}$ in the form%
\begin{align*}
&
O_{3,m,\delta}^{<}(r;W)=B_{m,\delta}(W)O_{1,m,\delta}^{<}(r;W)+Q_{3,m,%
\delta}(W)O_{4,m,\delta}^{<}(r;W), \\
& O_{4,m,\delta}^{<}(r;W)=\Gamma(\gamma_{2\delta})\left[ O_{2,m,%
\delta}^{<}(r;W)-A_{m,\delta}(W)O_{1,m,\delta}^{<}(r;W)\right] , \\
& Q_{3,m,\delta}(W)=\frac{\Gamma(\gamma_{3})\Gamma(2\mu_{\delta})}{%
\Gamma(\alpha_{1\delta})\Gamma(\beta_{1\delta})},\;A_{m,\delta}(W)=\frac{%
\Gamma(\alpha_{1})\Gamma(\beta_{1})}{\Gamma(\alpha_{2})\Gamma
(\beta_{2})\Gamma(\gamma_{1\delta})}, \\
& B_{m,\delta}(W)=\frac{\Gamma(\gamma_{3})\Gamma(-2\mu_{\delta})}{%
\Gamma(\alpha_{2\delta})\Gamma(\beta_{2\delta})}+\frac{\Gamma(\gamma
_{3})\Gamma(2\mu_{\delta})\Gamma(\gamma_{2\delta})}{\Gamma(\alpha_{1\delta
})\Gamma(\beta_{1\delta})}A_{m,\delta}(W)= \\
& \,=\frac{\Gamma(\gamma_{3})\Gamma(-2\mu_{\delta})}{\Gamma(\alpha_{2})%
\Gamma(\beta_{2})}\left[ \frac{\Gamma(\alpha_{2})\Gamma(\beta_{2})}{%
\Gamma(\alpha_{2\delta})\Gamma(\beta_{2\delta})}-\frac{\Gamma(\alpha
_{1})\Gamma(\beta_{1})}{\Gamma(\alpha_{1\delta})\Gamma(\beta_{1\delta})}%
\right] , \\
& \alpha_{1,2}=1/2\pm\mu+\nu+\sigma,\;\beta_{1,2}=1/2\pm\mu+\nu-\sigma
,\;\mu=|m|/2,
\end{align*}
where we used eq.9.131.2 in \cite{Grad-Ryzh}. The function $A_{m,\delta}(W)$
can be represented in the form%
\begin{align*}
& A_{m,\delta}(W)=\frac{1}{\Gamma(\gamma_{1\delta})}\mathcal{A}_{m}(W),\; \\
& \mathcal{A}_{m}(W)=\frac{\Gamma(\alpha_{1})\Gamma(\beta_{1})}{\Gamma
(\alpha_{2})\Gamma(\beta_{2})}=\prod_{k=1}^{|m|}(\alpha_{1}-k)(\beta_{1}-k)=
\\
& \,=\frac{1}{\Gamma(\gamma_{1\delta})}\left\{ 
\begin{array}{c}
(v^{2}-\sigma^{2})I_{p}(\nu,\sigma^{2})I_{p}(-\nu,\sigma^{2}),\;|m|=2p-1 \\ 
J_{p}(\nu,\sigma^{2})J_{p}(-\nu,\sigma^{2}),\;|m|=2p%
\end{array}
\right. , \\
& p=1,2,...,\;I_{1}(z)=1,\;I_{p}(z,\sigma^{2})=\prod_{k=1}^{p-1}\left[
(k+z)^{2}-\sigma^{2}\right] ,\;p\geq2, \\
& J_{p}(z)=\prod_{k=0}^{p-1}\left[ (k+1/2+z)^{2}-\sigma^{2}\right] ,
\end{align*}
i.e., $A_{m,\delta}(W)$ is a polynomial in $W$ with real coefficients, and,
therefore, $O_{4,m,\delta}^{<}(r;W)$ is real-entire in $W$.

We obtain the solution of eq.(\ref{Opso2.3.1.1}) as the limit $\delta
\rightarrow0$ of the solution of eq. (\ref{Opso2.3.1.2}):%
\begin{align*}
& O_{1,m}^{<}(r;W)=O_{1,m,0}^{<}(r;W)=\frac{R^{2}\sqrt{r}}{R^{2}-r^{2}}%
x^{\mu}(1-x)^{1/4+\nu}\mathcal{F}(\alpha_{1},\beta_{1};\gamma_{1};x)= \\
& (\mathrm{for}=\operatorname{Im}%
W>0)=Q_{1}(W)O_{3,m}^{<}(r;W)+Q_{2}(W)v_{m}^{<}(r;W),\;\gamma_{1}=1+2\mu, \\
& Q_{1}(W)=Q_{10}(W)=\frac{\Gamma(\gamma_{1})\Gamma(-2\nu)}{\Gamma(\alpha
_{4})\Gamma(\beta_{4})},\;Q_{2}(W)=Q_{20}(W)=\frac{\Gamma(\gamma_{1})%
\Gamma(2\nu)}{\Gamma(\alpha_{1})\Gamma(\beta_{1})}, \\
& v_{m}^{<}(r;W)=v_{m,0}^{<}(r;W)=\frac{R^{2}\sqrt{r}}{R^{2}-r^{2}}x^{\mu
}(1-x)^{1/4-\nu}\mathcal{F}(\alpha_{4},\beta_{4};\gamma_{4};1-x), \\
& O_{4,m}^{<}(r;W)=\lim_{\delta\rightarrow0}O_{4,m,\delta}^{<}(r;W), \\
& O_{3,m}^{<}(r;W)=\frac{R^{2}\sqrt{r}}{R^{2}-r^{2}}x^{\mu}(1-x)^{1/4+\nu }%
\mathcal{F}(\alpha_{1},\beta_{1};\gamma_{3};1-x)= \\
& =B_{m}(W)O_{1,m}^{<}(r;W)+C_{m}(W)O_{4,m}^{<}(r;W),\;C_{m}(W)=\frac{\Gamma
(\gamma_{3})\Gamma(|m|)}{\Gamma(\alpha_{1})\Gamma(\beta_{1})}, \\
& B_{m}(W)=B_{m,0}(W)=\frac{(-1)^{|m|+1}\Gamma(\gamma_{3})}{2\Gamma
(\alpha_{2})\Gamma(\beta_{2})\Gamma(\gamma_{1})}\left[ \psi(\alpha_{1})+%
\psi(\alpha_{2})+\psi(\beta_{1})+\psi(\beta_{2})\right] ,
\end{align*}
where we used relations.%
\begin{align*}
& |m+\delta|=|m|+\delta_{m},\;\delta_{m}=\delta\mathrm{sign}m,\;\Gamma
(-|m+\delta|)=\frac{(-1)^{|m|+1}}{\Gamma(\gamma_{1})}\frac{1}{\delta_{m}}, \\
& O_{2,m,0}^{<}(r;W)=\frac{\Gamma(\alpha_{1})\Gamma(\beta_{1})}{\Gamma
(\alpha_{2})\Gamma(\beta_{2})\Gamma(\gamma_{1})}O_{1,m}^{<}(r;W) \\
& \frac{\Gamma(\alpha_{1})}{\Gamma(\alpha_{1,\delta})}=1-\frac{1}{2}%
\delta_{m}\psi(\alpha_{1}),\;\frac{\Gamma(\alpha_{2})}{\Gamma(\alpha
_{2,\delta})}=1+\frac{1}{2}\delta_{m}\psi(\alpha_{2}), \\
& \frac{\Gamma(\beta_{1})}{\Gamma(\beta_{1,\delta})}=1-\frac{1}{2}\delta
_{m}\psi(\beta_{1}),\;\frac{\Gamma(\beta_{2})}{\Gamma(\beta_{2,\delta})}=1+%
\frac{1}{2}\delta_{m}\psi(\beta_{2}).
\end{align*}

Note that $O_{1,m}^{<}(r;W)$ and $O_{4,m}^{<}(r;W)$ are real-entire in $W$.

\subsubsection{Asymptotics, $r\rightarrow0$ ($x\rightarrow0$)}

We have%
\begin{align*}
& x=\frac{4r^{2}}{R^{2}}(1+O(r^{2})),\;p_{0}(r)=1+O(r^{2}), \\
& O_{1,m}^{<}(r;W)=(2/R)^{|m|}r^{1/2+|m|}(1+O(r^{2})), \\
& O_{4,m}^{<}(r;W)=(R/2)^{|m|}r^{1/2-|m|}\left( 1+\left\{ 
\begin{array}{c}
O(r^{2}),\;|m|\geq2 \\ 
O(r^{2}\ln r),\;|m|=1%
\end{array}
\right. \right) ,
\end{align*}

\begin{align*}
& O_{3,m}^{<}(r;W)=.\frac{\Gamma(\gamma_{3})\Gamma(2\mu)}{\Gamma(\alpha
_{1})\Gamma(\beta_{1})}(R/2)^{|m|}r^{1/2-|m|}\left( 1+\left\{ 
\begin{array}{c}
O(r^{2}),\;|m|\geq2 \\ 
O(r^{2}\ln r),\;|m|=1%
\end{array}
\right. \right) , \\
& \operatorname{Im}W>0\;\mathrm{or}\;W=0.
\end{align*}

\subsubsection{Asymptotics, $\Delta=R-r\rightarrow0$ ($\protect\delta%
=1-x\rightarrow0$)}

We have%
\begin{align*}
& \delta=\frac{\Delta^{2}}{R^{2}}(1+O(\Delta)),\;p_{0}(r)=\frac{4\Delta^{2}}{%
R^{2}}(1+O(\Delta)), \\
& O_{3,m}^{<}(r;W)=\frac{1}{2}R^{1-2\nu}\Delta^{-1/2+2\nu}(1+O(\Delta)),
\end{align*}

\begin{align*}
& O_{1,m}^{<}(r;W)=\frac{\Gamma(\gamma_{1})\Gamma(2\nu)}{2\Gamma(\alpha
_{1})\Gamma(\beta_{1})}R^{1+2\nu}\Delta^{-1/2-2\nu}(1+O(\Delta)), \\
& \operatorname{Im}W>0\;\mathrm{or}\;W=0.
\end{align*}

\subsubsection{Wronskian}

The Wronskian $\mathrm{Wr}(U,V)$ of two functions $U(r)$ and $V(r)$ is equal
to%
\begin{equation*}
\mathrm{Wr}(U,V)=p_{0}(r)[U(r)\partial_{r}V(r)-V(r)\partial_{r}U(r)]. 
\end{equation*}

We have

\begin{equation*}
\mathrm{Wr}(O_{1,m}^{<},O_{3,m}^{<})=-2\frac{\Gamma(\gamma_{1})\Gamma
(\gamma_{3})}{\Gamma(\alpha_{1})\Gamma(\beta_{1})}=-%
\omega_{m}(W)=-2|m|C_{m}(W). 
\end{equation*}

\subsubsection{Useful solutions, $r>R$}

We need solutions of eq. (\ref{Opso2.3.1.1}) for $r>R$. As two independent
solutions, we choose the following functions:%
\begin{align}
O_{1,m}^{>}(r;W) & =\frac{R}{r}O_{1,m}^{<}(R^{2}/r;W),  \label{Opso2.3.2.1a}
\\
O_{3,m}^{>}(r;W) & =\frac{R}{r}O_{3,m}^{<}(R^{2}/r;W)   \label{Opso2.3.2.1b}
\end{align}

According to eq. (\ref{Opso2.1.1.2.4b}), the functions (\ref{Opso2.3.2.1a})
and (\ref{Opso2.3.2.1b}) satisfy really eq. (\ref{Opso2.3.1.1}) for $r>R$.

\subsubsection{Asymptotics, $\Delta=r-R$ $\rightarrow0$}

We have%
\begin{equation*}
O_{3,m}^{>}(r;W)=\frac{1}{2}R^{1-2\nu}\Delta^{-1/2+2\nu}(1+O(\Delta
)),\;\Delta=R-r, 
\end{equation*}%
\begin{align*}
& O_{1,m}^{<}(r;W)=\frac{\Gamma(\gamma_{1})\Gamma(2\nu)}{2\Gamma(\alpha
_{1})\Gamma(\beta_{1})}R^{1+2\nu}\Delta^{-1/2-2\nu}(1+O(\Delta)), \\
& \operatorname{Im}W>0\;\mathrm{or}\;W=0.
\end{align*}

\subsubsection{Asymptotics, $r\rightarrow\infty$}

We have%
\begin{equation*}
O_{1,m}^{>}(r;W)=2^{|m|}R^{2+|m|}r^{-3/2-|m|}(1+O(r^{-2})), 
\end{equation*}

\begin{align*}
& O_{3,m}^{>}(r;W)=\frac{\Gamma(\gamma_{3})\Gamma(2\mu)}{\Gamma(\alpha
_{1})\Gamma(\beta_{1})}2^{-|m|}R^{2-|m|}r^{-3/2+|m|}\left( 1+\left\{ 
\begin{array}{c}
O(r^{-2}),\;|m|\geq2 \\ 
O(r^{-2}\ln r),\;|m|=1%
\end{array}
\right. \right) , \\
& \operatorname{Im}W>0\;\mathrm{or}\;W=0.
\end{align*}

We see that eq. (\ref{Opso2.3.1.1}) has no s.-integrable silutions for $%
\operatorname{Im}W>0$. This means that the deficiency indices of the symmetric
operator (see below) are equal to $(0,0)$.

\subsubsection{Symmetric operator $\hat{h}_{m,\protect\zeta}$}

For given a differential operation $\check{h}_{m}$, the symmetric operator $%
\hat{h}_{m,\zeta}$ is given by eq. \ref{Opso2.2.0}.

\subsubsection{Isometry}

We remind that the functions $\psi_{m,\zeta}(r)\in\mathfrak{h}%
_{m,\zeta}=L^{2}(\mathbb{R}_{+})$ have the structure, see eqs. (\ref%
{Opso2.1.2.2}), (\ref{Opso2.1.1.2.3}) and (\ref{Opso2.1.1.2.4a}), 
\begin{equation*}
\psi_{m,\zeta}(r)=\left( 
\begin{array}{c}
\psi_{m,\zeta}^{>}(\rho) \\ 
\psi_{m}^{<}(\xi)%
\end{array}
\right) ,\;\psi_{m,\zeta}^{>}(\rho)=\zeta\frac{\rho_{P}}{R}%
\psi_{m}^{<}(\rho_{P})=\zeta P_{\rho\xi}\frac{\xi}{R}\psi_{m}^{<}(\xi), 
\end{equation*}%
\begin{align*}
& \hat{h}_{m,\zeta}(r)\psi_{m,\zeta}(r)=\check{h}_{m}(r)\psi_{m,\zeta
}(r)=\left( 
\begin{array}{c}
\check{h}_{m}^{>}(\rho)\psi_{m,\zeta}^{>}(\rho) \\ 
\check{h}_{m}^{<}(\xi)\psi_{m}^{<}(\xi)%
\end{array}
\right) = \\
& \,=\left( 
\begin{array}{c}
\zeta P_{\rho\xi}\frac{\xi}{R}\check{h}_{m}^{<}(\xi)\psi_{m}^{<}(\xi) \\ 
\check{h}_{m}^{<}(\xi)\psi_{m}^{<}(\xi)%
\end{array}
\right) ,\;\psi_{m}^{<}(\xi)\in\mathcal{D}(0,R).
\end{align*}

Let us untroduce an isometric map $T$ ($\zeta$ is fixed),%
\begin{equation*}
\psi_{m,\zeta}(r)\in\mathfrak{h}_{m,\zeta}=L^{2,\zeta}(\mathbb{R}_{+})%
\overset{T}{\leftrightarrow}\psi_{T,m,\zeta}(\xi)=\sqrt{2}%
\psi_{m}^{<}(\xi)\in\mathfrak{h}_{m}^{(1/2)}=L^{2}(0,R), 
\end{equation*}%
\begin{equation*}
D_{h_{m,\zeta}}=\mathcal{D}_{\zeta}(\mathbb{R}_{+}\backslash\{R\})\overset{T}%
{\leftrightarrow}D_{h_{m}^{(1/2)}}=\mathcal{D}(0,R) 
\end{equation*}%
\begin{align*}
& \hat{h}_{m,\zeta}(r)\psi_{m,\zeta}(r)=\left( 
\begin{array}{c}
\zeta P_{\rho\xi}\frac{\xi}{R}\check{h}_{m}^{<}(\xi)\psi_{m,\zeta}^{<}(\xi)
\\ 
\check{h}_{m}^{<}(\xi)\psi_{m,\zeta}^{<}(\xi)%
\end{array}
\right) \overset{T}{\leftrightarrow}\hat{h}_{m,\zeta}^{(1/2)}(\xi
)\psi_{T,m,\zeta}(\xi)= \\
& \,=\sqrt{2}\hat{h}_{m}^{<}(\xi)\psi_{m}^{<}(\xi),\;\hat{h}_{m,\zeta
}^{(1/2)}=\hat{h}_{m}^{(1/2)}=\hat{h}_{m}^{<}(\xi),\;\psi_{m,\zeta}\in
D_{h_{m,\zeta}},\;\psi_{T,m,\zeta}\in D_{h_{m,\zeta}^{(1/2)}}.
\end{align*}

Let $\hat{h}_{\mathfrak{e}m,\zeta}^{(1/2)}$ be an s.a. extension of the
symmetric operator $\hat{h}_{m}^{(1/2)}$. Then the corresponding s.a.
operator $\hat{H}_{\mathfrak{e}m,\zeta}$, an s.a. extension of symmetric
operator $\hat{H}_{m,\zeta}$, can be reconstructed by the rule (\ref%
{Opso2.2.1}) with $\hat{h}_{\mathfrak{e}m,\zeta}$ be given by the expression
(see eq. (\ref{Opso2.1.1.2.3})),%
\begin{equation*}
\hat{h}_{\mathfrak{e}m,\zeta}(r)=.\left( 
\begin{array}{cc}
\hat{h}_{\mathfrak{e}m,\zeta}^{>}(\rho) & 0 \\ 
0 & \hat{h}_{\mathfrak{e}m,\zeta}^{(1/2)}(\xi)%
\end{array}
\right) ,\;\hat{h}_{\mathfrak{e}m,\zeta}^{>}(\rho)=\frac{R}{\rho }\hat{h}_{%
\mathfrak{e}m,\zeta}^{(1/2)}(\rho_{P})\frac{\rho}{R}. 
\end{equation*}

Pass to the construction of s.a. operators $\hat{h}_{\mathfrak{e}m,\zeta
}^{(1/2)}$ which are s.a. extensions of the symmetric operators $\hat{h}%
_{m,\zeta}^{(1/2)}=\hat{h}_{m,\zeta}^{<}(\xi)$ associated with the
differential operations $\check{h}_{m}^{<}(\xi)$ in the Hilbert space $%
\mathfrak{h}_{m,\zeta}^{(1/2)}=L^{2}(0,R)$.

\subsubsection{Adjoint operator $\hat{h}_{m}^{+}=\hat{h}_{m}^{\ast}$}

We will omit the superscript $(1/2)$.and will write $r$ for $\xi$, $0\leq
r\leq R$.

It is easy to prove by standard way that the adjoint operator $\hat{h}%
_{m}^{+}$ coincides with the operator $\hat{h}_{m}^{\ast}$,

\begin{equation*}
\hat{h}_{m}^{+}:\left\{ 
\begin{array}{l}
D_{h_{m}^{+}}=D_{\check{h}_{m}}^{\ast}(0,R)=\{\psi_{\ast},\psi_{\ast}^{%
\prime }\;\mathrm{are\;a.c.\;in}\mathcal{\;}(0,R),\;\psi_{\ast},\hat{h}%
_{m}^{+}\psi_{\ast}\in L^{2}(0,R)\} \\ 
\hat{h}_{m}^{+}\psi_{\ast}(r)=\check{h}_{m}\psi_{\ast}(r),\;\forall\psi_{%
\ast }\in D_{h_{m}^{+}}%
\end{array}
\right. . 
\end{equation*}

\subsubsection{Asymptotics}

Because $\check{h}_{m}\psi_{\ast}\in L^{2}(0,R)$, we have%
\begin{equation*}
\check{h}_{m}\psi_{\ast}(r)=\eta(r),\;\eta\in L^{2}(0,R), 
\end{equation*}
and we can represent $\psi_{\ast}$ in the form%
\begin{align*}
\psi_{\ast}(r) & =c_{1}O_{1,m}^{<}(r;0)+c_{2}O_{3,m}^{<}(r;0)+I(r), \\
\psi_{\ast}^{\prime}(r) &
=c_{1}\partial_{r}O_{1,m}^{<}(r;0)+c_{2}\partial_{r}O_{3,m}^{<}(r;0)+I^{%
\prime}(r),
\end{align*}
where%
\begin{align*}
I(r) & =\frac{O_{1,m}^{<}(r;0)}{\omega_{m}(0)}\int_{r}^{R}O_{3,m}^{<}(y;0)%
\eta(y)dy+\frac{O_{3,m}^{<}(r;0)}{\omega_{m}(0)}\int_{0}^{r}O_{1,m}^{<}(y;0)%
\eta(y)dy, \\
I^{\prime}(r) & =\frac{\partial_{r}O_{1,m}^{<}(r;0)}{\omega_{m}(0)}\int
_{r}^{R}O_{3,m}^{<}(y;0)\eta(y)dy+\frac{\partial_{r}O_{3,m}^{<}(r;0)}{%
\omega_{m}(0)}\int_{0}^{r}O_{1,m}^{<}(y;0)\eta(y)dy.
\end{align*}

I) $r\rightarrow0$

We obtain with the help of the Cauchy-Bunyakovskii inequality
(CB-inequality):%
\begin{equation*}
I(r)=\left\{ 
\begin{array}{c}
O(r^{3/2}),\;|m|\geq2 \\ 
O(r^{3/2}\sqrt{\ln r}),\;|m|=1%
\end{array}
\right. ,\;I^{\prime}(r)=\left\{ 
\begin{array}{c}
O(r^{1/2}),\;|m|\geq2 \\ 
O(r^{1/2}\sqrt{\ln r}),\;|m|=1%
\end{array}
\right. , 
\end{equation*}
such that we have%
\begin{align*}
& \psi_{\ast}(r)=c_{2}cr^{1/2-|m|}\left( 1+\left\{ 
\begin{array}{c}
O(r^{2}),\;|m|\geq2 \\ 
O(r^{2}\ln r),\;|m|=1%
\end{array}
\right. \right) + \\
& \,+\left\{ 
\begin{array}{c}
O(r^{3/2}),\;|m|\geq2 \\ 
O(r^{3/2}\sqrt{\ln r}),\;|m|=1%
\end{array}
\right. ,\;c=(R/2)^{|m|}\frac{\Gamma(\gamma_{3})\Gamma(2\mu)}{\Gamma
(\alpha_{1})\Gamma(\beta_{1})}.
\end{align*}
The condition $\psi_{\ast}\in L^{2}(0,R)$ gives $c_{2}=0$, such that we find
finally%
\begin{align}
& \psi_{\ast}(r)=\left\{ 
\begin{array}{c}
O(r^{3/2}),\;|m|\geq2 \\ 
O(r^{3/2}\sqrt{\ln r}),\;|m|=1%
\end{array}
\right. ,\;\psi_{\ast}^{\prime}(r)=\left\{ 
\begin{array}{c}
O(r^{1/2}),\;|m|\geq2 \\ 
O(r^{1/2}\sqrt{\ln r}),\;|m|=1%
\end{array}
\right. ,  \notag \\
& \,[\psi_{\ast},\chi_{\ast}]_{0}=0,\;\forall\psi_{\ast},\chi_{\ast}\in
D_{h_{m}^{+}}.   \label{Opso2.3.5.1.1}
\end{align}

II) $r\rightarrow R$

In this case, we prove that $[\psi_{\ast},\chi_{\ast}]^{R}=0$, $\forall
\psi_{\ast},\chi_{\ast}\in D_{h_{m}^{+}}$. Indeed, consider the Hilbert
space $\mathfrak{h}_{c,m}=L^{2}(c,R)$, $c$ is an interior point of the
interval $(0,R)$. and an symmetric operator $\hat{h}_{c,m}$, $D_{h_{c,m}}=%
\mathcal{D}(c,R)$, acting as $\check{h}_{m}$. We choose the functions $%
O_{1,m}^{<}(r;W)$ and $O_{3,m}^{<}(r;W)$ as the independent solutions of eq.
(\ref{Opso2.3.1.1}) for $\operatorname{Im}W>0$. The left end $c$ of the interval $%
(0,R)$ is regular and both solutions $O_{1,m}^{<}$ and $O_{3,m}^{<}$ are
s.-interable on the end $c$. The right end $R$ is singular. On the right end 
$R$, the solution $O_{3,m}^{<}$ is s.-integrable, but $O_{1,m}^{<}$ is not.
Thus, there is only one s. integrable solution of eq. (\ref{Opso2.3.1.1}) on
the interval $(c,R)$ for $\operatorname{Im}W>0$ and the deficient indexes of the
symmetric operator $\hat{h}_{c,m}$ are equal to $(1,1)$. In this case,
according to [\cite{Naima}, Lemma on the page 213], we have $%
[\psi_{\ast},\chi_{\ast }]^{R}=0$, $\forall\psi_{\ast},\chi_{\ast}\in
D_{h_{c,m}^{+}}$. Because the restriction $\psi_{c\ast}$ on the interval $%
(c,R)$ of any function $\psi _{\ast}\in D_{h_{m}^{+}}$ belongs to $%
D_{h_{c,m}^{+}}$, $\psi_{c\ast}\in D_{h_{c,m}^{+}}$. $\forall\psi_{\ast}\in
D_{h_{m}^{+}}$, we obtain that $[\psi_{\ast},\chi_{\ast}]^{R}=0$, $%
\forall\psi_{\ast},\chi_{\ast}\in D_{h_{m}^{+}}$.

\subsubsection{Self-adjoint hamiltonian $\hat{h}_{\mathfrak{e}m,\protect\zeta%
}$}

Because $\omega_{h_{m}^{+}}(\chi_{\ast},\psi_{\ast})=\Delta_{h_{m}^{+}}(%
\psi_{\ast})=0$ (and also because $O_{1,m}^{<}(r;W)$ and $O_{3,r}^{<}(u;W)$
and any their linear combinations are not s.-integrable on the interval $%
(0,R)$ for $\operatorname{Im}W\neq0$), the deficiency indices of initial symmetric
operator $\hat{h}_{m}$ are zero, which means that $\hat{h}_{\mathfrak{e}%
m,\zeta}=\hat{h}_{\mathfrak{e}m}=\hat{h}_{m}^{+}$ is a unique s.a. extension
of the initial symmetric operator $\hat{h}_{m}$:%
\begin{equation*}
\hat{h}_{\mathfrak{e}m}:\left\{ 
\begin{array}{l}
D_{h_{\mathfrak{e}m}}=D_{\check{h}_{m}}^{\ast}(0,R) \\ 
\hat{h}_{\mathfrak{e}m}\psi_{\ast}(r)=\check{h}_{m}\psi_{\ast}(r),\;\forall
\psi_{\ast}\in D_{h_{\mathfrak{e}m}}%
\end{array}
\right. . 
\end{equation*}

\subsubsection{The guiding functional $\Phi(\protect\xi;W)$}

As a guiding functional ( see \cite{{AkhGlaz},{Naima}})$\Phi(\xi;W)$ we
choose%
\begin{align}
& \Phi(\xi;W)=\int_{0}^{R}O_{1,m}^{<}(r;W)\xi(r)dr,\;\xi\in\mathbb{D}%
=D_{r}(0,R)\cap D_{h_{\mathfrak{e}m}}.  \label{Opso2.3.7.1} \\
& D_{r}(0,R)=\{\xi(u):\;\mathrm{supp}\xi\subseteq\lbrack0,\beta_{\xi
}],\;\beta_{\xi}<R\}.  \notag
\end{align}

The guiding functional $\Phi(\xi;W)$ is simple. It has, obviously, the
properties 1) and 3) and we should prove the properties 2) only. (see \cite%
{{Naima}}, pages 245-246). Let $\Phi(\xi_{0};E_{0})=0$, $\xi_{0}\in\mathbb{D}
$, $E_{0}\in\mathbb{R}$. As a solution $\psi(r)$ of equation 
\begin{equation*}
(\check{h}_{m}-E_{0})\psi(r)=\xi_{0}(r), 
\end{equation*}
we choose 
\begin{equation*}
\psi(r)=O_{1,m}^{<}(r;E_{0})\int_{r}^{R}U(r)\xi_{0}(r)dr+U(r)%
\int_{0}^{r}O_{1,m}^{<}(r;E_{0})\xi_{0}(r)dr, 
\end{equation*}
where $U(r)$ is any solution of eq. $(\check{h}_{m}-E_{0})U(r)=0$ satisfying
the condition $\mathrm{Wr}(O_{1,m}^{<},U)=-1$. Because $\xi_{0}\in D_{r}$,
the function $\psi(r)$ is well determined. Because $\xi_{0}\in D_{r}$ and $%
\int_{0}^{r}O_{1,m}^{<}(r;E_{0})\xi_{0}(r)=0$ for $r>\beta_{\xi_{0}}$, we
have $\psi(r)=0$ for $u>\beta_{\xi_{0}}$. Using the CB-inequality we show
that $\psi(r)$ satisfies the boundary condition (\ref{Opso2.3.5.1.1}), that
is, $\psi\in\mathbb{D}$. Thus, the guiding functional $\Phi(\xi;W)$ is
simple and the spectrum of $\hat{h}_{\mathfrak{e}m}$ is simple.

\subsubsection{Green function $G_{m}(r,y;W)$, spectral function $\protect%
\sigma_{m}(E)$}

We find the Green function $G_{m}(r,y;W)$ as the kernel of the integral
representation 
\begin{equation*}
\psi(r)=\int_{0}^{R}G_{m}(r,y;W)\eta(y)dy,\;\eta\in L^{2}(0,R), 
\end{equation*}
of unique solution of an equation%
\begin{equation}
(\hat{h}_{\mathfrak{e}m}-W)\psi(r)=\eta(r),\;\operatorname{Im}W>0, 
\label{Opso2.3.8.1}
\end{equation}
for $\psi\in D_{h_{\mathfrak{e}m}}$. General solution of eq. (\ref%
{Opso2.3.8.1}) can be represented in the form%
\begin{align*}
\psi(r) & =a_{1}O_{1,m}^{<}(r;W)+a_{3}O_{3,m}^{<}(r;W)+I(r), \\
I(r) & =\frac{O_{1,m}^{<}(r;W)}{\omega_{m}(W)}\int_{r}^{R}O_{3,m}^{<}(y;W)%
\eta(y)dy+\frac{O_{3,m}^{<}(r;W)}{\omega_{m}(W)}\int_{0}^{r}O_{1,m}^{<}(y;W)%
\eta(y)dy, \\
I(r) & =\left\{ 
\begin{array}{c}
O(r^{3/2}),\;|m|\geq2 \\ 
O(r^{3/2}\sqrt{\ln r}),\;|m|=1%
\end{array}
\right. ,\;r\rightarrow0,\;I(r)=O\left( \Delta^{-1/2}\right) ,\;r\rightarrow
R.
\end{align*}
A condition $\psi\in L^{2}(0,R)$ gives $a_{1}=a_{3}=0$, such that we find%
\begin{align}
& G_{m}(r,y;W)=\frac{1}{\omega_{m}(W)}\left\{ 
\begin{array}{l}
O_{3,m}^{<}(r;W)O_{1,m}^{<}(y;W),\;r>y \\ 
O_{1,m}^{<}(r;W)O_{3,m}^{<}(y;W).\;r<y%
\end{array}
\right. =  \notag \\
& \,=\pi\Omega_{m}(W)O_{1,m}^{<}(r;W)O_{1,m}^{<}(y;W)+\frac{1}{2|m|}\left\{ 
\begin{array}{c}
O_{4,m}^{<}(r;W)O_{1,m}^{<}(y;W),\;r>y \\ 
O_{1,m}^{<}(r;W)O_{4,m}^{<}(y;W),\;r<y%
\end{array}
\right. ,  \label{Opso2.3.8.2} \\
& \Omega_{m}(W)\equiv\frac{B_{m}(W)}{\pi\omega_{m}(W)}=\frac{%
(-1)^{|m|+1}[\psi(\alpha_{1})+\psi(\alpha_{2})+\psi(\beta_{1})+\psi(%
\beta_{2})]\mathcal{A}_{m}(W)}{4\pi\Gamma^{2}(\gamma_{1})}.  \notag
\end{align}
Note that the last term in the r.h.s. of eq. (\ref{Opso2.3.8.2}) is real for 
$W=E$. From the relation%
\begin{equation*}
\lbrack O_{1,m}^{<}(r_{0};E)]^{2}\sigma_{m}^{\prime}(E)=\frac{1}{\pi }\operatorname{%
Im}G_{m}(r_{0}-0,r_{0}+0;E+i0), 
\end{equation*}
where $f(E+i0)\equiv\lim_{\varepsilon\rightarrow+0}f(E+i\varepsilon)$, $%
\forall f(W)$, we find%
\begin{equation}
\sigma_{m}^{\prime}(E)=\operatorname{Im}\Omega_{m}(E+i0).   \label{Opso2.3.8.3}
\end{equation}

Consider $\Omega_{m}(W)$ in more details.

Using relations%
\begin{align*}
& \psi(\alpha_{2})=\psi(\alpha_{1})-T_{(\alpha)},\;T_{(\alpha)}=\sum
_{k=1}^{|m|}\frac{1}{\alpha_{1}-k}= \\
& \,=\sum_{k=1}^{|m|}\frac{1}{\left( \frac{1}{2}+\frac{|m|}{2}+\nu-k\right)
+\sigma}=\sum_{k=1}^{|m|}\frac{1}{\nu+\left( \frac{1}{2}+\frac{|m|}{2}%
+\sigma-k\right) },
\end{align*}%
\begin{align*}
& \psi(\beta_{2})=\psi(\beta_{1})-T_{(\beta)},\;T_{(\beta)}=\sum_{k=1}^{|m|}%
\frac{1}{\beta_{1}-k}= \\
& \,=\sum_{k=1}^{|m|}\frac{1}{\left( \frac{1}{2}+\frac{|m|}{2}+\nu-k\right)
-\sigma}=\sum_{k=1}^{|m|}\frac{1}{\nu-\left( \frac{1}{2}+\frac{|m|}{2}%
+\sigma-k\right) },
\end{align*}%
\begin{align*}
& T_{(\alpha,\beta)}(W)=T_{(\alpha)}+T_{(\beta)}=\sum_{k=1}^{|m|}\frac{%
1+|m|+2\nu-2k}{\left( \frac{1}{2}+\frac{|m|}{2}+\nu-k\right) ^{2}-\sigma^{2}}%
= \\
& \,=2\nu\sum_{k=1}^{|m|}\frac{1}{\nu^{2}-\left( \frac{1}{2}+\frac{|m|}{2}%
+\sigma-k\right) ^{2}}\Longrightarrow \\
& \,\Longrightarrow T_{(\alpha,\beta)}(W)=\frac{2\nu\mathcal{B}_{m}(W)}{%
\mathcal{A}_{m}(W)},
\end{align*}
where $\mathcal{B}_{m}(W)$ is an polinomoal in $\nu$ and $\sigma$, even in
both $\nu$ and $\sigma$, and therefore, is a real-entire polynomial in $W$,
we can represent $\Omega_{m}(W)$ in the form%
\begin{align*}
& \Omega_{m}(W)=\Omega_{1m}(W)+\Omega_{2m}(W), \\
& \Omega_{1m}(W)=\frac{(-1)^{|m|+1}\mathcal{A}_{m}(W)}{2\pi\Gamma^{2}(%
\gamma_{1})}\left[ \psi(\alpha_{1})+\psi(\beta_{1})\right] , \\
& \Omega_{2m}(W)=-\nu\tilde{\Omega}_{2m}(W),\;\tilde{\Omega}_{2m}(W)=\frac{%
(-1)^{|m|+1}\mathcal{B}_{m}(W)}{2\pi\Gamma^{2}(\gamma_{1})}.
\end{align*}

\subsubsection{Spectrum}

\subsubsection{$w=R^{2}E>q$}

In this case, we have $\alpha_{1}=1/2+|m|/2+\sigma-i\sqrt{w-q}/4$, $\beta
_{1}=1/2+|m|/2-\sigma-i\sqrt{w-q}/4$ and it is easy to prove that $\alpha
_{1},\beta_{1}\notin\mathbb{Z}_{-}$, such that $\Omega_{m}(E)$ is finite
complex function of $E$ and we have 
\begin{equation}
\sigma_{m}^{\prime}(E)=\operatorname{Im}\Omega_{m}(E)\equiv\varrho_{m}^{2}(E)>0. 
\label{Opso2.3.9.1.1.1}
\end{equation}
The spectrum of $\hat{h}_{\mathfrak{e}m}$\ is simple and continuous, $%
\mathrm{spec}\hat{h}_{\mathfrak{e}m}=[q/R^{2},\infty)$. Note that%
\begin{equation}
\lim_{\Delta\rightarrow+0}\varrho_{m}(E)=\left\{ 
\begin{array}{c}
0,\;q\neq q_{m,k} \\ 
O(\Delta^{1/4}),\;q=q_{m,k},%
\end{array}
\right. ,   \label{Ops02.3.9.1.1.2}
\end{equation}
$\Delta=E-q/R^{2}\rightarrow+0$, $q_{m,k}=4N_{m,k}^{2}$, $N_{m,k}=1+|m|+2k$.

\subsubsection{$w=R^{2}E\leq q$}

\paragraph{$q>0$, $\protect\sigma>0$}

In this case, we have $\left. \operatorname{Im}\nu\right| _{W=E}=\left. \operatorname{Im}%
\alpha_{1}\right| _{W=E}=\left. \operatorname{Im}\beta _{1}\right| _{W=E}=0$,

$\operatorname{Im}\left. \psi(\alpha_{1})\right| _{W=E}=0$, and%
\begin{equation*}
\sigma_{m}^{\prime}(E)=\frac{(-1)^{|m|+1}\mathcal{A}_{m}(E)}{2\pi\Gamma
^{2}(\gamma_{1})}\operatorname{Im}\left. \psi(\beta_{1})\right| _{W=E+i0}. 
\end{equation*}
$\sigma_{m}^{\prime}(E)$ can be different from zero in the points $E_{m,n}$
when $\beta_{1}=-n$, $n=0,1,2,...$, i.e., $E_{m,n}$ satisfy the equations%
\begin{align}
& \sqrt{q-w_{m,n}}=\sqrt{q}-2N_{m,n},\;N_{m,n}=1+|m|+2n\;\Longrightarrow
\label{Opso2.3.9.2.1.1} \\
& \;\Longrightarrow\;E_{m,n}=\frac{q}{R^{2}}-\frac{\left( \sqrt{q}%
-2N_{m,n}\right) ^{2}}{R^{2}}=\frac{4\sqrt{q}N_{m,n}}{R^{2}}-\frac{%
4N_{m,n}^{2}}{R^{2}}.  \notag
\end{align}

Eq. (\ref{Opso2.3.9.2.1.1}) has solutions only if $q\geq q_{m,0}$, $%
q_{m,0}=4N_{m,0}^{2}=4(1+|m|)^{2}$. With property (\ref{Opso2.3.9.2.1.1})
taken into account, we have: $n=0,1,...,n_{\max}$, $E_{m,n}>E_{m,n-1}$, $%
n=1,...,n_{\max}$, $0<E_{m,n}\leq q/R^{2}$, where%
\begin{align*}
& \,\left\{ 
\begin{array}{l}
q\leq q_{m,0},\;\mathrm{no\;levels} \\ 
n_{\max}=k,\;\sqrt{q}=2(1+|m|+2(k+\delta)%
\end{array}
\right. , \\
& k=0.1,...,\;0<\delta\leq1,
\end{align*}%
\begin{equation*}
\sigma_{m}^{\prime}(E)=\sum_{n=0}^{n_{\max}}Q_{m,n}^{2}\delta(E-E_{m,n}),%
\;Q_{m,n}=\frac{2\sqrt{(-1)^{|m|}\mathcal{A}_{m}(E_{m,n})\sqrt{q-w_{m,n}}}}{%
R\Gamma(\gamma_{1})}.. 
\end{equation*}
We note that the nonexistence of the level $E_{m,n_{\max}+1}=E_{m,k+1}=$ $%
q/R^{2}$ for $q=q_{m,k+1}$.follows from the fact that $Q_{m,k+1}=0$. Thus,
the discrete part of the spectrum of $\hat{h}_{\mathfrak{e}m}$\ is simple
and has the form%
\begin{align*}
& \mathrm{spec}\hat{h}_{\mathfrak{e}m}=\{E_{m,n},\;0<E_{m,n}<q/R^{2},%
\;n=0,1,...,n_{\max}\}, \\
& n_{\max}=k\;\mathrm{for}\;\frac{1}{4}\sqrt{q}=\frac{1+|m|}{2}%
+k+\delta,\;0<\delta\leq1,\;k\in\mathbb{Z}_{+}
\end{align*}
The discrete part of the spectrum is absent for $q\leq q_{m,0}$.

\paragraph{$q=0$, $\protect\sigma=0$}

We have in this case for $W=E$: $\alpha_{1}=\beta_{1}=1/2+|m|+\sqrt{|w|}$, $%
\operatorname{Im}\nu=0$, $\operatorname{Im}\alpha_{1}=0$, $\alpha_{1}>0$, , and $%
\sigma_{m}^{\prime}(E)=0$, the spectrum points are absent.

\paragraph{$q<0$, $\protect\sigma=i\varkappa$, $\varkappa=\protect\sqrt{|q|}$%
}

In this case, we have for $W=E$: $\nu=\sqrt{|w|-|q|}>0$, $\alpha
_{1}=1/2+|m|/2+\nu+i\varkappa$, $\beta_{1}=1/2+|m|/2+\nu-i\varkappa =%
\overline{\alpha_{1}}$, such that $\left[ \operatorname{Im}\psi(\alpha _{1})+\operatorname{Im%
}\psi(\beta_{1})\right] _{W=E}=0$, and%
\begin{equation*}
\sigma_{m}^{\prime}(E)=0. 
\end{equation*}

Finally, we find for fixed $m$, $|m|\geq1$:

The spectrum of $\hat{h}_{\mathfrak{e}m}$\ is simple, $\mathrm{spec}\hat{h}_{%
\mathfrak{e}m}=[q/R^{2},\infty)\cup\{E_{m,n},\;n=0,1,...n_{\max}\}$, the
discrete part of spectrum is present for $q>q_{m,0}$. The set of functions 
\begin{equation*}
\left\{ U_{m}(r;E)=\varrho_{m}(E)O_{1m}^{<}(r;E),\;E\geq
q/R^{2};\;U_{m,n}(r)=Q_{m,n}O_{1m}^{<}(r;E_{m,n}),\;n=0,1,...n_{\max}\right%
\} 
\end{equation*}
forms a complete orthogonalized system in $L^{2}(0,R)$.

\subsection{$m=0$}

In this case, we have: $\mu_{\delta}=\delta/2$, $\delta\geq0$; $\mu=0$; $%
\alpha_{1}=1/2+\nu+\sigma$, $\beta_{1}=1/2+\nu-\sigma$.

\subsubsection{ Useful solutions, $r<R$}

We need solutions of an equation%
\begin{equation}
(\check{h}_{0}-W)\psi_{0}^{<}(r)=0,\;   \label{Opso2.4.1.1}
\end{equation}
where $\check{h}_{0}$ is given by eq. (\ref{Opso2.1.1.1}).

It is convenient for our aims first to consider solutions more general
equation%
\begin{equation}
(\check{h}_{0,\delta}-W)\psi_{\delta}(r)=0,\;\check{h}_{0,\delta}=\left. 
\check{h}_{m}\right| _{m\rightarrow\delta}.   \label{Opso2.4.1.2}
\end{equation}

Introduce a new function $\phi_{\xi_{\mu}\delta}(x)$,%
\begin{equation*}
\psi_{\delta}(r)=\frac{R^{2}\sqrt{r}}{R^{2}-r^{2}}x^{\xi_{\mu}\delta
/2}(1-x)^{1/4+\nu}\phi_{\xi_{\mu}\delta}(x). 
\end{equation*}
Then we obtain%
\begin{align}
& [x(1-x)\partial_{x}^{2}+(\gamma_{\xi_{\mu}\delta}-(1+\alpha_{\xi_{\mu
}\delta}+\beta_{\xi_{\mu}\delta})x)\partial_{x}-\alpha_{\xi_{\mu}\delta}%
\beta_{\xi_{\mu}\delta}]\phi_{\xi_{\mu}\delta}(x)=0,  \label{Opso2.4.1.3} \\
& \alpha_{\xi_{\mu}\delta}=1/2+\xi_{\mu}\delta/2+\nu+\sigma,\;\beta_{\xi
_{\mu}\delta}=1/2+\xi_{\mu}\delta/2+\nu-\sigma,\;\gamma_{\xi_{\mu}\delta
}=1+\xi_{\mu}\delta.  \notag
\end{align}

Eq. (\ref{Opso2.4.1.3}) is the equation for hypergeometric functions, in the
terms of which we can express solutions of eq. (\ref{Opso2.4.1.2}). We will
use the following solutions:%
\begin{align*}
& O_{1,0,\delta}^{<}(r;W)=\frac{R^{2}\sqrt{r}}{R^{2}-r^{2}}x^{\delta
/2}(1-x)^{1/4+\nu}\mathcal{F}(\alpha_{1\delta},\beta_{1\delta};\gamma
_{1\delta};x)= \\
& \,=\frac{\Gamma(\gamma_{1\delta})\Gamma(-2\nu)}{\Gamma(\alpha_{4\delta
})\Gamma(\beta_{4\delta})}O_{3,0,\delta}^{<}(r;W)+\frac{\Gamma(\gamma
_{1\delta})\Gamma(2\nu)}{\Gamma(\alpha_{1\delta})\Gamma(\beta_{1\delta})}%
v_{0,\delta}^{<}(r;W)= \\
& \,=\left. O_{1,0,\delta}^{<}(r;W)\right| _{\sigma\rightarrow-\sigma
}=\left. O_{1,m,\delta}^{<}(r;W)\right| _{\nu\rightarrow-\nu}, \\
& O_{2,0,\delta}^{<}(r;W)=\frac{R^{2}\sqrt{r}}{R^{2}-r^{2}}x^{-\delta
/2}(1-x)^{1/4+\nu}\mathcal{F}(\alpha_{2\delta},\beta_{2\delta};\gamma
_{2\delta};x)= \\
& \,=O_{1,0,-\delta}^{<}(r;W)=\left. O_{2,0,\delta}^{<}(r;W)\right|
_{\sigma\rightarrow-\sigma}=\left. O_{2,0,\delta}^{<}(r;W)\right|
_{\nu\rightarrow-\nu},
\end{align*}%
\begin{align*}
& O_{3,0,\delta}^{<}(r;W)=\frac{R^{2}\sqrt{r}}{R^{2}-r^{2}}x^{\delta
/2}(1-x)^{1/4+\nu}\mathcal{F}(\alpha_{1\delta},\beta_{1\delta};\gamma
_{3};1-x)=\frac{R^{2}\sqrt{r}}{R^{2}-r^{2}}x^{\delta/2}(1-x)^{1/4+\nu}\times
\\
& \times\left[ \frac{\Gamma(\gamma_{3})\Gamma(-\delta)}{\Gamma
(\alpha_{2\delta})\Gamma(\beta_{2\delta})}\mathcal{F}(\alpha_{1\delta},%
\beta_{1\delta};\gamma_{1\delta};x)+x^{-\delta}\frac{\Gamma(\gamma_{3})%
\Gamma(\delta)}{\Gamma(\alpha_{1\delta})\Gamma(\beta_{1\delta})}\mathcal{F}%
(\alpha_{2\delta},\beta_{2\delta};\gamma_{2\delta};x)\right] = \\
& =\frac{\Gamma(\gamma_{3})\Gamma(\gamma_{1\delta})}{\Gamma(\alpha_{2\delta
})\Gamma(\beta_{2\delta})}O_{1,0,\delta}^{<}(r;W)+\frac{\Gamma(\gamma
_{3})\Gamma(\delta)}{\Gamma(\alpha_{1\delta})\Gamma(\beta_{1\delta})}%
O_{2,0,\delta}^{<}(r;W), \\
& v_{0,\delta}^{<}(r;W)=\frac{R^{2}\sqrt{r}}{R^{2}-r^{2}}x^{\delta
/2}(1-x)^{1/4-\nu}\mathcal{F}(\alpha_{4\delta},\beta_{4\delta};\gamma
_{4};1-x)=\left. v_{0,\delta}(r;W)\right| _{\nu\rightarrow-\nu},
\end{align*}%
\begin{align*}
& \alpha_{1\delta,2\delta}=1/2\pm\delta/2+\nu+\sigma,\;\beta_{1\delta
,2\delta}=1/2\pm\delta/2+\nu-\sigma, \\
& \alpha_{4\delta}=1/2+\delta/2-\nu+\sigma,\;\beta_{4\delta}=1/2+\delta
/2-\nu-\sigma, \\
& \gamma_{1\delta,2\delta}=1\pm\delta,\;\gamma_{3,4}=1\pm2\nu.
\end{align*}
Note that $O_{1,0,\delta}^{<}(r;W)$ and $O_{2,0,\delta}^{<}(r;W)$ are
real-entire in $W$ solutions of eq. (\ref{Opso2.4.1.2}).

Represent $O_{3,0,\delta}^{<}$ in the form%
\begin{align*}
& O_{3,0,\delta}^{<}(r;W)=B_{0,\delta}(W)O_{1,0,\delta}^{<}(r;W)-\frac{%
\Gamma (\gamma_{3})}{\Gamma(\alpha_{1\delta})\Gamma(\beta_{1\delta})}%
O_{4,0,\delta }^{<}(r;W), \\
& O_{4,0,\delta}^{<}(r;W)=\Gamma(\delta)\left[ O_{1,0,%
\delta}^{<}(r;W)-O_{1,0,-\delta}^{<}(r;W)\right] , \\
& B_{0,\delta}(W)=\frac{\Gamma(\gamma_{3})\Gamma(-\delta)}{\Gamma
(\alpha_{2\delta})\Gamma(\beta_{2\delta})}+\frac{\Gamma(\gamma_{3})\Gamma(%
\delta)}{\Gamma(\alpha_{1\delta})\Gamma(\beta_{1\delta})}.
\end{align*}

We obtain the solution of eq.(\ref{Opso2.4.1.1}) as the limit $\delta
\rightarrow0$ of the solution of (\ref{Opso2.4.1.2}):%
\begin{align*}
& O_{1,0}^{<}(r;W)=O_{1,0,0}^{<}(r;W)=\frac{R^{2}\sqrt{r}}{R^{2}-r^{2}}%
(1-x)^{1/4+\nu}\mathcal{F}(\alpha_{1},\beta_{1};1;x)= \\
& (\mathrm{for}\;\operatorname{Im}W>0)=\frac{\Gamma(-2\nu)}{\Gamma(\alpha
_{4})\Gamma(\beta_{4})}O_{3,0}^{<}(r;W)+\frac{\Gamma(2\nu)}{\Gamma(\alpha
_{1})\Gamma(\beta_{1})}v_{0}^{<}(r;W), \\
& v_{0}^{<}(r;W)=v_{0,0}^{<}(r;W)=\frac{R^{2}\sqrt{r}}{R^{2}-r^{2}}%
(1-x)^{1/4-\nu}\mathcal{F}(\alpha_{4},\beta_{4};\gamma_{4};1-x), \\
& O_{4,0}^{<}(r;W)=\lim_{\delta\rightarrow0}O_{4,0,\delta}^{<}(r;W)=2\lim
_{\delta\rightarrow0}\partial_{\delta}O_{1,0,\delta}^{<}(r;W), \\
& O_{3,0}^{<}(r;W)=\frac{R^{2}\sqrt{r}}{R^{2}-r^{2}}(1-x)^{1/4+\nu }\mathcal{%
F}(\alpha_{1},\beta_{1};\gamma_{3};1-x)= \\
& =B_{0}(W)O_{1,0}(r;W)-C_{0}(W)O_{4,0}(r;W),\;C_{0}(W)=\frac{\Gamma
(\gamma_{3})}{\Gamma(\alpha_{1})\Gamma(\beta_{1})}, \\
& B_{0}(W)=B_{0,0}(W)=C_{0}(W)f_{0}(W),\;f_{0}(W)=\left[ 2\psi
(1)-\psi(\alpha_{1})-\psi(\beta_{1})\right] .
\end{align*}

Note that $O_{1,0}^{<}(r;W)$ and $O_{4,0}^{<}(r;W)$ are real-entire in $W$.

\subsubsection{Asymptotics, $r\rightarrow0$ ($x\rightarrow0$)}

We have%
\begin{align*}
& O_{1,0}^{<}(r;W)=u_{1\mathrm{as}}(r)(1+O(r^{2})),\;O_{4,0}^{<}(r;W)=u_{4%
\mathrm{as}}(r)\left( 1+O(r^{2})\right) , \\
& u_{1\mathrm{as}}(r)=r^{1/2},\;u_{4\mathrm{as}}(r)=r^{1/2}\ln\left( \frac{%
4r^{2}}{R^{2}}\right) .
\end{align*}%
\begin{align*}
& O_{3,0}^{<}(r;W)=.C_{0}(W)\left[ f_{0}(W)u_{1\mathrm{as}}(r)-C_{0}(W)u_{4%
\mathrm{as}}(r)\right] \left( 1+O(r^{2})\right) , \\
& \operatorname{Im}W>0\;\mathrm{or}\;W=0.
\end{align*}

\subsubsection{Asymptotics, $\Delta=R-r\rightarrow0$ ($\protect\delta%
=1-x\rightarrow0$)}

We have%
\begin{equation*}
O_{3,0}^{<}(r;W)=\frac{1}{2}R^{1-2\nu}\Delta^{-1/2+2\nu}(1+O(\Delta)), 
\end{equation*}

\begin{align*}
& O_{1,0}^{<}(r;W)=\frac{\Gamma(2\nu)}{2\Gamma(\alpha_{1})\Gamma(\beta_{1})}%
R^{1+2\nu}\Delta^{-1/2-2\nu}(1+O(\Delta)), \\
& \operatorname{Im}W>0\;\mathrm{or}\;W=0.
\end{align*}

\subsubsection{Wronskian}

We have

\begin{align*}
& \mathrm{Wr}(O_{1,0}^{<},O_{4,0}^{<})=2, \\
& \mathrm{Wr}(O_{1,0}^{<},O_{4,0}^{<})=-\frac{2\Gamma(\gamma_{3})}{%
\Gamma(\alpha_{1})\Gamma(\beta_{1})}=-2C_{0}(W).
\end{align*}

Note that any solutions of eq. (\ref{Opso2.4.1.1}) are s.-integrable in the
origin and only one solution ($O_{3,0}^{<}$) is s.-integrable on the right
end $R$ (for $\operatorname{Im}W>0$), such that there is one solution ($O_{3,0}^{<}$%
) belonging to $L^{2}(\mathbb{R}_{+})$ for $\operatorname{Im}W>0$ and the deficiency
indexes of the symmetric operator $\check{h}_{0}$ (see below) are equal to $%
(1,1)$.

\subsubsection{Symmetric operator $\hat{h}_{0,\protect\zeta}$}

For given a differential operation $\check{h}_{0}$, the symmetric operator $%
\hat{h}_{0,\zeta}$ is given by eq. \ref{Opso2.2.0} with substitution $0$ for 
$m$.

\subsubsection{Isometry}

All considerations of subsec 2.4 retain hold with substitution $0$ for $m$.
Further, we consider the interval $(0,R)$.

\subsubsection{Adjoint operator $\hat{h}_{0}^{+}=\hat{h}_{0}^{\ast}$}

We will omit the superscript $(1/2)$.and will write $r$ for $\xi$, $0\leq
r\leq R$.

It is easy to prove by standard way that the adjoint operator $\hat{h}%
_{0}^{+}$ coincides with the operator $\hat{h}_{0}^{\ast}$,

\begin{equation*}
\hat{h}_{0}^{+}:\left\{ 
\begin{array}{l}
D_{h_{0}^{+}}=D_{\check{h}_{0}}^{\ast}(0,R)=\{\psi_{\ast},\psi_{\ast}^{%
\prime }\;\mathrm{are\;a.c.\;in}\mathcal{\;}(0,R),\;\psi_{\ast},\hat{h}%
_{0}^{+}\psi_{\ast}\in L^{2}(0,R)\} \\ 
\hat{h}_{0}^{+}\psi_{\ast}(r)=\check{h}_{0}\psi_{\ast}(r),\;\forall\psi_{%
\ast }\in D_{h_{0}^{+}}%
\end{array}
\right. . 
\end{equation*}

\subsubsection{Asymptotics}

Because $\check{h}_{0}\psi_{\ast}\in L^{2}(0,R)$, we have%
\begin{equation*}
\check{h}_{0}\psi_{\ast}(r)=\eta(r),\;\eta\in L^{2}(0,R), 
\end{equation*}
and we can represent $\psi_{\ast}$ in the form%
\begin{align*}
\psi_{\ast}(r) & =c_{1}O_{1,0}^{<}(r;0)+c_{2}O_{3,0}^{<}(r;0)+I(r), \\
\psi_{\ast}^{\prime}(r) &
=c_{1}\partial_{r}O_{1,0}^{<}(r;0)+c_{2}\partial_{r}O_{3,0}^{<}(r;0)+I^{%
\prime}(r),
\end{align*}
where%
\begin{align*}
I(r) & =\frac{O_{3,0}^{<}(r;0)}{2C_{0}(0)}\int_{0}^{r}O_{1,0}^{<}(y;0)%
\eta(y)dy-\frac{O_{1,0}^{<}(r;0)}{2C_{0}(0)}\int_{0}^{r}O_{3,0}^{<}(y;0)%
\eta(y)dy, \\
I^{\prime}(r) & =\frac{\partial_{r}O_{3,0}^{<}(r;0)}{2C_{0}(0)}%
\int_{0}^{r}O_{1,0}^{<}(y;0)\eta(y)dy-\frac{\partial_{r}O_{1,0}^{<}(r;0)}{%
2C_{0}(0)}\int_{0}^{r}O_{3,0}^{<}(y;0)\eta(y)dy.
\end{align*}

I) $r\rightarrow0$

We obtain with the help of the CB-inequality:%
\begin{equation*}
I(r)=O(r^{3/2}\ln r),\;I^{\prime}(r)=O(r^{1/2}\ln r), 
\end{equation*}
such that we have%
\begin{align*}
& \psi_{\ast}(r)=c_{1}u_{1\mathrm{as}}(r)+c_{2}u_{4\mathrm{as}%
}(r)+O(r^{3/2}\ln r), \\
& \psi_{\ast}^{\prime}(r)=c_{1}u_{1\mathrm{as}}^{\prime}(r)+c_{2}u_{4\mathrm{%
as}}^{\prime}(r)+O(r^{1/2}\ln r).
\end{align*}

II) $r\rightarrow R$

In this case, we prove that $[\psi_{\ast},\chi_{\ast}]^{R}=0$, $\forall
\psi_{\ast},\chi_{\ast}\in D_{h_{m}^{+}}$.

\subsubsection{Self-adjoint hamiltonians $\hat{h}_{0\protect\theta}$}

The calculation of the asymetry form $\Delta_{h_{0}^{+}}(\psi_{\ast})$ gives%
\begin{align*}
& \Delta_{h_{m}^{+}}(\psi_{\ast})=-[\psi_{\ast},\psi_{\ast}]_{0}=\left.
p_{0}(r)\left[ \overline{\psi_{\ast}(r)}\psi_{\ast}^{\prime}(r)\overline {%
-\psi_{\ast}^{\prime}(r)}\psi_{\ast}(r)\right] \right| _{r\rightarrow0}= \\
& \,=2(\overline{c_{1}}c_{2}-\overline{c_{2}}c_{1})=-i(.\overline{c_{+}}%
c_{+}-\overline{c_{-}}c_{-}),\;c_{\pm}=c_{1}\pm ic_{2}.
\end{align*}

The condition $\Delta_{h_{0}^{+}}(\psi)=0$ gives%
\begin{align*}
& c_{-}=-e^{2i\theta}c_{+},\;|\theta|\leq\pi/2,\;\theta=-\pi/2\sim\theta
=\pi/2,\;\Longrightarrow \\
& \Longrightarrow\;c_{1}\cos\theta=c_{2}\sin\theta,
\end{align*}
or%
\begin{align}
& \psi(r)=C\psi_{\theta\mathrm{as}}(r)+O(r^{3/2}\ln r),\;\psi^{\prime
}(r)=C\psi_{\theta\mathrm{as}}^{\prime}(r)+O(r^{1/2}\ln r), 
\label{Opso2.4.5.1} \\
& \psi_{\theta\mathrm{as}}(r)=u_{1\mathrm{as}}(r)\sin\theta+u_{4\mathrm{as}%
}(r)\cos\theta  \notag
\end{align}
(in this section we write $\theta$ instead of more right but more cumbersome 
$\theta_{\zeta}$). We thus have a family of s.a. hamiltohians $\hat{h}%
_{0\theta}$,%
\begin{equation}
\hat{h}_{0\theta}:\left\{ 
\begin{array}{l}
D_{h_{0\theta}}=\{\psi\in D_{h_{0}^{+}},\;\psi \;\mathrm{satisfy\;the%
\;boundary\;condition\;(\ref{Opso2.4.5.1})} \\ 
\hat{h}_{0\theta}\psi=\check{h}_{0}\psi,\;\forall\psi\in D_{h_{0\theta}}%
\end{array}
\right. .   \label{Opso2.4.5.2}
\end{equation}

\subsubsection{The guiding functional}

As a guiding functional $\Phi_{\theta}(\xi;W)$ we choose%
\begin{align}
& \Phi_{0\theta}(\xi;W)=\int_{0}^{R}U_{0\theta}(r;W)\xi(r)dr,\;\xi \in%
\mathbb{D}_{\theta}=D_{r}(0,R)\cap D_{h_{0\theta}}.  \label{Osc2.5.2.5.1} \\
& .U_{0\theta}(r;W)=O_{1,0}^{<}(r;W)\sin\theta+O_{4,0}^{<}(r;W)\cos \theta, 
\notag
\end{align}
$U_{0\theta}(r;W)$ is real-entire solution of eq. (\ref{Opso2.4.1.1})
satisfying the boundary condition (\ref{Opso2.4.5.1}).

The guiding functional $\Phi_{0\theta}(\xi;W)$ is simple and the spectrum of 
$\hat{h}_{0\theta}$ is simple.

\subsubsection{Green function $G_{0\protect\theta}(r,y;W)$, spectral
function $\protect\sigma_{0\protect\theta}(E)$}

We find the Green function $G_{0\theta}(r,y;W)$ as the kernel of the
integral representation 
\begin{equation*}
\psi(r)=\int_{0}^{\infty}G_{0\theta}(r,y;W)\eta(y)dy,\;\eta\in L^{2}(\mathbb{%
R}_{+}), 
\end{equation*}
of unique solution of an equation%
\begin{equation}
(\hat{h}_{0\theta}-W)\psi(r)=\eta(r),\;\operatorname{Im}W>0,   \label{Opso2.4.6.1}
\end{equation}
for $\psi\in D_{h_{0\theta}}$. General solution of eq. (\ref{Opso2.4.6.1})
(under condition $\psi\in L^{2}(0,R)$) can be represented in the form%
\begin{align*}
\psi(r) & =aO_{3,0}^{<}(r;W)+\frac{O_{1,0}^{<}(r;W)}{2C_{0}(W)}\eta
_{3}(W)+I(r),\;\eta_{3}(W)=\int_{0}^{R}O_{3,0}^{<}(y;W)\eta(y)dy, \\
I(r) & =\frac{O_{3,0}^{<}(r;W)}{2C_{0}(W)}\int_{0}^{r}O_{1,0}^{<}(y;W)%
\eta(y)dy-\frac{O_{1,0}^{<}(r;W)}{2C_{0}(W)}\int_{0}^{r}O_{3,0}^{<}(y;W)%
\eta(y)dy, \\
I(r) & =O\left( r^{3/2}\ln r\right) ,\;r\rightarrow0.
\end{align*}
A condition $\psi\in D_{h_{0\theta}\text{ , }}$(i.e.$\psi$ satisfies the
boundary condition (\ref{Opso2.4.5.1})) gives%
\begin{equation*}
a=-\frac{\cos\theta}{2C_{0}^{2}(W)\omega_{\theta}(W)}\eta_{3}(W),\;\omega
_{\theta}(W)=f_{0}(W)\cos\zeta+\sin\zeta, 
\end{equation*}%
\begin{align}
& G_{0\theta}(r,y;W)=\pi\Omega_{0\theta}(W)U_{0\theta}(r;W)U_{0\theta }(y;W)+
\notag \\
& \,+\frac{1}{2}\left\{ 
\begin{array}{c}
\tilde{U}_{0\theta}(r;W)U_{0\theta}(y;W),\;r>y \\ 
U_{0\theta}(r;W)\tilde{U}_{0\theta}(y;W),\;r<y%
\end{array}
\right. ,  \label{Opso2.4.6.2} \\
& \Omega_{0\theta}(W)=\frac{\tilde{\omega}_{\theta}(W)}{2\pi\omega_{\theta
}(W)},\;\tilde{\omega}_{\theta}(W)=f_{0}(W)\sin\theta-\cos\zeta,  \notag \\
& U_{0\theta}(y;W)=O_{1,0}^{<}(r;W)\sin\theta+O_{4,0}^{<}(r;W)\cos \theta, 
\notag \\
& \tilde{U}_{0\theta}(r;W)=O_{1,0}^{<}(r;W)\cos\theta-O_{4,0}^{<}(r;W)\sin%
\theta,  \notag
\end{align}
where we used an equality%
\begin{equation*}
O_{3,0}^{<}(r;W)=C_{0}(W)[\tilde{\omega}_{\theta}(W)U_{0\theta}(r;W)+\omega
_{\theta}(W)\tilde{U}_{0\theta}(r;W)]. 
\end{equation*}
Note that the functions $U_{0\theta}(r;W)$ and $\tilde{U}_{0\theta}(r;W)$
are real-entire in $W$ and the last term in the r.h.s. of eq. (\ref%
{Opso2.4.6.2}) is real for $W=E$. For $\sigma_{0\theta}^{\prime}(E)$, we find%
\begin{equation}
\sigma_{0\theta}^{\prime}(E)=\operatorname{Im}\Omega_{0\theta}(E+i0). 
\label{Opso2.4.6.3}
\end{equation}

\subsubsection{Spectrum}

\subsubsection{$\protect\theta=\protect\pi/2$}

First we consider the case $\theta=\pi/2$.

In this case, we have $U_{0\pi/2}(r;W)=O_{1,0}^{<}(r;W)$ and%
\begin{equation*}
\sigma_{0\pi/2}^{\prime}(E)=\frac{1}{2\pi}\operatorname{Im}\left. f(W)\right|
_{W=E+i0}. 
\end{equation*}

\paragraph{$w=R^{2}E>q$}

In this case, we have $\alpha_{1}=1/2+\sigma-i\sqrt{w-q}$, $\beta
_{1}=1/2-\sigma-i\sqrt{w-q}$, and it is easy to prove that $%
\alpha_{1},\beta_{1}\notin\mathbb{Z}_{-}$, such that $f(E)$ is finite
complex function of $E$, $f(E)=\mathcal{U}(E)+i\mathcal{V}(E)$, $\mathcal{U}%
(E)=\operatorname{Re}f(E)$, $\mathcal{V}(E)=\operatorname{Im}f(E)>0$, and we find%
\begin{align*}
& \sigma_{0\pi/2}^{\prime}(E)=\frac{1}{2\pi}\mathcal{V}(E)\equiv\rho_{0\pi
/2}^{2}(E)>0 \\
& \mathrm{spec}\hat{h}_{0\pi/2}=[q/R^{2},\infty),
\end{align*}
where $f_{0}(E)=\mathcal{U}_{0}(E)+i\mathcal{V}_{0}(E)$, $\mathcal{U}_{0}(E)=%
\operatorname{Re}f_{0}(E)$, $\mathcal{V}_{0}(E)=\operatorname{Im}f_{0}(E)>0$. Note that $%
\lim_{\Delta\rightarrow0}\varrho_{0\pi/2}(E)=0$ for $q\neq q_{0,k}$ and $%
\varrho_{0\pi/2}(E)=O(\Delta^{1/4})$ for $q=q_{0,k}$, $\Delta=E-q/R^{2}%
\rightarrow+0$, $q_{0,k}=4N_{0,k}^{2}$, $N_{0,k}=1+2k$

\paragraph{$w=R^{2}E=q+\Delta$, $\Delta\sim0$}

In this case, we have $\nu=\sqrt{-\Delta}$,

\subparagraph{$q\geq0$}

In this case, we have%
\begin{equation*}
\operatorname{Im}\psi(\alpha_{1})=\left\{ 
\begin{array}{l}
-O(\sqrt{\Delta}).\;\Delta>0 \\ 
0,\;\Delta<0%
\end{array}
\right. . 
\end{equation*}

i) $q\neq q_{0,k}$ 
\begin{equation*}
\operatorname{Im}\psi(\beta_{1})=\left\{ 
\begin{array}{l}
-O(\sqrt{\Delta}).\;\Delta>0 \\ 
0,\;\Delta<0%
\end{array}
\right. \;\Longrightarrow 
\end{equation*}%
\begin{equation*}
\sigma_{0\pi/2}^{\prime}(E)=\left\{ 
\begin{array}{l}
O(\sqrt{\Delta}).\;\Delta>0 \\ 
0,\;\Delta<0%
\end{array}
\right. . 
\end{equation*}

ii) $q=q_{0,k}$, $\beta_{1}=-k+\sqrt{-\Delta}/4$, $k=0,1,...$%
\begin{equation*}
\operatorname{Im}\psi(\beta_{1})=\left\{ 
\begin{array}{l}
-\frac{1}{2R\sqrt{\Delta}}+O(\sqrt{\Delta}).\;\Delta>0 \\ 
0,\;\Delta<0%
\end{array}
\right. \;\Longrightarrow 
\end{equation*}%
\begin{equation*}
\sigma_{0\pi/2}^{\prime}(E)=\left\{ 
\begin{array}{l}
-\frac{1}{2R\sqrt{\Delta}}+O(\sqrt{\Delta}).\;\Delta>0 \\ 
0,\;\Delta<0%
\end{array}
\right. . 
\end{equation*}

\subparagraph{$q<0$}

In this case, we have $\alpha_{1}=1/2+\sqrt{-\Delta}+i\varkappa$, $\beta
_{1}=1/2+\sqrt{-\Delta}-i\varkappa$, and%
\begin{equation*}
\sigma_{0\pi/2}^{\prime}(E)=\left\{ 
\begin{array}{l}
O(\sqrt{\Delta}).\;\Delta>0 \\ 
0,\;\Delta<0%
\end{array}
\right. . 
\end{equation*}

\paragraph{$w=R^{2}E<q$}

In this case, we have $\left. \operatorname{Im}\nu\right| _{W=E}=0$, .$\left.
\nu\right| _{W=E}>0$.

\subparagraph{$q>0$}

In this case, we have $\left. \operatorname{Im}\alpha_{1}\right| _{W=E}=\left. \operatorname{%
Im}\beta_{1}\right| _{W=E}=0$, $\operatorname{Im}\left. \psi(\alpha_{1})\right|
_{W=E}=0$, and%
\begin{equation*}
\sigma_{0\pi/2}^{\prime}(E)=-\frac{1}{2\pi}\operatorname{Im}\left. \psi
(\beta_{1})\right| _{W=E+i0}. 
\end{equation*}
$\sigma_{0\pi/2}^{\prime}(E)$ can be different from zero in the points $%
\mathcal{E}_{0n}$ when $\beta_{1}=-n$, $n=0,1,2,...$, i.e., $\mathcal{E}_{0n}
$ satisfy the equations%
\begin{align}
& \sqrt{q-w_{0\pi/2,n}}=\sqrt{q}-2N_{0,n},\;N_{0,n}=1+2n\;\Longrightarrow
\label{Opso2.4.8.1.3.1.1} \\
& \;\Longrightarrow\;\mathcal{E}_{0n}=\frac{q}{R^{2}}-\frac{\left( \sqrt {q}%
-2N_{0,n}\right) ^{2}}{R^{2}}=\frac{4\sqrt{q}N_{0,n}}{R^{2}}-\frac{%
4N_{0,n}^{2}}{R^{2}}.  \notag
\end{align}

Eq. (\ref{Opso2.4.8.1.3.1.1}) has solutions only if $q\geq q_{0,0}$, $%
q_{0,0}=4N_{0,0}^{2}=q$. Then we have: $n=0,1,...,n_{\max}$, $\mathcal{E}%
_{0n}>\mathcal{E}_{0n-1}$, $n=1,...,n_{\max}$, $0<\mathcal{E}_{0n}\leq
q/R^{2}$, where%
\begin{equation*}
\sigma_{0\pi/2}^{\prime}(E)=\sum_{n=0}^{n_{\max}}Q_{0\pi/2,n}^{2}\delta(E-%
\mathcal{E}_{n}),\;Q_{0\pi/2,n}=2R^{-1}\left( q-w_{0\pi/2,n}\right) ^{1/4}.. 
\end{equation*}%
\begin{equation*}
\,\left\{ 
\begin{array}{l}
q\leq q_{0,0},\;\mathrm{no\;levels} \\ 
n_{\max}=k,\;\sqrt{q}=4[1/2+k+\delta)%
\end{array}
\right. ,\;k=0.1,...,\;0<\delta\leq1, 
\end{equation*}
The level $\mathcal{E}_{0n_{\max}+1}=\mathcal{E}_{0k+1}$ would be equal to $%
q/R^{2}$ for $q=q_{0,k+1}=4N_{0,k+1}^{2}$. But, in this case, we have $%
Q_{0\pi/2,k+1}=0$, such that the level $\mathcal{E}_{0n_{\max}+1}=\mathcal{E}%
_{0k+1}$ is factually absent. Analogously, we see that the level $\mathcal{E}%
_{00}\;$for $q=q_{0,0}$ is factually absent. Thus, the discrete part of the
spectrum of $\hat{h}_{0\pi/2}$\ is simple and has the form%
\begin{align*}
& \mathrm{spec}\hat{h}_{0\pi/2}=\{\mathcal{E}_{0n},\;0<\mathcal{E}%
_{0n}<q/R^{2},\;n=0,1,...,n_{\max}\}, \\
& n_{\max}=k\;\mathrm{for}\;\sqrt{q}=4[\frac{1}{2}+k+\delta],\;0<\delta
\leq1,\;k\in\mathbb{Z}_{+}
\end{align*}
The discrete part of the spectrum is absent for $q\leq q_{0,0}$.

\subparagraph{$q\leq0$}

In this case, we have $\alpha_{1}=1/2+\nu+i\varkappa$, .$\beta_{1}=1/2+\nu-i%
\varkappa$, and

\begin{equation*}
\sigma_{0\pi/2}^{\prime}(E)=0. 
\end{equation*}

Finally, we find.

The spectrum of $\hat{h}_{0\pi/2}$\ is simple, $\mathrm{spec}\hat{h}_{0\pi
/2}=[q/R^{2},\infty)\cup\{\mathcal{E}_{0n},\;n=0,1,...n_{\max}\}$, the
discrete part of spectrum is present for $q>q_{0,0}$. The set of functions 
\begin{equation*}
\left\{ U_{0\pi/2}(r;E)=\varrho_{0\pi/2}(E)O_{10}^{<}(r;E),\;E\geq
q/R^{2};\;U_{0\pi/2,n}(r)=Q_{0\pi/2,n}O_{10}^{<}(r;\mathcal{E}%
_{0n}),\;n=0,1,...n_{\max}\right\}
\end{equation*}
forms a complete orthogonalized system in $L^{2}(\mathbb{R}_{+})$.

Note that these results for spectrum and the set of eigenfunctions can be
obtained from the corresponding results of sec. 3 by formal substitution $%
|m|\rightarrow0$.

The same results we obtain for the case $\zeta=-\pi/2$.

\subsubsection{$|\protect\theta|<\protect\pi/2$}

Now we consider the case $|\theta|<\pi/2$.

In this case, we can represent $\sigma_{\theta}^{\prime}(E)$ in the form%
\begin{equation*}
\sigma_{0\theta}^{\prime}(E)=-\frac{1}{2\pi\cos^{2}\theta}\operatorname{Im}\frac{1}{%
f_{\theta}(E+i0)},\;f_{0\theta}(W)=f_{0}(W)+\tan\theta. 
\end{equation*}

\paragraph{$w>q$}

In this case, we have%
\begin{equation*}
\sigma_{0\theta}^{\prime}(E)=\frac{1}{2\pi}\frac{\mathcal{V}_{0}(E)}{[%
\mathcal{U}_{0}(E)\cos\theta+\sin\theta]^{2}+\mathcal{V}_{0}^{2}(E)\cos^{2}%
\theta}\equiv\rho_{0\theta}^{2}(E). 
\end{equation*}
The spectrum of $\hat{h}_{0\theta}$ is simple and continuous, $\mathrm{spec}%
\hat{h}_{0\theta}=[q/R^{2},\infty)$.

\paragraph{$w=q+\tilde{\Delta}$, $\tilde{\Delta}=\Delta+i\protect\varepsilon%
\sim0$, $\operatorname{Im}\Delta=0$}

In this case, we have $\alpha_{1}=1/2+\sigma+\sqrt{-\Delta}/4$, $\beta
_{1}=1/2-\sigma+\sqrt{-\Delta}/4$,%
\begin{equation*}
\sqrt{-\Delta}=\left\{ 
\begin{array}{c}
-i\sqrt{\Delta},\;\Delta\geq0 \\ 
\sqrt{|\Delta|},\;\Delta\leq0%
\end{array}
\right. ,\;\sigma=\left\{ 
\begin{array}{l}
\sqrt{q}/4,\;q\geq0 \\ 
i\sqrt{|q|}/4,\;q\leq0%
\end{array}
\right. . 
\end{equation*}
A direct estimation gives%
\begin{equation*}
\sigma_{0\theta}^{\prime}(E)=\left\{ 
\begin{array}{l}
\left\{ 
\begin{array}{c}
O(\sqrt{\Delta}),\;q=q_{0k}\;\mathrm{or}\;q\neq q_{0k},\theta\neq\theta_{0}
\\ 
O(1/\sqrt{\Delta}),\;\theta=\theta_{0},q\neq q_{0k}%
\end{array}
\right. ,\;\Delta>0 \\ 
0,\;\Delta<0\;%
\end{array}
\right. , 
\end{equation*}
where $\tan\theta_{0}=\psi(\alpha_{10})+\psi(\beta_{10})-2\psi(1)$, $%
\alpha_{10}=1/2+\sigma$, $\beta_{10}=1/2-\sigma$

\paragraph{$w<q$}

In this case, we have $\left. \operatorname{Im}\nu\right| _{W=E}=0$, $\left.
\nu\right| _{W=E}>0$,%
\begin{align*}
\alpha_{1} & =1/2+\nu+\sqrt{q}/4,\;\beta_{1}=1/2-\nu+\sqrt{q}/4,\;q\geq0, \\
\alpha_{1} & =1/2+\nu+i\sqrt{|q|}/4,\;\beta_{1}=\overline{\alpha_{1}}%
,\;q\leq0.
\end{align*}

Thus, we have: the function $[f_{0\theta}(E)]^{-1}$ is real except the
points $E_{0n}(\theta)$,%
\begin{equation}
f_{0\theta}(E_{0n}(\theta))=0,   \label{Opso2.4.8.3.1}
\end{equation}
such that we obtain%
\begin{align}
\sigma_{0\theta}^{\prime}(E) & =\sum_{n\in\mathcal{N}}Q_{0\theta,n}^{2}%
\delta(E-E_{0n}(\theta)),\;Q_{0\theta,n}=\frac{1}{\sqrt{2\partial
_{E}f_{0\theta}(E_{0n}(\theta))}\cos\theta},  \notag \\
\partial_{E}f(E) & =\frac{R^{2}[\psi^{\prime}(\alpha_{1})+\psi^{\prime}(%
\beta_{1})]}{8\sqrt{q-w}}>0,   \label{Opso2.4.8.3.2}
\end{align}
where $\mathcal{N}$ is a subset of $\mathbb{Z}$ to be described below.
Furthermore, we find 
\begin{equation*}
\partial_{\theta}E_{0n}(\theta)=-[\partial_{E}f_{0\theta}(E_{0n}(\theta))%
\cos^{2}\theta]^{-1}<0. 
\end{equation*}

\subparagraph{$q_{0,k}<q\leq q_{0,k+1}$, $\protect\sqrt{q}/4=1/2+k+\protect%
\delta$, $0<\protect\delta\leq1$, $k\in\mathbb{Z}_{+}$}

In this case,\ the function $f_{0}(E)$ has the properties: $%
f_{0}(E)\rightarrow-\infty$ as $E\rightarrow-\infty$; $f_{0}(\mathcal{E}%
_{0n}\pm0)=\mp\infty$ $\ n=0,1,...,n_{\max}=k$; $f_{0}(E)\rightarrow-\tan%
\theta _{0}=2\psi(1)-\psi(1+k+\delta)-\psi(-k-\delta)-0$ as $E\rightarrow
q/R^{2}-0$. We find: in each interval $(\mathcal{E}_{0n-1},\mathcal{E}_{0n})$%
, $n=0,...,k$ (we set $\mathcal{E}_{0-1}=-\infty$), for any fixed $%
\theta\in(-\pi/2,\pi/2)$, there is one level $E_{0n}(\theta)$ which run
monotonocally from $\mathcal{E}_{0n-1}+0$ to $\mathcal{E}_{0n}-0$ as $\theta$
run from $\pi/2-0$ to -$\pi/2+0$; in the interval $(\mathcal{E}_{0k},q/R^{2})
$, for any fixed $\theta\in(\theta_{0},\pi/2)$, there is one level $%
E_{0k+1}(\theta)$ which run monotonocally from $\mathcal{E}_{0k}+0$ to $%
q/R^{2}-0$ as $\theta$ run from $\pi/2-0$ to $\theta_{0}+0$; eq. (\ref%
{Opso2.4.8.3.1}) has no solutions in the interval $(\mathcal{E}_{0k},q/R^{2})
$ for $\theta\in(-\pi/2,\theta_{0})$. Formally, eq. (\ref{Opso2.4.8.3.1})
has solution $E_{0k+1}(\theta_{0})=q/R^{2}$ for $\theta=\theta_{0}$.
However, in this case $Q_{0\theta_{0},k}=0$, as it is follows from eq. (\ref%
{Opso2.4.8.3.2}).We find also%
\begin{equation*}
\mathcal{N}=\mathcal{N}_{0k}(\theta)=\left\{ 
\begin{array}{l}
\{0,1,...,k\},\;\theta\in(-\pi/2,\theta_{0}] \\ 
\{0,1,...,k+1\},\;\theta\in(\theta_{0},\pi/2)%
\end{array}
\right. . 
\end{equation*}

\subparagraph{.$q\leq q_{0,0}=4$}

In this case, the function $f_{0}(E)$ has the properties: $%
f_{0}(E)\rightarrow-\infty$ as $E\rightarrow-\infty$; $f_{0}(E)\rightarrow
-\tan\theta_{0}-0=2\psi(1)-\psi(\alpha_{1})-\psi(\beta_{1})-0$ as $%
E\rightarrow q/R^{2}-0$;.$f(E)$ increases monotonocally on the interval $%
(-\infty,q/R^{2})$. Then we find: in the interval $(-\infty,q/R^{2}]$, for
any fixed $\theta\in(\theta_{0},\pi/2)$, there is one level $E_{0-1}(\theta)$
which run monotonocally from $-\infty$ to $q/R^{2}-0$ as $\theta$ run from $%
\pi/2-0$ to $\theta_{0}+0$; there are no discrete levels on the interval $%
(-\infty,q/R^{2}]$ for $\theta\in(-\pi/2,\theta_{0}]$. We find also%
\begin{equation*}
\mathcal{N}=\mathcal{N}_{0-1}(\theta)=\left\{ 
\begin{array}{l}
\varnothing,\;\theta\in(-\pi/2,\theta_{0}] \\ 
\{-1\},\;\theta\in(\theta_{0},\pi/2)%
\end{array}
\right. . 
\end{equation*}

Finally, we obtain. The spectrum of $\hat{h}_{0\theta}$ is simple and $%
\mathrm{spec}\hat{h}_{0\theta}=[q/R^{2},\infty)\cup\{E_{0n}(\theta ),\;n\in%
\mathcal{N}\}$. The set of functions 
\begin{equation*}
\left\{ U_{0\theta }(r;E)=\varrho_{0\theta}(E)O_{10}^{<}(r;E),\;E\geq
q/R^{2};\;U_{0\theta ,n}(r)=Q_{0\theta,n}U_{\theta}(r;E_{n}(\theta),),\;n\in%
\mathcal{N}\right\} 
\end{equation*}
forms a complete orthogonalized system in $L^{2}(0,R)$.

\subsection{S.a. operators $\hat{H}_{\mathfrak{e}m,\protect\zeta}$}

Then s.a. operator $\hat{H}_{\mathfrak{e}m,\zeta}$can be reconstructed by
the rules of subsec. 4 Isometry of sec. 3 from operators $\hat{h}_{m,\zeta
}^{(1/2)}$ which are equal to the operators $\hat{h}_{\mathfrak{e}m}$ of
sec. 3 for $|m|\geq1$ and to the operators $\hat{h}_{0\theta}$ of sec. 4 for 
$m=0$. We do not write explicitly the corresponding formulas and note only
that the parameter $\theta$ of s.a. hamiltonians $\hat{h}_{0\theta}$ should
be dependent of $\zeta$, $\theta=\theta(\zeta)$. Note also that a QM-systems
obtained can be considered (for fixed $\mathfrak{H}_{m,\zeta}$ and $%
\theta_{\zeta}$) as union of two isomorphic noninteracting systems placed on
two sheets of the huperboloid. The absent of interaction is due to the
vanishind of the current density through the boundary (through the points $%
r=0,R-0,R+0,r=\infty$).

\section{Quantum two-dimensional Coulomb-like interaction on pseudosphere}

Consider a space with coordinates $\rho$, $0<\rho<\infty$, and $\phi$, $%
0\leq\phi\leq4\pi$, $\phi=0\sim\phi=4\pi$. Let a differential operation $%
\check{H}_{C}=\check{H}$,%
\begin{align}
& \check{H}=-\Delta_{BL\rho}-\Delta_{BL\phi}+V_{C},\;\Delta_{BL\rho}=\frac{%
(R_{C}^{2}-\rho^{2})^{2}}{R_{C}^{4}}\frac{1}{\rho}\partial_{\rho}\rho%
\partial_{\rho},  \label{Cps2.1.1} \\
& \Delta_{BL\phi}=\frac{(R_{C}^{2}-\rho^{2})^{2}}{R_{C}^{4}}\frac{1}{\rho
^{2}}\partial_{\phi}^{2},\;V_{C}=\frac{g(R_{C}+\rho)^{2}}{R_{C}^{2}\rho }, 
\notag
\end{align}
be given. Consider the equation%
\begin{align}
& \,(\check{H}-W_{C})\Psi(\rho,\phi)=0,\;\rho<R_{C},\;  \notag \\
& \Psi(\rho,\phi)=\sum_{m\in\mathbb{Z}}\Lambda_{m}(\rho,\phi),\;\Lambda
_{m}(\rho,\phi)=\frac{1}{\sqrt{4\pi}}e^{im\phi/2}\Psi_{m}(\rho
)\;\Longrightarrow  \notag \\
& \left( \check{H}_{Cm}-W_{C}\right) \Psi_{m}(\rho)=0,\;\check{H}_{Cm}=%
\check{H}_{m}=-\Delta_{BL\rho}+\frac{m^{2}(R_{C}^{2}-\rho^{2})^{2}}{%
4R_{C}^{4}\rho^{2}}+V_{C}   \label{Cps2.1.2}
\end{align}

Represent $\Psi_{m}(\rho)$ in the form%
\begin{equation*}
\Psi_{m}(\rho)=x^{\tilde{\mu}}(1-x)^{1/4+\tilde{\nu}}\Phi_{m,\xi_{\mu},\xi_{%
\nu}}(x),\;\tilde{\mu}=\xi_{\mu}\mu,\;\tilde{\nu}=\xi_{\nu}\nu
,\;\xi_{\mu},\xi_{\nu}=\pm1, 
\end{equation*}
where%
\begin{align*}
x_{C} & =x=\frac{4R_{C}\rho}{(R_{C}+\rho)^{2}},\;\rho=R_{C}\frac {(1-\sqrt{%
1-x})^{2}}{x},\;\frac{\partial\rho}{\partial x}=\frac{\rho}{x\sqrt{1-x}}, \\
\rho\partial_{\rho} & =x\sqrt{1-x}\partial_{x},\;\frac{(R_{C}^{2}-\rho
^{2})^{2}}{R_{C}^{4}\rho^{2}}=\frac{16(1-x)}{R_{C}^{2}x^{2}},\;V_{C}=\frac{%
16(1-x)}{R_{C}^{2}x}\frac{gR_{C}}{4(1-x)}, \\
\Delta_{BL\rho} & =\frac{16(1-x)^{3/2}}{R_{C}^{2}x}\partial_{x}x\sqrt {1-x}%
\partial_{x}=\frac{1}{x}\Delta_{BLr},\;\Delta_{BL\phi}=\frac {16(1-x)}{%
R_{C}^{2}x}\left( -\frac{(m+\delta)^{2}}{4x}\right) .
\end{align*}
Then we have%
\begin{align*}
& \left[ x(1-x)\partial_{x}^{2}+(\gamma_{\xi_{\mu}}-(1+\alpha_{\xi_{\mu},%
\xi_{\nu}}+\beta_{\xi_{\mu},\xi_{\nu}})x)\partial_{x}-\alpha_{\xi_{\mu},%
\xi_{\nu}}\beta_{\xi_{\mu},\xi_{\nu}}\right] \Phi_{m,\xi_{\mu},\xi_{\nu}}=0,
\\
& \alpha_{\xi_{\mu},\xi_{\nu}}=1/2+\tilde{\mu}+\tilde{\nu}+\sigma
,\;\beta_{\xi_{\mu},\xi_{\nu}}=1/2+\tilde{\mu}+\tilde{\nu}-\sigma
,\;\gamma_{\xi_{\mu}}=1+2\tilde{\mu}, \\
& \mu=\frac{|m|}{2},\;\nu=\frac{1}{4}\sqrt{1+4R_{C}g-w_{C}},\;\sigma=\frac {1%
}{4}\sqrt{1-w_{C}}, \\
& w_{C}=R_{C}^{2}W_{C}.
\end{align*}

It is convenient to solve first more general equation%
\begin{equation}
\left( \check{H}_{Cm,\delta}-W_{C}\right) \Psi_{Cm,\delta}(\rho)=0.\check {H}%
_{Cm,\delta}=\check{H}_{m,\delta}=-\Delta_{BL\rho}+\frac{(m+\delta
)^{2}(R_{C}^{2}-\rho^{2})^{2}}{4R_{C}^{4}\rho^{2}}+V_{C}   \label{Cps2.1.3}
\end{equation}
Represent $\Psi_{Cm,\delta}(\rho)=\Psi_{m,\delta}(\rho)$ in the form%
\begin{equation*}
\Psi_{m,\delta}(\rho)=x^{\tilde{\mu}_{\delta}}(1-x)^{1/4+\tilde{\nu}%
}\Phi_{m,\xi_{\mu},\xi_{\nu},\delta}(x),\;\tilde{\mu}=\xi_{\mu}\mu,\;\tilde {%
\nu}=\xi_{\nu}\nu,\;\xi_{\mu},\xi_{\nu}=\pm1, 
\end{equation*}
where%
\begin{equation*}
\tilde{\mu}_{\delta}=\xi_{\mu}\mu_{\delta},\;\mu_{\delta}=\frac{|m+\delta|}{2%
}. 
\end{equation*}
Then we have%
\begin{align*}
& \left[ x(1-x)\partial_{x}^{2}+(\gamma_{\xi_{\mu},\delta}-(1+\alpha
_{\xi_{\mu},\xi_{\nu},\delta}+\beta_{\xi_{\mu},\xi_{\nu},\delta})x)%
\partial_{x}-\alpha_{\xi_{\mu},\xi_{\nu},\delta}\beta_{\xi_{\mu},\xi_{\nu
},\delta}\right] \Phi_{m,\xi_{\mu},\xi_{\nu},\delta}=0, \\
& \alpha_{\xi_{\mu},\xi_{\nu},\delta}=1/2+\tilde{\mu}_{\delta}\tilde{\nu }%
+\sigma,\;\beta_{\xi_{\mu},\xi_{\nu},\delta}=1/2+\tilde{\mu}_{\delta}+\tilde{%
\nu}-\sigma,\;\gamma_{\xi_{\mu},\delta}=1+2\tilde{\mu}_{\delta}.
\end{align*}

\subsection{Scalar product}

Differential operation (\ref{Cps2.1.1}) is s.a. for the standard scalar
product on the pseudosphere%
\begin{equation*}
d\Lambda(\rho,\phi)=d\omega(\rho)d\phi,\;d\omega(r)=\frac{R_{C}^{4}\rho }{%
(R_{C}^{2}-\rho^{2})^{2}}d\rho, 
\end{equation*}%
\begin{align*}
& \left( \Lambda_{1m_{1}},\Lambda_{2m_{2}}\right) =\delta_{m_{1}m_{2}}\left(
\Lambda_{1m_{1}}\Lambda_{2m_{1}}\right) ,\;\left( \Lambda
_{1m},\Lambda_{2m}\right) = \\
& =\left( \Psi_{1m},\Psi_{2m}\right) =\int_{0}^{R_{C}}\overline{\Psi
_{1m}(\rho)}\Psi_{2m}(\rho)d\omega(\rho)
\end{align*}

Represent $\Psi_{m}(r)$ in the form%
\begin{equation*}
\Psi_{m}(r)=\frac{R_{C}^{2}-\rho^{2}}{R_{C}^{2}\sqrt{\rho}}\psi_{m}(\rho). 
\end{equation*}
Then we have%
\begin{align}
& \left( \Lambda_{1m},\Lambda_{2m}\right) =\langle\psi_{1m},\psi
_{2m}\rangle=\int_{0}^{R_{C}}\overline{\psi_{1m}(\rho)}\psi_{2m}(\rho )d\rho,
\notag \\
& \check{H}_{m}\Psi_{m}(r)=\frac{R_{C}^{2}-\rho^{2}}{R_{C}^{2}\sqrt{\rho}}%
\check{h}_{m}\psi_{m}(\rho),  \notag \\
& \check{h}_{m}=\frac{\sqrt{\rho}}{R_{C}^{2}-\rho^{2}}\check{H}_{m}\frac{%
R_{C}^{2}-\rho^{2}}{\sqrt{\rho}}=-\partial_{\rho}p_{0}(\rho )\partial_{\rho}+
\notag \\
& +\frac{10R_{C}^{2}\rho^{2}-9\rho^{4}-R_{C}^{4}+m^{2}(R_{C}^{2}-\rho
^{2})^{2}}{4R_{C}^{4}r^{2}}+V(\rho),\;p_{0}(\rho)=\frac{(R_{C}^{2}-\rho
^{2})^{2}}{R_{C}^{4}}.   \label{Cps2.2.0}
\end{align}
We introduce also the differential operation $\check{h}_{m,\delta}$,%
\begin{equation*}
\check{h}_{m,\delta}=\frac{\sqrt{\rho}}{R_{C}^{2}-\rho^{2}}\check{H}%
_{m,\delta}\frac{R_{C}^{2}-\rho^{2}}{\sqrt{\rho}}=\left. \check{h}_{m,\delta
}\right\vert _{m\rightarrow m+\delta}, 
\end{equation*}
and corresponding the Shroedinger equations%
\begin{equation}
(\check{h}_{m}-W)\psi_{m}(\rho)=0,   \label{Cps2.2.1}
\end{equation}%
\begin{equation}
(\check{h}_{m,\delta}-W)\psi_{m,\delta}(\rho)=0.   \label{Cps2.2.2}
\end{equation}

\subsection{Connections and coincidences}

In this section, we describe a connection of the oscillator and Coulomb
problems on the two-dimensional pseudospheres.

The duality of these theories was demonstrated in \cite{Ners-Pog}.

The coordinates of two pseudospheres are connected by so called
Levi-Civita--Bohlini transformation

\begin{equation*}
\rho=\kappa_{0}r^{2},\;r=\sqrt{\rho/\kappa_{0}},\;R_{C}=\kappa_{0}R^{2},\;%
\phi=2\varphi, 
\end{equation*}
and the parameters of two problems satisfy the relations%
\begin{equation*}
W_{O}=-4\kappa_{0}g,\;g=-\frac{W_{O}}{4\kappa_{0}},\;q=1-R_{C}^{2}W_{C},%
\;W_{C}=\frac{1-q}{R_{C}^{2}}=\frac{1}{R_{C}^{2}}-\frac{\lambda}{%
4\kappa_{0}^{2}}. 
\end{equation*}
We have%
\begin{align*}
& \Delta_{BLr}=\mathcal{R}\Delta_{BL\rho},\;\mathcal{R}=\frac{%
4\kappa_{0}^{2}R^{4}r^{2}}{(R^{2}+r^{2})^{2}}=\frac{4\kappa_{0}R_{C}^{2}\rho%
}{(R_{C}+\rho)^{2}},\;\Delta_{BL\varphi}=\mathcal{R}\Delta_{BL\phi}, \\
& W_{O}-V_{O}(r)=\mathcal{R}\left[ W_{C}-V_{C}(\rho)\right] ,\;\check {H}%
_{O}-W_{O}=\mathcal{R}\left[ \check{H}_{C}-W_{C}\right] , \\
& \check{H}_{Om}-W_{O}=\mathcal{R}\left[ \check{H}_{Cm}-W_{C}\right] ,\;%
\check{H}_{Om,\delta}-W_{O}=\mathcal{R}\left[ \check{H}_{Cm,\delta}-W_{C}%
\right] ,
\end{align*}

\begin{align*}
& 4gR_{C}=-R^{2}W_{O}=-w_{O},\;w_{C}=1-q, \\
&
\mu_{C}=\mu_{O},\;\mu_{C,\delta}=\mu_{O,\delta},\;\nu_{C}=\nu_{O},\;%
\sigma_{C}=\sigma_{O}, \\
& \alpha_{C\xi_{\mu},\xi_{\nu},\delta}=\alpha_{O,\xi_{\mu},\xi_{\nu},\delta
},\;\beta_{C,\xi_{\mu},\xi_{\nu},\delta}=\beta_{O,\xi_{\mu},\xi_{\nu},\delta
},\;\gamma_{C\xi_{\mu},\delta}=\gamma_{O,\xi_{\mu},\delta}, \\
& x_{C}=\frac{4R_{C}\rho}{(R_{C}+\rho)^{2}}=\frac{4R^{2}r^{2}}{%
(R^{2}+r^{2})^{2}}=x_{O}.
\end{align*}%
\begin{align*}
A_{Cm,\delta}(W_{C}) &
=A_{Om,\delta}(W_{O}),\;B_{Cm}(W_{C})=B_{Om}(W_{O}),\;|m|\geq1, \\
C_{Cm}(W_{C}) &
=C_{Om}(W_{O}),\;\omega_{Cm}(W_{C})=\omega_{Om}(W_{O}),\;|m|\geq1.
\end{align*}

Connection of functions $C(\rho)$ and $O(r)$:%
\begin{align}
C(\rho) & =A(\rho)O(r),\;A(\rho)=\frac{R_{C}^{2}\rho^{1/2}}{%
R_{C}^{2}-\rho^{2}}\frac{R^{2}-r^{2}}{R^{2}r^{1/2}}=\frac{%
R^{2}(\kappa_{0}r)^{1/2}}{R^{2}+r^{2}}=\frac{R_{C}(\kappa_{0}\rho)^{1/4}}{%
R_{C}+\rho},   \label{Cps2.3.1} \\
O(r) & =B(r)C(\rho),\;B(r)=\frac{R^{2}r^{1/2}}{R^{2}-r^{2}}\frac{%
R_{C}^{2}-\rho^{2}}{R_{C}^{2}\rho^{1/2}}=\frac{R_{C}+\rho}{%
R_{C}(\kappa_{0}\rho)^{1/4}}=\frac{R^{2}+r^{2}}{R^{2}(\kappa_{0}r)^{1/2}}, 
\notag
\end{align}%
\begin{equation*}
\mathrm{Wr}_{\rho}(C_{1},C_{2})=\frac{1}{2}\mathrm{Wr}_{r}(O_{1},O_{2}) 
\end{equation*}

Connections of asymptotics of functions $C(\rho)$ and $O(r)$:

i) $r\rightarrow0$, $\rho\rightarrow0$%
\begin{equation*}
C(\rho)=(\kappa_{0}r)^{1/2}O(r),\;O(r)=(\kappa_{0}\rho)^{-1/4}C(\rho). 
\end{equation*}

ii) $R-r=\Delta_{O}\rightarrow0$, $R_{C}-\rho=\Delta_{C}\rightarrow0$%
\begin{align*}
\Delta_{C} &
=2\kappa_{0}R\Delta_{O},\;\Delta_{O}=(4\kappa_{0}R_{C})^{-1/2}\Delta_{C}, \\
C(\rho) & =\frac{1}{2}(\kappa_{0}R)^{1/2}O(r),\;O(r)=2(%
\kappa_{0}R_{C})^{-1/4}C(\rho).
\end{align*}

One of the results of these considerations is the following.

Let $\Psi_{Om}(r)=x_{O}^{\tilde{\mu}_{O}}(1-x_{O})^{1/4+\tilde{\nu}%
_{O}}\Phi_{m}(x_{O})$ is the solution of equation%
\begin{equation*}
(\check{H}_{Om}-W_{O})\Psi_{Om}(r)=0. 
\end{equation*}
Then the function $\Psi_{Cm}(\rho)=x_{C}^{\tilde{\mu}_{C}}(1-x_{C})^{1/4+%
\tilde{\nu}_{C}}\Phi_{m}(x_{C})$ is the solution of eq. (\ref{Cps2.1.2}) and
reverse.

Let $\Psi_{Om\delta}(r)=x_{O}^{\tilde{\mu}_{O\delta}}(1-x_{O})^{1/4+\tilde {%
\nu}_{O}}\Phi_{m\delta}(x_{O})$ is the solution of eq. (\ref{Opso2.3.1.2a}).
Then the function $\Psi_{Cm\delta}(\rho)=x_{C}^{\tilde{\mu}%
_{C\delta}}(1-x_{C})^{1/4+\tilde{\nu}_{C}}\Phi_{m\delta}(x_{C})$ is the
solution of eq. (\ref{Cps2.1.3}) and reverse.

Let $\psi_{Om}(r)$,%
\begin{equation}
\psi_{Om}(r)=\frac{R^{2}\sqrt{r}}{R^{2}-r^{2}}\Psi_{Om}(r)=\frac{R^{2}\sqrt {%
r}}{R^{2}-r^{2}}x_{O}^{\tilde{\mu}_{O}}(1-x_{O})^{1/4+\tilde{\nu}%
_{O}}\Phi_{m}(x_{O}),   \label{Cps2.3.2a}
\end{equation}
is the solution of eq. (\ref{Opso2.3.1.1}). Then the function $%
\psi_{Cm}(\rho)$,%
\begin{equation}
\psi_{Cm}(\rho)=\frac{R_{C}^{2}\sqrt{\rho}}{R_{C}^{2}-\rho^{2}}%
\Psi_{Cm}(\rho)=\frac{R_{C}^{2}\sqrt{\rho}}{R_{C}^{2}-\rho^{2}}x_{C}^{\tilde{%
\mu}_{C}}(1-x_{C})^{1/4+\tilde{\nu}_{C}}\Phi_{m}(x_{C}),   \label{Cps2.3.2b}
\end{equation}
is the solution of eq. (\ref{Cps2.2.1}) and reverse.

Let $\psi_{Om\delta}(r)=\frac{R^{2}\sqrt{r}}{R^{2}-r^{2}}\Psi_{Om\delta }(r)=%
\frac{R^{2}\sqrt{r}}{R^{2}-r^{2}}x_{O}^{\tilde{\mu}_{O\delta}}(1-x_{O})^{1/4+%
\tilde{\nu}_{O}}\Phi_{m\delta}(x_{O})$ is the solution of eq. (\ref%
{Opso2.3.1.2}). Then the function $\psi_{Cm\delta}(\rho)=\frac{R_{C}^{2}%
\sqrt{\rho}}{R_{C}^{2}-\rho^{2}}\Psi_{Cm\delta}(\rho)=\frac{R_{C}^{2}\sqrt{%
\rho}}{R_{C}^{2}-\rho^{2}}x_{C}^{\tilde{\mu}_{C}}(1-x_{C})^{1/4+\tilde {\nu}%
_{C}}\Phi_{m\delta}(x_{C})$ is the solution of eq. (\ref{Cps2.2.2}) and
reverse.

\subsection{$|m|\geq2$}

\subsubsection{Useful solutions, $\protect\rho<R_{C}$}

We construct the solutions of eq. (\ref{Cps2.2.1}) with the help of
correspondence formulas (\ref{Cps2.3.1}), (\ref{Cps2.3.2a}), and (\ref%
{Cps2.3.2b}). We have%
\begin{align*}
& C_{1,m}(\rho;W)=\frac{R_{C}^{2}\sqrt{\rho}}{R_{C}^{2}-\rho^{2}}x^{\mu
}(1-x)^{1/4+\nu}\mathcal{F}(\alpha_{1},\beta_{1};\gamma_{1};x)= \\
& (\mathrm{for}~\operatorname{Im}W>0)=Q_{1}(W)C_{3,m}(\rho;W)+Q_{2}(W)v_{m}(\rho;W),%
\;\gamma_{1}=1+2\mu, \\
& Q_{1}(W)=\frac{\Gamma(\gamma_{1})\Gamma(-2\nu)}{\Gamma(\alpha_{4})\Gamma(%
\beta_{4})},\;Q_{2}(W)=\frac{\Gamma(\gamma_{1})\Gamma(2\nu)}{%
\Gamma(\alpha_{1})\Gamma(\beta_{1})},
\end{align*}%
\begin{equation*}
v_{m}(\rho;W)=\frac{R_{C}^{2}\sqrt{\rho}}{R_{C}^{2}-\rho^{2}}x^{\mu
}(1-x)^{1/4-\nu}\mathcal{F}(\alpha_{4},\beta_{4};\gamma_{4};1-x), 
\end{equation*}%
\begin{align*}
& C_{4,m}(\rho;W)=\frac{R_{C}^{2}\sqrt{\rho}}{R_{C}^{2}-\rho^{2}}\frac {%
R^{2}-r^{2}}{R^{2}r^{1/2}}O_{4,m}^{<}(r;W)= \\
& \,=\frac{R_{C}^{2}\sqrt{\rho}}{R_{C}^{2}-\rho^{2}}(1-x)^{1/4+\nu}\lim_{%
\delta\rightarrow0}[x^{-\mu_{\delta}}\mathcal{F}(\alpha_{2\delta},\beta_{2%
\delta};\gamma_{2\delta};x)- \\
& \,-A_{m,\delta}(W)\Gamma(\gamma_{2\delta})x^{\mu_{\delta}}\mathcal{F}%
(\alpha_{1\delta},\beta_{1\delta};\gamma_{1\delta};x)], \\
& A_{m,\delta}(W)=\frac{\Gamma(\alpha_{1})\Gamma(\beta_{1})}{\Gamma
(\alpha_{2})\Gamma(\beta_{2})\Gamma(\gamma_{1\delta})},
\end{align*}%
\begin{align*}
& C_{3,m}(r;W)=\frac{R_{C}^{2}\sqrt{\rho}}{R_{C}^{2}-\rho^{2}}x^{\mu
}(1-x)^{1/4+\nu}\mathcal{F}(\alpha_{1},\beta_{1};\gamma_{3};1-x)= \\
& =B_{m}(W)C_{1,m}(\rho;W)+C_{m}(W)C_{4,m}(\rho;W),\;C_{m}(W)=\frac {%
\Gamma(\gamma_{3})\Gamma(|m|)}{\Gamma(\alpha_{1})\Gamma(\beta_{1})}, \\
& B_{m}(W)=\frac{(-1)^{|m|+1}\Gamma(\gamma_{3})}{2\Gamma(\alpha_{2})\Gamma(%
\beta_{2})\Gamma(\gamma_{1})}\left[ \psi(\alpha_{1})+\psi(\alpha
_{2})+\psi(\beta_{1})+\psi(\beta_{2})\right] ,
\end{align*}%
\begin{align*}
& \alpha_{1,2}=1/2\pm\mu+\nu+\sigma,\;\beta_{1,2}=1/2\pm\mu+\nu-\sigma, \\
& \alpha_{4}=1/2+\mu-\nu+\sigma,\;\beta_{4}=1/2+\mu-\nu-\sigma, \\
& \gamma_{1}=1+2\mu,\;\gamma_{3,4}=1\pm2\nu,\;\mu=|m|/2,\;\sigma=\frac{1}{4}%
\sqrt{1-w_{C}}, \\
& \nu=\frac{1}{4}\sqrt{1+4gR_{C}-w_{C}},\;w_{C}=R_{C}^{2}W_{C}.
\end{align*}

$C_{1,m}(\rho;W)$ and $C_{4,m}(\rho;W)$ are real-entire in $W$.

\subsubsection{Asymptotics, $\protect\rho\rightarrow0$ ($x\rightarrow0$)}

We have%
\begin{align*}
& x=\frac{4\rho}{R_{C}}(1+O(\rho)),\;p_{0}(\rho)=1+O(\rho^{2}), \\
& C_{1,m}(\rho;W)=(4/R_{C})^{|m|/2}\rho^{1/2+|m|/2}(1+O(\rho)), \\
& C_{4,m}(\rho;W)=(R_{C}/4)^{|m|/2}\rho^{1/2-|m|/2}\left( 1+O(\rho)\right) ,
\end{align*}

\begin{align*}
& C_{3,m}(\rho;W)=.\frac{\Gamma(\gamma_{3})\Gamma(|m|)}{\Gamma(\alpha
_{1})\Gamma(\beta_{1})}(R_{C}/4)^{|m|/2}\rho^{1/2-|m|/2}\left( 1+O(\rho
)\right) , \\
& \operatorname{Im}W>0\;\mathrm{or}\;W=0.
\end{align*}

\subsubsection{Asymptotics, $\Delta=R_{C}-\protect\rho\rightarrow0$ ($%
\protect\delta =1-x\rightarrow0$)}

We have%
\begin{align*}
& \delta=\frac{\Delta^{2}}{4R_{C}^{2}}(1+O(\Delta)),\;p_{0}(r)=\frac {%
4\Delta^{2}}{R_{C}^{2}}(1+O(\Delta)), \\
& C_{3,m}(\rho;W)=2^{-3/2-2\nu}R_{C}^{1-2\nu}\Delta^{-1/2+2\nu}(1+O(\Delta)),
\end{align*}

\begin{align*}
& C_{1,m}(\rho;W)=\frac{\Gamma(\gamma_{1})\Gamma(2\nu)}{\Gamma(\alpha
_{1})\Gamma(\beta_{1})}2^{-3/2+2\nu}R_{C}^{1+2\nu}\Delta^{-1/2-2\nu
}(1+O(\Delta)), \\
& \operatorname{Im}W>0\;\mathrm{or}\;W=0.
\end{align*}

\subsubsection{Wronskian}

\begin{equation*}
\mathrm{Wr}(C_{1,m},C_{3,m})=-\frac{\Gamma(\gamma_{1})\Gamma(\gamma_{3})}{%
\Gamma(\alpha_{1})\Gamma(\beta_{1})}=-\omega_{m}(W)=-|m|C_{m}(W). 
\end{equation*}
We see that eq. (\ref{Cps2.2.1}) has no s.-integrable silutions for $\operatorname{Im%
}W>0$. This means that the deficiency indices of the symmetric operator (see
below) are equal to $(0,0)$.

\subsubsection{Symmetric operator $\hat{h}_{m}$}

A symmetric operator $\hat{h}_{m}$ is defined in the Hilbert space $%
\mathfrak{h}_{m}=L^{2.}(0,R_{C})$ by equation 
\begin{equation}
\hat{h}_{m}:\left\{ 
\begin{array}{l}
D_{h_{m}}=\mathcal{D}(0,R_{C}) \\ 
\hat{h}_{m}\psi_{m}=\check{h}_{m}\psi_{m},\;\forall\psi_{m}\in D_{h_{m}},%
\end{array}
\right. ,   \label{Cps2.4.2.1}
\end{equation}
the differential operation $\check{h}_{m}$ is given by eq. (\ref{Cps2.2.0}).

\subsubsection{Adjoint operator $\hat{h}_{m}^{+}=\hat{h}_{m}^{\ast}$}

It is easy to prove by standard way that the adjoint operator $\hat{h}%
_{m}^{+}$ coincides with the operator $\hat{h}_{m}^{\ast}$,

\begin{equation*}
\hat{h}_{m}^{+}:\left\{ 
\begin{array}{l}
D_{h_{m}^{+}}=D_{\check{h}_{m}}^{\ast}(0,R_{C})=\{\psi_{\ast},\psi_{\ast
}^{\prime}\;\mathrm{are\;a.c.\;in}\mathcal{\;}(0,R_{C}),\; \\ 
\psi_{\ast},\hat{h}_{m}^{+}\psi_{\ast}\in L^{2}(0,R_{C})\} \\ 
\hat{h}_{m}^{+}\psi_{\ast}(\rho)=\check{h}_{m}\psi_{\ast}(\rho),\;\forall
\psi_{\ast}\in D_{h_{m}^{+}}%
\end{array}
\right. . 
\end{equation*}

\subsubsection{Asymptotics}

Because $\check{h}_{m}\psi_{\ast}\in L^{2}(0,R_{C})$, we have%
\begin{equation*}
\check{h}_{m}\psi_{\ast}(\rho)=\eta(\rho),\;\eta\in L^{2}(0,R_{C}), 
\end{equation*}
and we can represent $\psi_{\ast}$ in the form%
\begin{align*}
\psi_{\ast}(\rho) & =c_{1}C_{1,m}(\rho;0)+c_{2}C_{3,m}(\rho;0)+I(\rho), \\
\psi_{\ast}^{\prime}(\rho) & =c_{1}\partial_{\rho}C_{1,m}(\rho
;0)+c_{2}\partial_{\rho}C_{3,m}(\rho;0)+I^{\prime}(\rho),
\end{align*}
where%
\begin{align*}
I(\rho) & =\frac{C_{1,m}(\rho;0)}{\omega_{m}(0)}\int_{%
\rho}^{R_{C}}C_{3,m}(y;0)\eta(y)dy+\frac{C_{3,m}(\rho;0)}{\omega_{m}(0)}%
\int_{0}^{\rho }C_{1,m}(y;0)\eta(y)dy, \\
I^{\prime}(\rho) & =\frac{\partial_{\rho}C_{1,m}(\rho;0)}{\omega_{m}(0)}%
\int_{\rho}^{R}C_{3,m}(y;0)\eta(y)dy+\frac{\partial_{\rho}C_{3,m}(\rho ;0)}{%
\omega_{m}(0)}\int_{0}^{\rho}C_{1,m}(y;0)\eta(y)dy.
\end{align*}

I) $\rho\rightarrow0$

We obtain with the help of the Cauchy-Bunyakovskii inequality
(CB-inequality):%
\begin{equation*}
I(\rho)=\left\{ 
\begin{array}{c}
O(\rho^{3/2}),\;|m|\geq3 \\ 
O(\rho^{3/2}\sqrt{\ln\rho}),\;|m|=2%
\end{array}
\right. ,\;I^{\prime}(\rho)=\left\{ 
\begin{array}{c}
O(\rho^{1/2}),\;|m|\geq3 \\ 
O(\rho^{1/2}\sqrt{\ln\rho}),\;|m|=2%
\end{array}
\right. , 
\end{equation*}
such that we have%
\begin{align*}
& \psi_{\ast}(\rho)=c_{2}c\rho^{1/2-|m|/2}\left( 1+O(\rho)\right) +\left\{ 
\begin{array}{c}
O(\rho^{3/2}),\;|m|\geq3 \\ 
O(\rho^{3/2}\sqrt{\ln\rho}),\;|m|=2%
\end{array}
\right. , \\
& c=(R_{C}/4)^{|m|/2}\frac{\Gamma(\gamma_{3})\Gamma(2\mu)}{\Gamma(\alpha
_{1})\Gamma(\beta_{1})}.
\end{align*}
The condition $\psi_{\ast}\in L^{2}(0,R_{C})$ gives $c_{2}=0$, such that we
find finally%
\begin{align*}
& \psi_{\ast}(\rho)=\left\{ 
\begin{array}{c}
O(\rho^{3/2}),\;|m|\geq3 \\ 
O(\rho^{3/2}\sqrt{\ln\rho}),\;|m|=2%
\end{array}
\right. ,\;\psi_{\ast}^{\prime}(\rho)=\left\{ 
\begin{array}{c}
O(\rho^{1/2}),\;|m|\geq3 \\ 
O(\rho^{1/2}\sqrt{\ln\rho}),\;|m|=2%
\end{array}
\right. , \\
& \,[\psi_{\ast},\chi_{\ast}]_{0}=0,\;\forall\psi_{\ast},\chi_{\ast}\in
D_{h_{m}^{+}}.
\end{align*}

II) $\rho\rightarrow R_{C}$

In this case, we prove that $[\psi_{\ast},\chi_{\ast}]^{R_{C}}=0$, $%
\forall\psi_{\ast},\chi_{\ast}\in D_{h_{m}^{+}}$. Indeed, consider the
Hilbert space $\mathfrak{h}_{c,m}=L^{2}(c,R_{C})$, $c$ is an interior point
of the interval $(0,R_{C})$. and an symmetric operator $\hat{h}_{c,m}$, $%
D_{h_{c,m}}=\mathcal{D}(c,R_{C})$, acting as $\check{h}_{m}$. We choose the
functions $C_{1,m}(\rho;W)$ and $C_{3,m}(\rho;W)$ as the independent
solutions of eq. (\ref{Cps2.2.1}) for $\operatorname{Im}W>0$. The left end $c$ of
the interval $(0,R_{C})$ is regular and both solutions $C_{1,m}$ and $C_{3,m}
$ are s.-interable on the end $c$. The right end $R_{C}$ is singular. On the
right end $R_{C}$, the solution $C_{3,m}$ is s.-integrable, but $C_{1,m}$ is
not. Thus, there is only one s. integrable solution of eq. (\ref{Cps2.2.1})
on the interval $(c,R_{C})$ for $\operatorname{Im}W>0$ and the deficient indexes of
the symmetric operator $\hat{h}_{c,m}$ are equal to $(1,1)$. In this case,
according to [\cite{Naima}, Lemma on the page 213], we have $%
[\psi_{\ast},\chi_{\ast }]^{R_{C}}=0$, $\forall\psi_{\ast},\chi_{\ast}\in
D_{h_{c,m}^{+}}$. Because the restriction $\psi_{c\ast}$ on the interval $%
(c,R_{C})$ of any function $\psi_{\ast}\in D_{h_{m}^{+}}$ belongs to $%
D_{h_{c,m}^{+}}$, $\psi_{c\ast}\in D_{h_{c,m}^{+}}$. $\forall\psi_{\ast}\in
D_{h_{m}^{+}}$, we obtain that $[\psi_{\ast},\chi_{\ast}]^{R_{C}}=0$, $%
\forall\psi_{\ast},\chi_{\ast}\in D_{h_{m}^{+}}$.

\subsubsection{Self-adjoint hamiltonian $\hat{h}_{\mathfrak{e}m}$}

Because $\omega_{h_{m}^{+}}(\chi_{\ast},\psi_{\ast})=\Delta_{h_{m}^{+}}(%
\psi_{\ast})=0$ (and also because $C_{1,m}(\rho;W)$ and $C_{3,r}(u;W)$ and
any their linear combinations are not s.-integrable on the interval $%
(0,R_{C})$ for $\operatorname{Im}W\neq0$), the deficiency indices of initial
symmetric operator $\hat{h}_{m}$ are zero, which means that $\hat {h}_{%
\mathfrak{e}m}=\hat{h}_{m}^{+}$ is a unique s.a. extension of the initial
symmetric operator $\hat{h}_{m}$:%
\begin{equation*}
\hat{h}_{\mathfrak{e}m}:\left\{ 
\begin{array}{l}
D_{h_{\mathfrak{e}m}}=D_{\check{h}_{m}}^{\ast}(0,R_{C}) \\ 
\hat{h}_{\mathfrak{e}m}\psi_{\ast}(\rho)=\check{h}_{m}\psi_{\ast}(\rho),\;%
\forall\psi_{\ast}\in D_{h_{\mathfrak{e}m}}%
\end{array}
\right. . 
\end{equation*}

\subsubsection{The guiding functional $\Phi(\protect\xi;W)$}

As a guiding functional $\Phi(\xi;W)$ we choose%
\begin{align*}
& \Phi(\xi;W)=\int_{0}^{R_{C}}C_{1,m}(y;W)\xi(y)dy,\;\xi\in\mathbb{D}%
=D_{r}(0,R_{C})\cap D_{h_{\mathfrak{e}m}}. \\
& D_{r}(0,R_{C})=\{\xi(u):\;\mathrm{supp}\xi\subseteq\lbrack0,\beta_{\xi
}],\;\beta_{\xi}<R_{C}\}.
\end{align*}

The guiding functional $\Phi(\xi;W)$ is simple and the spectrum of $\hat {h}%
_{\mathfrak{e}m}$ is simple.

\subsubsection{Green function $G_{m}(\protect\rho,y;W)$, spectral function $%
\protect\sigma _{m}(E)$}

We find the Green function $G_{m}(\rho,y;W)$ as the kernel of the integral
representation 
\begin{equation*}
\psi(\rho)=\int_{0}^{R_{C}}G_{m}(\rho,y;W)\eta(y)dy,\;\eta\in
L^{2}(0,R_{C}), 
\end{equation*}
of unique solution of an equation%
\begin{equation}
(\hat{h}_{\mathfrak{e}m}-W)\psi(\rho)=\eta(\rho),\;\operatorname{Im}W>0, 
\label{Cps2.4.6.1}
\end{equation}
for $\psi\in D_{h_{\mathfrak{e}m}}$. General solution of eq. (\ref%
{Cps2.4.6.1}) can be represented in the form%
\begin{align*}
\psi(\rho) & =a_{1}C_{1,m}(\rho;W)+a_{3}C_{3,m}(\rho;W)+I(\rho), \\
I(\rho) & =\frac{C_{1,m}(\rho;W)}{\omega_{m}(W)}\int_{%
\rho}^{R_{C}}C_{3,m}(y;W)\eta(y)dy+\frac{C_{3,m}(\rho;W)}{\omega_{m}(W)}%
\int_{0}^{\rho }C_{1,m}(y;W)\eta(y)dy, \\
I(\rho) & =\left\{ 
\begin{array}{c}
O(\rho^{3/2}),\;|m|\geq3 \\ 
O(\rho^{3/2}\sqrt{\ln\rho}),\;|m|=2%
\end{array}
\right. ,\;\rho\rightarrow0,\;I(\rho)=O\left( \Delta^{-1/2}\right)
,\;\rho\rightarrow R_{C}.
\end{align*}
A condition $\psi\in L^{2}(0,\rho)$ gives $a_{1}=a_{3}=0$, such that we find%
\begin{align}
& G_{m}(\rho,y;W)=\frac{1}{\omega_{m}}\left\{ 
\begin{array}{l}
C_{3,m}(\rho;W)C_{1,m}(y;W),\;\rho>y \\ 
C_{1,m}(\rho;W)C_{3,m}(y;W).\;\rho<y%
\end{array}
\right. =  \notag \\
& \,=\pi\Omega_{m}(W)C_{1,m}(\rho;W)C_{1,m}(y;W)+\frac{1}{|m|}\left\{ 
\begin{array}{c}
C_{4,m}(\rho;W)C_{1,m}(y;W),\;\rho>y \\ 
C_{1,m}(\rho;W)C_{4,m}(y;W),\;\rho<y%
\end{array}
\right. ,  \label{Cps2.4.6.2} \\
& \Omega_{m}(W)\equiv\frac{B_{m}(W)}{\pi\omega_{m}(W)}=\frac{%
(-1)^{|m|+1}[\psi(\alpha_{1})+\psi(\alpha_{2})+\psi(\beta_{1})+\psi(%
\beta_{2})]\mathcal{A}_{m}(W)}{2\pi\Gamma^{2}(\gamma_{1})},  \notag \\
& \mathcal{A}_{m}(W)=\frac{\Gamma(\alpha_{1})\Gamma(\beta_{1})}{\Gamma
(\alpha_{2})\Gamma(\beta_{2})}.  \notag
\end{align}
Note that the last term in the r.h.s. of eq. (\ref{Cps2.4.6.2}) is real for $%
W=E$. From the relation%
\begin{equation*}
\lbrack C_{1,m}(\rho_{0};E)]^{2}\sigma_{m}^{\prime}(E)=\frac{1}{\pi }\operatorname{Im%
}G_{m}(\rho_{0}-0,\rho_{0}+0;E+i0), 
\end{equation*}
where $f(E+i0)\equiv\lim_{\varepsilon\rightarrow+0}f(E+i\varepsilon)$, $%
\forall f(W)$, we find%
\begin{equation*}
\sigma_{m}^{\prime}(E)=\operatorname{Im}\Omega_{m}(E+i0). 
\end{equation*}

Consider $\Omega_{m}(W)$ in more details.

Using relations%
\begin{equation*}
\psi(\alpha_{2})=\psi(\alpha_{1})-T_{(\alpha)},\;T_{(\alpha)}=%
\sum_{k=1}^{|m|}\frac{1}{\alpha_{1}-k}, 
\end{equation*}%
\begin{equation*}
\psi(\beta_{2})=\psi(\beta_{1})-T_{(\beta)},\;T_{(\beta)}=\sum_{k=1}^{|m|}%
\frac{1}{\beta_{1}-k}, 
\end{equation*}%
\begin{equation*}
T_{(\alpha,\beta)}(W)=T_{(\alpha)}+T_{(\beta)}=\frac{2\nu\mathcal{B}_{m}(W)}{%
\mathcal{A}_{m}(W)}, 
\end{equation*}
where $\mathcal{B}_{m}(W)$ is an polinomial in $\nu$ and $\sigma$, even in
both $\nu$ and $\sigma$, and therefore, is a real-entire polynomial in $W$,
we can represent $\Omega_{m}(W)$ in the form%
\begin{align*}
& \Omega_{m}(W)=\Omega_{1m}(W)+\Omega_{2m}(W), \\
& \Omega_{1m}(W)=\frac{(-1)^{|m|+1}\mathcal{A}_{m}(W)}{\pi\Gamma^{2}(%
\gamma_{1})}\left[ \psi(\alpha_{1})+\psi(\beta_{1})\right] , \\
& \Omega_{2m}(W)=-\nu\tilde{\Omega}_{2m}(W),\;\tilde{\Omega}_{2m}(W)=\frac{%
(-1)^{|m|+1}\mathcal{B}_{m}(W)}{\pi\Gamma^{2}(\gamma_{1})}.
\end{align*}

\subsubsection{Spectrum}

\subsubsection{$W=(1+4R_{C}g)/R_{C}^{2}+\tilde{\Delta}$, $\tilde{\Delta}\sim
0$}

A direct estimation gives 
\begin{equation*}
\Omega_{m}(W)=\left\{ 
\begin{array}{c}
\Omega_{m}(W_{0})+O(\sqrt{\tilde{\Delta}}),\;g\neq g_{m,k} \\ 
O(1/\sqrt{\tilde{\Delta}}),\;g=g_{m,k},%
\end{array}
\right. , 
\end{equation*}
$R_{C}g_{m,k}=-N_{m,k}^{2}$, $N_{m,k}=1+|m|+2k,W_{0}=(1+4R_{C}g)/R_{C}^{2},%
\operatorname{Im}[\Omega_{m}(W_{0})]=0$. This result means that the levels with $%
E=E_{0}$ are absent.

\subsubsection{$w=R_{C}^{2}E>1+4R_{C}g$}

In this case, we have $\alpha_{1}=1/2+|m|/2+\sigma-i\sqrt{w-1-4R_{C}g}/4$, $%
\beta_{1}=1/2+|m|/2-\sigma-i\sqrt{w-1-4R_{C}g}/4$ and it is easy to prove
that $\alpha_{1},\beta_{1}\notin\mathbb{Z}_{-}$, such that $\Omega_{m}(E)$
is finite complex function of $E$ and we have 
\begin{equation*}
\sigma_{m}^{\prime}(E)=\operatorname{Im}\Omega_{m}(E)\equiv\varrho_{m}^{2}(E)>0. 
\end{equation*}
The spectrum of $\hat{h}_{\mathfrak{e}m}$\ is simple and continuous, $%
\mathrm{spec}\hat{h}_{\mathfrak{e}m}=[(1+4R_{C}g)/R_{C}^{2},\infty)$. Note
that%
\begin{equation}
\left. \sigma_{m}^{\prime}(E)\right\vert _{\Delta\rightarrow+0}=\left\{ 
\begin{array}{c}
O(\sqrt{\tilde{\Delta}}),\;g\neq g_{m,k} \\ 
O(1/\sqrt{\tilde{\Delta}}),\;g=g_{m,k}%
\end{array}
\right. ,   \label{Cps2.4.7.1.1}
\end{equation}
$\Delta=E-(1+4R_{C}g)/R_{C}^{2}\rightarrow+0$.

\subsubsection{$w=R_{C}^{2}E\leq1+4R_{C}g$}

\paragraph{$E<1/R_{C}^{2}$}

In this case, we have $\left. \operatorname{Im}\nu\right| _{W=E}=\left. \operatorname{Im}%
\sigma\right| _{W=E}=\left. \operatorname{Im}\alpha _{1}\right| _{W=E}=\left. \operatorname{%
Im}\beta_{1}\right| _{W=E}=0$, $\operatorname{Im}\left. \psi(\alpha_{1})\right|
_{W=E}=0$, and%
\begin{equation*}
\sigma_{m}^{\prime}(E)=\frac{(-1)^{|m|+1}\mathcal{A}_{m}(E)}{\pi\Gamma
^{2}(\gamma_{1})}\operatorname{Im}\left. \psi(\beta_{1})\right| _{W=E+i0}. 
\end{equation*}
$\sigma_{m}^{\prime}(E)$ can be different from zero in the points $E_{m,n}$
when $\beta_{1}=-n$, $n=0,1,2,...$, i.e., $E_{m,n}$ satisfy the equations%
\begin{equation}
\sqrt{1-w_{m,n}}=\sqrt{1-w_{m,n}+4R_{C}g}+2N_{m,n},\;N_{m,n}=1+|m|+2n. 
\label{Cps2.4.7.2.1}
\end{equation}

Eq. (\ref{Cps2.4.7.2.1}) has solutions only if $g<0$ and only one inequality%
\begin{equation}
w_{m,n}\leq1-4R_{C}|g|   \label{Cps2.4.7.2.2}
\end{equation}
must be satisfied \ It follows from eq.(\ref{Cps2.4.7.2.1}) that%
\begin{equation}
N_{m,n}\sqrt{1-w_{m,n}-4R_{C}|g|}=R_{C}|g|-N_{m,n}^{2}, 
\label{Cps2.4.7.2.3}
\end{equation}
and we obtain one more inequality%
\begin{equation}
R_{C}|g|\geq N_{m,n}^{2}.   \label{Cps2.4.7.2.4}
\end{equation}
It follows from eq.(\ref{Cps2.4.7.2.3}) that%
\begin{align*}
& w_{m,n}=1-(R_{C}|g|/N_{m,n}+N_{m,n})^{2}= \\
& \,=1-4R_{C}|g|-R_{C}|g|\left[ (a+a^{-1})^{2}-4\right] , \\
& E_{m,n}=w_{m,n}/R_{C}^{2},\;a=N_{m,n}/\sqrt{R_{C}|g|},
\end{align*}
so that inequality (\ref{Cps2.4.7.2.2}) is satisfied. It follows from
inequality (\ref{Cps2.4.7.2.4}) and eq. (\ref{Cps2.4.7.1.1}) that there are
no levels for $g\geq g_{m,0}=-N_{m,0}^{2}/R_{C}=-(1+|m|)^{2}/R_{C}$ and
there are $n_{\max}+1$ levels $n=0,1,...,n_{\max}=k$ for $\sqrt{R_{C}|g|}%
=1+|m|+2(k+\delta)$, $k=0,1,...,$ $0<\delta\leq1$. Note that $%
1-4R_{C}|g|>E_{m,n}>E_{m,n-1}$, $n=1,...,n_{\max}$. \ 

We have%
\begin{equation*}
\sigma_{m}^{\prime}(E)=\sum_{n=0}^{n_{\max}}Q_{m,n}^{2}\delta(E-E_{m,n}),%
\;Q_{m,n}=\sqrt{\frac{(-1)^{|m|}4\mathcal{A}%
_{m}(E_{m,n})(R_{C}^{2}g^{2}-N_{m,n}^{4})}{\Gamma^{2}(%
\gamma_{1})N_{m,n}^{3}R_{C}^{2}}}.. 
\end{equation*}
The discrete part of the spectrum of $\hat{h}_{\mathfrak{e}m}$\ is simple
and has the form%
\begin{equation*}
\mathrm{spec}\hat{h}_{\mathfrak{e}m}=\{E_{m,n},\;E_{m,n}<1-4R_{C}|g|,%
\;n=0,1,...,n_{\max}\}. 
\end{equation*}
The discrete part of the spectrum is absent for $g\geq g_{m,0}$.

\paragraph{$w=1$, $\protect\sigma=0$}

We have in this case for $W=E=R_{C}^{-2}$: $g\geq0$, $\alpha_{1}=\beta
_{1}=1/2+|m|/2+\sqrt{R_{C}g}/2$, $\operatorname{Im}\nu=0$, $\operatorname{Im}\alpha_{1}=0$, $%
\alpha_{1}>0$, , and $\sigma_{m}^{\prime}(R_{C}^{-2})=0$.

\paragraph{$w>1$, $\protect\sigma=i\varkappa/4$, $\varkappa=\protect\sqrt{w-1%
}$}

In this case, we have for $W=E$:$\operatorname{Im}\nu=0$,$\;\alpha
_{1,2}=1/2\pm|m|/2+\nu+i\varkappa/4$, $\beta_{1,2}=1/2\pm|m|/2+\nu
-i\varkappa/4=\overline{\alpha_{1}}$, such that $\left[ \operatorname{Im}%
\psi(\alpha_{1,2})+\operatorname{Im}\psi(\beta_{1,2})\right] _{W=E}=0$, and%
\begin{equation*}
\sigma_{m}^{\prime}(E)=0. 
\end{equation*}

Finally, we find for fixed $m$, $|m|\geq2$:

The spectrum of $\hat{h}_{\mathfrak{e}m}$\ is simple, $\mathrm{spec}\hat {h}%
_{\mathfrak{e}m}=[1+4R_{C}g)/R_{C}^{2},\infty)\cup\{E_{m,n},\;n=0,1,...n_{%
\max}\}$, the discrete part of spectrum is present for $g<g_{m,0}$. The set
of functions 
\begin{equation*}
\left\{
U_{m}(\rho;E)=\varrho_{m}(E)C_{1m}(\rho;E),\;E\geq1+4R_{C}g)/R_{C}^{2};%
\;U_{m,n}(\rho)=Q_{m,n}C_{1m}(\rho;E_{m,n}),\;n=0,1,...n_{\max }\right\} 
\end{equation*}
forms a complete orthogonalized system in $L^{2}(0,R_{C})$.

\subsection{$m=1$}

\subsubsection{Useful solutions, $\protect\rho<R_{C}$}

We construct again the solutions of eq. (\ref{Cps2.2.1}) with the help of
correspondence formulas (\ref{Cps2.3.1}), (\ref{Cps2.3.2a}), and (\ref%
{Cps2.3.2b}). We have%
\begin{align*}
& C_{1,1}(\rho;W)=\frac{R_{C}^{2}\sqrt{\rho}}{R_{C}^{2}-\rho^{2}}%
x^{1/2}(1-x)^{1/4+\nu}\mathcal{F}(\alpha_{1},\beta_{1};2;x)= \\
& (\mathrm{for\ \ }\operatorname{Im}W>0)=Q_{1}(W)C_{3,1}(\rho;W)+Q_{2}(W)v_{1}(%
\rho;W), \\
& Q_{1}(W)=\frac{\Gamma(-2\nu)}{\Gamma(\alpha_{4})\Gamma(\beta_{4})}%
,\;Q_{2}(W)=\frac{\Gamma(2\nu)}{\Gamma(\alpha_{1})\Gamma(\beta_{1})},
\end{align*}%
\begin{equation*}
v_{1}(\rho;W)=\frac{R_{C}^{2}\sqrt{\rho}}{R_{C}^{2}-\rho^{2}}%
x^{1/2}(1-x)^{1/4-\nu}\mathcal{F}(\alpha_{4},\beta_{4};\gamma_{4};1-x), 
\end{equation*}%
\begin{align*}
& C_{4,1}(\rho;W)=\frac{R_{C}^{2}\sqrt{\rho}}{R_{C}^{2}-\rho^{2}}\frac {%
R^{2}-r^{2}}{R^{2}r^{1/2}}O_{4,1}^{<}(r;W)= \\
& \,=\frac{R_{C}^{2}\sqrt{\rho}}{R_{C}^{2}-\rho^{2}}(1-x)^{1/4+\nu}\lim_{%
\delta\rightarrow0}[x^{-\mu_{\delta}}\mathcal{F}(\alpha_{2\delta},\beta_{2%
\delta};\gamma_{2\delta};x)- \\
& \,-A_{1,\delta}(W)\Gamma(\gamma_{2\delta})x^{\mu_{\delta}}\mathcal{F}%
(\alpha_{1\delta},\beta_{1\delta};\gamma_{1\delta};x)], \\
& A_{1,\delta}(W)=\frac{\Gamma(\alpha_{1})\Gamma(\beta_{1})}{\Gamma
(\alpha_{2})\Gamma(\beta_{2})\Gamma(\gamma_{1\delta})},
\end{align*}%
\begin{align*}
& C_{3,1}(r;W)=\frac{R_{C}^{2}\sqrt{\rho}}{R_{C}^{2}-\rho^{2}}%
x^{1/2}(1-x)^{1/4+\nu}\mathcal{F}(\alpha_{1},\beta_{1};\gamma_{3};1-x)= \\
& =B_{1}(W)C_{1,1}(\rho;W)+C_{1}(W)C_{4,1}(\rho;W),\;C_{1}(W)=\frac {%
\Gamma(\gamma_{3})}{\Gamma(\alpha_{1})\Gamma(\beta_{1})}, \\
& B_{1}(W)=C_{1}(W)f_{1}(W),\;f_{1}(W)=\frac{R_{C}g}{4}\left[ \psi
(\alpha_{1})+\psi(\beta_{1})\right] -\nu,
\end{align*}%
\begin{align*}
&
\alpha_{1,2\delta}=1/2\pm(1+\delta)/2+\nu+\sigma,\;\beta_{1,2\delta}=1/2%
\pm(1+\delta)/2+\nu-\sigma, \\
& \gamma_{1,2\delta}=1\pm(1+\delta) \\
& \alpha_{1,2}=1/2\pm1/2+\nu+\sigma,\;\beta_{1,2}=1/2\pm1/2+\nu-\sigma, \\
& \alpha_{4}=1-\nu+\sigma,\;\beta_{4}=1-\nu-\sigma,\;\gamma_{3,4}=1\pm2\nu.
\end{align*}

$C_{1,1}(\rho;W)$ and $C_{4,1}(\rho;W)$ are real-entire in $W$.

\subsubsection{Asymptotics, $\protect\rho\rightarrow0$ ($x\rightarrow0$)}

We have%
\begin{align*}
& C_{1,1}(\rho;W)=C_{1,1\mathrm{as}}(\rho)(1+O(\rho)), \\
& C_{4,1}(\rho;W)=C_{4,1\mathrm{as}}(\rho)\left( 1+O(\rho\ln\rho)\right) ,
\end{align*}

\begin{align*}
& C_{3,1}(\rho;W)=C_{1}(W)[f_{1}(W)C_{1,1\mathrm{as}}(\rho)+C_{4,1\mathrm{as}%
}(\rho)](1+O(\rho\ln\rho), \\
& \operatorname{Im}W>0\;\mathrm{or}\;W=0.
\end{align*}%
\begin{align*}
& C_{1,1\mathrm{as}}(\rho)=(4/R_{C})^{1/2}\rho, \\
& C_{4,1\mathrm{as}}(\rho)=(R_{C}/4)^{1/2}\left[ 1+g\rho\left( \ln
(4\rho/R_{C})+2\mathbf{C}\right) \right]
\end{align*}

\subsubsection{Asymptotics, $\Delta=R_{C}-\protect\rho\rightarrow0$ ($%
\protect\delta =1-x\rightarrow0$)}

We have%
\begin{equation*}
C_{3,1}(\rho;W)=2^{-3/2-2\nu}R_{C}^{1-2\nu}\Delta^{-1/2+2\nu}(1+O(\Delta)), 
\end{equation*}

\begin{align*}
& C_{1,1}(\rho;W)=\frac{\Gamma(2\nu)}{\Gamma(\alpha_{1})\Gamma(\beta_{1})}%
2^{-3/2+2\nu}R_{C}^{1+2\nu}\Delta^{-1/2-2\nu}(1+O(\Delta)), \\
& \operatorname{Im}W>0\;\mathrm{or}\;W=0.
\end{align*}

\subsubsection{Wronskians}

\begin{align*}
\mathrm{Wr}(C_{1,1},C_{4,1}) & =-1, \\
.\mathrm{Wr}(C_{1,1},C_{3,1}) & =-\frac{\Gamma(\gamma_{3})}{\Gamma
(\alpha_{1})\Gamma(\beta_{1})}=-\omega_{1}(W)=-C_{1}(W)
\end{align*}
We see that any solutions of eq. (\ref{Cps2.2.1}) are s.-integrable at $%
\rho=0$ and there is one (and only one) silution, $C_{3,1}$, which is
s.-integrable on the interval $(0,R_{C})$ for $\operatorname{Im}W>0$. This means
that the deficiency indices of the symmetric operator (see below) are equal
to $(1,1)$.

\subsubsection{Symmetric operator $\hat{h}_{1}$}

A symmetric operator $\hat{h}_{1}$ is defined by eq. (\ref{Cps2.4.2.1}) with 
$m=1$.

\subsubsection{Adjoint operator $\hat{h}_{1}^{+}=\hat{h}_{1}^{\ast}$}

It is easy to prove by standard way that the adjoint operator $\hat{h}%
_{1}^{+}$ coincides with the operator $\hat{h}_{1}^{\ast}$,

\begin{equation*}
\hat{h}_{1}^{+}:\left\{ 
\begin{array}{l}
D_{h_{1}^{+}}=D_{\check{h}_{1}}^{\ast}(0,R_{C})=\{\psi_{\ast},\psi_{\ast
}^{\prime}\;\mathrm{are\;a.c.\;in}\mathcal{\;}(0,R_{C}),\; \\ 
\psi_{\ast},\hat{h}_{1}^{+}\psi_{\ast}\in L^{2}(0,R_{C})\} \\ 
\hat{h}_{1}^{+}\psi_{\ast}(\rho)=\check{h}_{1}\psi_{\ast}(\rho),\;\forall
\psi_{\ast}\in D_{h_{1}^{+}}%
\end{array}
\right. . 
\end{equation*}

\subsubsection{Asymptotics}

Because $\check{h}_{1}\psi_{\ast}\in L^{2}(0,R_{C})$, we have%
\begin{equation*}
\check{h}_{1}\psi_{\ast}(\rho)=\eta(\rho),\;\eta\in L^{2}(0,R_{C}), 
\end{equation*}
and we can represent $\psi_{\ast}$ in the form%
\begin{align*}
\psi_{\ast}(\rho) & =c_{1}C_{1,1}(\rho;0)+c_{2}C_{4,1}(\rho;0)+I(\rho), \\
\psi_{\ast}^{\prime}(\rho) & =c_{1}\partial_{\rho}C_{1,1}(\rho
;0)+c_{2}\partial_{\rho}C_{4,1}(\rho;0)+I^{\prime}(\rho),
\end{align*}
where%
\begin{align*}
I(\rho) &
=C_{4,1}(\rho;0)\int_{0}^{\rho}C_{1,1}(y;0)\eta(y)dy-C_{1,1}(\rho;0)%
\int_{0}^{_{\rho}}C_{4,1}(y;0)\eta(y)dy, \\
I^{\prime}(\rho) &
=\partial_{\rho}C_{4,1}(\rho;0)\int_{0}^{\rho}C_{1,1}(y;0)\eta(y)dy-%
\partial_{\rho}C_{1,1}(\rho;0)\int_{0}^{\rho}C_{4,1}(y;0)\eta(y)dy.
\end{align*}

I) $\rho\rightarrow0$

We obtain with the help of the Cauchy-Bunyakovskii inequality
(CB-inequality):%
\begin{equation*}
I(\rho)=O(\rho^{3/2}),\;I^{\prime}(\rho)=O(\rho^{1/2}), 
\end{equation*}
such that we have%
\begin{align*}
& \psi_{\ast}(\rho)=c_{1}C_{1,1\mathrm{as}}(\rho)+c_{2}C_{4,1\mathrm{as}%
}(\rho)+O(\rho^{3/2}), \\
& \psi_{\ast}^{\prime}(\rho)=c_{1}C_{1,1\mathrm{as}}^{\prime}(\rho
)+c_{2}C_{4,1\mathrm{as}}^{\prime}(\rho)+O(\rho^{1/2}).
\end{align*}

II) $\rho\rightarrow R_{C}$

In this case, we prove that $[\psi_{\ast},\chi_{\ast}]^{R_{C}}=0$, $%
\forall\psi_{\ast},\chi_{\ast}\in D_{h_{m}^{+}}$, such that we obtain%
\begin{equation*}
\Delta_{h_{1}^{+}}=\overline{c_{2}}c_{1}-\overline{c_{1}}c_{2})=\frac{i}{2}(%
\overline{c_{+}}c_{+}-\overline{c_{-}}c_{-}),\;c_{\pm}=c_{1}\pm ic_{2}. 
\end{equation*}

\subsubsection{Self-adjoint hamiltonians $\hat{h}_{1\protect\zeta}$}

The condition $\Delta_{h_{1}^{+}}(\psi)=0$ gives%
\begin{align*}
& c_{-}=-e^{2i\zeta}c_{+},\;|\zeta|\leq\pi/2,\;\zeta=-\pi/2\sim\zeta
=\pi/2,\;\Longrightarrow \\
& \Longrightarrow\;c_{1}\cos\zeta=c_{2}\sin\zeta,
\end{align*}
or%
\begin{align}
& \psi(\rho)=C\psi_{\zeta\mathrm{as}}(\rho)+O(\rho^{3/2}),\;\psi^{\prime
}(\rho)=C\psi_{\zeta\mathrm{as}}^{\prime}(\rho)+O(\rho^{1/2}),
\label{Cps2.5.4.1} \\
& \psi_{\zeta\mathrm{as}}(\rho)=C_{1,1\mathrm{as}}(\rho)\sin\zeta +C_{4,1%
\mathrm{as}}(\rho)\cos\zeta.  \notag
\end{align}
Thus we have a family of s.a. hamiltohians $\hat{h}_{1\zeta}$,%
\begin{equation}
\hat{h}_{1\zeta}:\left\{ 
\begin{array}{l}
D_{h_{1\zeta}}=\{\psi\in D_{h_{1}^{+}},\;\psi \;\mathrm{satisfy\;the%
\;boundary\;condition\;(\ref{Cps2.5.4.1})} \\ 
\hat{h}_{1\zeta}\psi=\check{h}_{1}\psi,\;\forall\psi\in D_{h_{1\zeta}}%
\end{array}
\right. .   \label{Cps2.5.4.2}
\end{equation}

\subsubsection{The guiding functional}

As a guiding functional $\Phi_{1\zeta}(\xi;W)$ we choose%
\begin{align*}
& \Phi_{1\zeta}(\xi;W)=\int_{0}^{R_{C}}U_{1\zeta}(\rho;W)\xi(\rho)d\rho
,\;\xi\in\mathbb{D}_{1\zeta}=D_{r}(0,R_{C})\cap D_{h_{1\zeta}}. \\
& .U_{1\zeta}(\rho;W)=C_{1,1}(\rho;W)\sin\zeta+C_{4,1}(\rho;W)\cos\zeta,
\end{align*}
$U_{1\zeta}(\rho;W)$ is real-entire solution of eq. (\ref{Cps2.2.1})
satisfying the boundary condition (\ref{Cps2.5.4.1}).

The guiding functional $\Phi_{1\zeta}(\xi;W)$ is simple and the spectrum of $%
\hat{h}_{1\zeta}$ is simple.

\subsubsection{Green function $G_{1\protect\zeta}(\protect\rho,y;W)$,
spectral function $\protect\sigma_{1\protect\zeta}(E)$}

We find the Green function $G_{1\zeta}(\rho,y;W)$ as the kernel of the
integral representation 
\begin{equation*}
\psi(\rho)=\int_{0}^{\infty}G_{1\zeta}(\rho,y;W)\eta(y)dy,\;\eta\in L^{2}(%
\mathbb{R}_{+}), 
\end{equation*}
of unique solution of an equation%
\begin{equation}
(\hat{h}_{1\zeta}-W)\psi(\rho)=\eta(\rho),\;\operatorname{Im}W>0, 
\label{Cps2.5.6.1}
\end{equation}
for $\psi\in D_{h_{1\zeta}}$. General solution of eq. (\ref{Cps2.5.6.1})
(under condition $\psi\in L^{2}(0,R)$) can be represented in the form%
\begin{align*}
& \psi(\rho)=aC_{3,1}(\rho;W)+\frac{C_{1,1}(\rho;W)}{C_{1}(W)}\eta
_{3}(W)+I(\rho),\;\eta_{3}(W)=\int_{0}^{R}C_{3,1}(y;W)\eta(y)dy, \\
& I(\rho)=\frac{C_{3,1}(\rho;W)}{C_{1}(W)}\int_{0}^{\rho}C_{1,1}(y;W)%
\eta(y)dy-\frac{C_{1,1}(\rho;W)}{C_{1}(W)}\int_{0}^{\rho}C_{3,1}(y;W)%
\eta(y)dy, \\
& I(\rho)=O\left( \rho^{3/2}\right) ,\;\rho\rightarrow0.
\end{align*}
A condition $\psi\in D_{h_{1\zeta}\text{ , }}$(i.e.$\psi$ satisfies the
boundary condition (\ref{Cps2.5.4.1})) gives%
\begin{equation*}
a=-\frac{\cos\zeta}{C_{1}^{2}(W)\omega_{\zeta}(W)}\eta_{3}(W),\;\omega_{%
\zeta }(W)=f_{1}(W)\cos\zeta-\sin\zeta, 
\end{equation*}%
\begin{align}
& G_{1\zeta}(\rho,y;W)=\pi\Omega_{1\zeta}(W)U_{1\zeta}(\rho;W)U_{1\zeta
}(y;W)-  \notag \\
& \,-\left\{ 
\begin{array}{c}
\tilde{U}_{1\zeta}(\rho;W)U_{1\zeta}(y;W),\;\rho>y \\ 
U_{1\zeta}(\rho;W)\tilde{U}_{1\zeta}(y;W),\;\rho<y%
\end{array}
\right. ,  \label{Cps2.5.6.2} \\
& \Omega_{1\zeta}(W)=-\frac{\tilde{\omega}_{\zeta}(W)}{\pi\omega_{\zeta}(W)}%
,\;\tilde{\omega}_{\zeta}(W)=f_{1}(W)\sin\zeta+\cos\zeta,  \notag \\
& U_{1\zeta}(y;W)=C_{1,1}(\rho;W)\sin\zeta+C_{4,1}(\rho;W)\cos\zeta ,  \notag
\\
& \tilde{U}_{1\zeta}(\rho;W)=C_{1,1}(\rho;W)\cos\zeta-C_{4,1}(\rho
;W)\sin\zeta,  \notag
\end{align}
where we used an equality%
\begin{equation*}
C_{3,1}(\rho;W)=\tilde{\omega}_{\zeta}(W)U_{\zeta}(\rho;W)+\omega_{\zeta }(W)%
\tilde{U}_{1\zeta}(\rho;W). 
\end{equation*}
Note that the functions $U_{1\zeta}(\rho;W)$ and $\tilde{U}_{1\zeta}(\rho;W)$
are real-entire in $W$ and the last term in the r.h.s. of eq. (\ref%
{Cps2.5.6.2}) is real for $W=E$. For $\sigma_{1\zeta}^{\prime}(E)$, we find%
\begin{equation*}
\sigma_{1\zeta}^{\prime}(E)=\operatorname{Im}\Omega_{1\zeta}(E+i0). 
\end{equation*}

\subsubsection{Spectrum}

\subsubsection{$W=(1+4R_{C}g)/R_{C}^{2}+\tilde{\Delta}$, $\tilde{\Delta}$ $%
\sim0$}

A direct estimation gives 
\begin{equation*}
\Omega_{1\zeta}(W)=\left\{ 
\begin{array}{l}
\Omega_{1\zeta}(W_{0})+O(\sqrt{\tilde{\Delta}}),\;g=g_{1k}\;\mathrm{or}%
\;g\neq g_{1k},\;\zeta\neq\zeta_{1} \\ 
O(1/\sqrt{\tilde{\Delta}}),\;g\neq g_{1k},\;\zeta=\zeta_{1}%
\end{array}
\right. , 
\end{equation*}
$R_{C}g_{1,k}=-N_{1,k}^{2}$, $N_{1,k}=2(1+k)$, $\operatorname{Im}[\Omega_{1%
\zeta}(W_{0})]=0,\tan\zeta_{1}=f_{1|0}=\left. f_{1}(W)\right\vert _{\tilde{%
\Delta}=0}$. This result means that the levels with $E=$ $E_{0}$ are absent.

\subsubsection{$\protect\zeta=\protect\pi/2$}

First we consider the case $\zeta=\pi/2$.

In this case, we have $U_{1\pi/2}(\rho;W)=C_{1,1}(\rho;W)$ and%
\begin{equation*}
\sigma_{1\pi/2}^{\prime}(E)=\frac{1}{\pi}\operatorname{Im}\left. f_{1}(W)\right|
_{W=E+i0}. 
\end{equation*}
All results for spectrum ang spectral function can be obtained from the
corresponding results of previous section by setting there $m=1$.

\paragraph{$w=R_{C}^{2}E>1+4R_{C}g$}

In this case, $f_{1}(E)$ is a finite complex function and we find%
\begin{equation*}
\sigma_{1\pi/2}^{\prime}(E)=\frac{1}{\pi}\operatorname{Im}\mathcal{V}%
_{1}(E)\equiv\varrho_{1\pi/2}^{2}(E)>0. 
\end{equation*}
where $f_{1}(E)=\mathcal{U}_{1}(E)+i\mathcal{V}_{1}(E)$, $\mathcal{U}_{1}(E)=%
\operatorname{Re}f_{1}(E)$, $\mathcal{V}_{1}(E)=\operatorname{Im}f_{1}(E)>0$. The spectrum
of $\hat{h}_{1\pi/2}$\ is simple and continuous, $\mathrm{spec}\hat{h}%
_{1\pi/2}=[(1+4R_{C}g)/R_{C}^{2},\infty)$, and%
\begin{equation*}
\sigma_{1\pi/2}^{\prime}(E)=\left\{ 
\begin{array}{c}
O(\sqrt{\Delta}),\;g\neq g_{1,k} \\ 
O(1/\sqrt{\Delta}),\;g=g_{1,k}%
\end{array}
\right. ,\;\Delta\rightarrow+0, 
\end{equation*}
$\Delta=E-(1+4R_{C}g)/R_{C}^{2}$.

\paragraph{$w=R_{C}^{2}E\leq1+4R_{C}g$, $w<1$}

In this case, we have%
\begin{align*}
\sigma_{1\pi/2}^{\prime}(E) & =\sum_{n=0}^{n_{\max}}Q_{1\pi/2,n}^{2}\delta(E-%
\mathcal{E}_{1,n}),\;Q_{1\pi/2,n}=\sqrt{\frac{|g|(R_{C}^{2}g^{2}-N_{1,n}^{4})%
}{N_{1,n}^{3}R_{C}}}, \\
\mathcal{E}_{1,n} & =.\frac{1-(R_{C}|g|/N_{1,n}+N_{1,n})^{2}}{R_{C}^{2}}%
,\;R_{C}g<R_{C}g_{1,0}=-N_{1,0}^{2}=-4.
\end{align*}
The spectrum of $\hat{h}_{1\pi/2}$\ is discrete and simple and has the form%
\begin{equation*}
\mathrm{spec}\hat{h}_{1\pi/2}=\{\mathcal{E}_{1,n},\;\mathcal{E}%
_{1,n}<1-4R_{C}|g|,\;n=0,1,...,n_{\max}\}, 
\end{equation*}
$n_{\max}=k$ for $\sqrt{R_{C}|g|}=2(1+k+\delta)$, $0<\delta\leq1$. The
discrete part of the spectrum is absent for $g\geq
g_{1,0}=-N_{1,0}^{2}/R_{C}=-4/R_{C}$.

\paragraph{$w=1$, $\protect\sigma=0$}

We have in this case for $W=E=R_{C}^{-2}$: $g\geq0$, $\alpha_{1}=\beta
_{1}=1+\sqrt{R_{C}g}/2$, $\operatorname{Im}\nu=0$, $\operatorname{Im}\alpha_{1}=0$, $%
\alpha_{1}>0$, , and $\sigma_{m}^{\prime}(R_{C}^{-2})=0$.

\paragraph{$w>1$, $\protect\sigma=i\varkappa/4$, $\varkappa=\protect\sqrt{w-1%
}$}

In this case, we have for $W=E$:$\operatorname{Im}\nu=0$,$\;\alpha_{1}=1+\nu+i%
\varkappa/4$, $\beta_{1}=1+\nu-i\varkappa/4=\overline{\alpha_{1}}$, such
that $\left[ \operatorname{Im}\psi(\alpha_{1})+\operatorname{Im}\psi(\beta_{1})\right]
_{W=E}=0$, and%
\begin{equation*}
\sigma_{m}^{\prime}(E)=0. 
\end{equation*}

Finally, we find.

The spectrum of $\hat{h}_{1\pi/2}$\ is simple, $\mathrm{spec}\hat{h}_{1\pi
/2}=[1+4R_{C}g)/R_{C}^{2},\infty)\cup\{\mathcal{E}_{1,n},\;n=0,1,...n_{\max
}\}$, the discrete part of spectrum is present for $g<g_{1,0}$. The set of
functions 
\begin{equation*}
U_{1\pi/2}(\rho;E)=\varrho_{1,\pi/2}(E)C_{1,1}(\rho;E),\;E%
\geq1+4R_{C}g)/R_{C}^{2}; 
\end{equation*}
\begin{equation*}
{U_{1\pi/2,n}(\rho)=Q_{1\pi/2,n}C_{1,1}(\rho;\mathcal{E}_{1,n}),%
\;n=0,1,...n_{\max}} 
\end{equation*}
forms a complete orthogonalized system in $L^{2}(0,R_{C})$.

The same results for spectrum and eigenfunctions are obtained for the case $%
\zeta=-\pi/2$.

\subsubsection{$|\protect\zeta|<\protect\pi/2$}

Now we consider the case $|\zeta|<\pi/2$.

In this case, we can represent $\sigma_{1\zeta}^{\prime}(E)$ in the form%
\begin{equation*}
\sigma_{1\zeta}^{\prime}(E)=-\frac{1}{\pi\cos^{2}\zeta}\operatorname{Im}\frac{1}{%
f_{1\zeta}(E+i0)},\;f_{1\zeta}(W)=f_{1}(W)-\tan\zeta. 
\end{equation*}

\paragraph{$E>(1+4R_{C}g)/R_{C}^{2}$}

In this case, we have%
\begin{equation*}
\sigma_{1\zeta}^{\prime}(E)=\frac{1}{\pi}\frac{\mathcal{V}_{1}(E)}{[\mathcal{%
U}_{1}(E)\cos\zeta-\sin\zeta]^{2}+\mathcal{V}_{1}^{2}(E)\cos ^{2}\zeta}%
\equiv\rho_{0\zeta}^{2}(E). 
\end{equation*}
The spectrum of $\hat{h}_{1\zeta}$ is simple and continuous, $\mathrm{spec}%
\hat{h}_{1\zeta}=[(1+4R_{C}g)/R_{C}^{2},\infty)$.

\paragraph{$w=1+4R_{C}g+\Delta$, $\Delta\sim0$}

In this case, we have%
\begin{equation*}
4\nu=\sqrt{-\Delta}=\left\{ 
\begin{array}{c}
-i\sqrt{\Delta},\;\Delta\geq0 \\ 
\sqrt{|\Delta|},\;\Delta\leq0%
\end{array}
\right. ,\;\sigma=\left\{ 
\begin{array}{l}
-i\sqrt{R_{C}g}/2+O(\Delta),\;g>0 \\ 
\nu,\;g=0 \\ 
\sqrt{R_{C}|g|}/2+O(\Delta),,\;g<0%
\end{array}
\right. , 
\end{equation*}%
\begin{align*}
& \alpha_{1}=1+\left\{ 
\begin{array}{l}
-i\sqrt{R_{C}g}/2+\nu+O(\Delta),\;g>0 \\ 
2\nu,\;g=0 \\ 
\sqrt{R_{C}|g|}/2+\nu+O(\Delta),,\;g<0%
\end{array}
\right. , \\
& \beta_{1}=1+\left\{ 
\begin{array}{l}
i\sqrt{R_{C}g}/2+\nu+O(\Delta),\;g>0 \\ 
0,\;g=0 \\ 
-\sqrt{R_{C}|g|}/2+\nu+O(\Delta),,\;g<0%
\end{array}
\right. .
\end{align*}
A direct estimation gives%
\begin{equation*}
\sigma_{1\zeta}^{\prime}(E)=\left\{ 
\begin{array}{l}
\left\{ 
\begin{array}{l}
O(\sqrt{\Delta}),\;g=g_{1k}\;\mathrm{or}\;g\neq g_{1k},\;\zeta\neq\zeta_{1}
\\ 
O(1/\sqrt{\Delta}),\;g\neq g_{1k},\;\zeta=\zeta_{1}%
\end{array}
\right. ,\;\Delta>0 \\ 
0,\;\Delta<0\;%
\end{array}
\right. , 
\end{equation*}
where $\tan\zeta_{1}=f_{1|0}=\left. f_{1}(E)\right\vert _{\Delta=0}$. Note
that the discrete level with $E=(1+4R_{C}g)R_{C}^{2}$ is absent.

\paragraph{$E<(1+4R_{C}g)R_{C}^{2}$}

In this case, we have $\left. \operatorname{Im}\nu\right| _{W=E}=0$, $\left.
\nu\right| _{W=E}>0$,%
\begin{align*}
\alpha_{1} & =1+\nu-i\sqrt{w-1}/4,\;\beta_{1}=\overline{\alpha_{1}},\;w\geq1,
\\
\alpha_{1} & =1+\nu+\sqrt{1-w}/4,\;\beta_{1}=1+\nu-\sqrt{1-w}/4,\;w<1.
\end{align*}

Thus, we have: the function $[f_{1\zeta}(E)]^{-1}$ is real except the points 
$E_{1n}(\zeta)$,%
\begin{equation}
f_{1\zeta}(E_{1n}(\zeta))=0,   \label{Cps2.5.7.2.3.1}
\end{equation}
such that we obtain%
\begin{align*}
\sigma_{1\zeta}^{\prime}(E) & =\sum_{n\in\mathcal{N}_{1}}Q_{1\zeta,n}^{2}%
\delta(E-E_{1n}(\zeta)),\;Q_{1\zeta,n}=\frac{1}{\sqrt{\partial
_{E}f_{1\zeta}(E_{1n}(\zeta))}\cos\zeta}, \\
\partial_{E}f_{1}(E) & >0,
\end{align*}
where $\mathcal{N}_{1}$ is a subset of $\mathbb{Z}$ to be described below.
Furthermore, we find 
\begin{equation*}
\partial_{\zeta}E_{1n}(\zeta)=[\partial_{E}f_{1}(E_{n}(\zeta))\cos^{2}%
\zeta]^{-1}>0. 
\end{equation*}

\subparagraph{$g\geq g_{1,0}=-4/R_{C}$}

In this case, the function $f_{1}(E)$ has the properties: $%
f_{1}(E)\rightarrow-\infty$ as $E\rightarrow-\infty$; $f_{1}(E)\rightarrow%
\tan \zeta_{1}-0=f_{1|0}-0$ as $E\rightarrow(1+4R_{C}g)R_{C}^{2}-0$;.$%
f_{1}(E)$ increases monotonocally on the interval $(-%
\infty,(1+4R_{C}g)R_{C}^{2})$. Then we find: in the interval $%
(-\infty,(1+4R_{C}g)R_{C}^{2})$, for any fixed $\zeta\in(-\pi/2,\zeta_{1})$,
there is one level $E_{1,-1}(\zeta)$ which run monotonocally from $-\infty$
to $(1+4R_{C}g)R_{C}^{2}-0$ as $\zeta$ run from $-\pi/2-0$ to $\zeta_{1}-0$;
there are no discrete levels on the interval $(-\infty,(1+4R_{C}g)R_{C}^{2})$
for $\zeta\in(\zeta_{1},\pi/2)$. Formally, eq. (\ref{Cps2.5.7.2.3.1}) has
solution $E_{1,-1}(\zeta_{1})=(1+4R_{C}g)R_{C}^{2}$ for $\zeta=\zeta_{1}$.
However, as was noted above, such levels are absent We find also%
\begin{equation*}
\mathcal{N}_{1}=\mathcal{N}_{1,-1}(\zeta)=\left\{ 
\begin{array}{l}
\varnothing,\;\zeta\in\lbrack\zeta_{1},\pi/2) \\ 
\{-1\},\;\zeta\in(-\pi/2,\zeta_{1})%
\end{array}
\right. . 
\end{equation*}

\subparagraph{$g_{1,k+1}\leq g<g_{1,k}$, $\protect\sqrt{R_{C}|g|}=2(1+k+%
\protect\delta)$, $0<\protect\delta\leq1$, $k\in\mathbb{Z}_{+}$}

In this case,\ the function $f_{1}(E)$ has the properties: $%
f_{1}(E)\rightarrow-\infty$ as $E\rightarrow-\infty$; $f(\mathcal{E}_{1n}\pm
0)=\mp\infty$ $\ n=0,1,...,n_{\max}=k$; $f_{1}(E)\rightarrow\tan\zeta
_{1}-0=f_{1|0}(E)-0$ as $E\rightarrow(1+4R_{C}g)/R_{C}^{2}-0$. We find: in
each interval $(\mathcal{E}_{1n-1},\mathcal{E}_{1n})$, $n=0,...,k$ (we set $%
\mathcal{E}_{1,-1}=-\infty$), for any fixed $\zeta\in(-\pi/2,\pi/2)$, there
is one level $E_{1n}(\zeta)$ which run monotonocally from $\mathcal{E}%
_{n-1}+0$ to $\mathcal{E}_{n}-0$ as $\zeta$ run from $-\pi/2+0$ to $\pi/2-0$%
; in the interval $(\mathcal{E}_{k},(1+4R_{C}g)R_{C}^{2})$; for any fixed $%
\zeta\in(-\pi/2,\zeta_{1})$, there is one level $E_{k+1}(\zeta)$ which run
monotonocally from $\mathcal{E}_{k}+0$ to $(1+4R_{C}g)/R_{C}^{2}-0$ as $\zeta
$ run from $-\pi/2+0$ to $\zeta_{1}-0$; thare are no levels in the interval $%
(\mathcal{E}_{k},(1+4R_{C}g)/R_{C}^{2})$ for $\zeta\in(\zeta_{1},\pi/2)$.
Formally, eq. (\ref{Cps2.5.7.2.3.1}) has solution $E_{1,k+1}(\zeta
_{1})=(1+4R_{C}g)/R_{C}^{2}$ for $\zeta=\zeta_{1}$. However, as was noted
above, such levels are absent.. We find also%
\begin{equation*}
\mathcal{N}_{1}=\mathcal{N}_{1,k}(\zeta)=\left\{ 
\begin{array}{l}
\{0,1,...,k\},\;\zeta\in\lbrack\zeta_{1},\pi/2) \\ 
\{0,1,...,k+1\},\;\zeta\in(-\pi/2,\zeta_{1})%
\end{array}
\right. . 
\end{equation*}

Finally, we obtain. The spectrum of $\hat{h}_{1\zeta}$ is simple and $%
\mathrm{spec}\hat{h}_{1\zeta}=[(1+4R_{C}g)/R_{C}^{2},\infty)\cup
\{E_{1n}(\zeta),\;n\in\mathcal{N}_{1}\}$. The set of functions 
\begin{equation*}
\left\{ U_{1\zeta,E}(\rho)=\varrho_{1\zeta}(E)U_{1\zeta}(\rho;E),\;E\geq
(1+4R_{C}g)/R_{C}^{2};\;U_{1\zeta,n}(\rho)=Q_{1\zeta,n}U_{1\zeta}(\rho
;E_{1n}(\zeta),),\;n\in\mathcal{N}\right\} 
\end{equation*}
forms a complete orthogonalized system in $L^{2}(0,R_{C})$.

\bigskip

\subsection{$m=-1$}

Only modification which we must do is the following: the extension parameter
for the case $m=-1$ should be considered as indendent of the extension
parameter for the case $m=1$. It is convenient to denote the extension
parameter for the case $m=1$ as $\zeta_{(1)}$ and \ for the case $m=-1$ as $%
\zeta_{(-1)}$.

\subsection{$m=0$}

In this case, we have $\mu=0$; $\alpha_{1}=1/2+\nu+\sigma$, $%
\beta_{1}=1/2+\nu-\sigma$.

\subsubsection{ Useful solutions}

We need solutions of an equation%
\begin{equation}
(\check{h}_{0}-W)\psi_{0}^{<}(\rho)=0,\;   \label{Cps2.6.1.1}
\end{equation}
where $\check{h}_{0}$ is given by eq. (\ref{Cps2.1.2}) with $m=0$.

We use the following solution of eq.(\ref{Cps2.6.1.1}) 
\begin{align*}
& C_{1,0}(\rho;W)=\frac{R_{C}^{2}\sqrt{\rho}}{R_{C}^{2}-\rho^{2}}%
(1-x)^{1/4+\nu}\mathcal{F}(\alpha_{1},\beta_{1};1;x)= \\
& (\mathrm{for}\;\operatorname{Im}W>0)=\frac{\Gamma(-2\nu)}{\Gamma(\alpha
_{4})\Gamma(\beta_{4})}C_{3,0}(\rho;W)+\frac{\Gamma(2\nu)}{\Gamma(\alpha
_{1})\Gamma(\beta_{1})}v_{0}(\rho;W), \\
& v_{0}(\rho;W)=\frac{R_{C}^{2}\sqrt{\rho}}{R_{C}^{2}-\rho^{2}}%
(1-x)^{1/4-\nu}\mathcal{F}(\alpha_{4},\beta_{4};\gamma_{4};1-x), \\
& C_{4,0}(\rho;W)=2\lim_{\delta\rightarrow0}\partial_{\delta}C_{1,0,\delta
}^{<}(\rho;W), \\
& C_{1,0,\delta}(r;W)=\frac{R_{C}^{2}\sqrt{r}}{R_{C}^{2}-r^{2}}x^{\delta
/2}(1-x)^{1/4+\nu}\mathcal{F}(\alpha_{1\delta},\beta_{1\delta};\gamma
_{1\delta};x), \\
& C_{3,0}(\rho;W)=\frac{R_{C}^{2}\sqrt{\rho}}{R_{C}^{2}-\rho^{2}}%
(1-x)^{1/4+\nu}\mathcal{F}(\alpha_{1},\beta_{1};\gamma_{3};1-x)= \\
& =B_{0}(W)C_{1,0}(\rho;W)-C_{0}(W)C_{4,0}(\rho;W),\;C_{0}(W)=\frac {%
\Gamma(\gamma_{3})}{\Gamma(\alpha_{1})\Gamma(\beta_{1})}, \\
& B_{0}(W)=C_{0}(W)f_{0}(W),\;f_{0}(W)=\left[ 2\psi(1)-\psi(\alpha_{1})-%
\psi(\beta_{1})\right] .
\end{align*}%
\begin{align*}
& \alpha_{1\delta}=1/2+\delta/2+\nu+\sigma,\;\beta_{1\delta}=1/2+\delta
/2+\nu-\sigma,\;\gamma_{1\delta}=1+\delta \\
& \alpha_{1}=1/2+\nu+\sigma,\;\beta_{1}=1/2+\nu-\sigma, \\
& \alpha_{4}=1/2-\nu+\sigma,\;\beta_{4}=1/2-\nu-\sigma.
\end{align*}

Note that $C_{1,0}^{<}(\rho;W)$ and $C_{4,0}^{<}(\rho;W)$ are real-entire in 
$W$.

\subsubsection{Asymptotics, $\protect\rho\rightarrow0$ ($x\rightarrow0$)}

We have%
\begin{align*}
& C_{1,0}(\rho;W)=C_{1,0\mathrm{as}}(\rho)(1+O(\rho)),\;C_{4,0}(%
\rho;W)=C_{4,0\mathrm{as}}(\rho)\left( 1+O(\rho)\right) , \\
& C_{1,0\mathrm{as}}(\rho)=\rho^{1/2},\;C_{4,0\mathrm{as}}(\rho)=\rho
^{1/2}\ln\left( \frac{4\rho}{R_{C}}\right) .
\end{align*}%
\begin{align*}
& C_{3,0}(\rho;W)=.C_{0}(W)\left[ f_{0}(W)C_{1,0\mathrm{as}}(\rho )-C_{4,0%
\mathrm{as}}(\rho)\right] \left( 1+O(\rho^{2})\right) , \\
& \operatorname{Im}W>0\;\mathrm{or}\;W=0.
\end{align*}

\subsubsection{Asymptotics, $\Delta=R-\protect\rho\rightarrow0$ ($\protect%
\delta =1-x\rightarrow0$)}

We have%
\begin{equation*}
C_{3,0}(\rho;W)=2^{-3/2-2\nu}R_{C}^{1-2\nu}\Delta^{-1/2+2\nu}(1+O(\Delta)), 
\end{equation*}

\begin{align*}
& C_{1,0}(\rho;W)=\frac{\Gamma(2\nu)}{\Gamma(\alpha_{1})\Gamma(\beta_{1})}%
2^{-3/2+2\nu}R_{C}^{1+2\nu}\Delta^{-1/2-2\nu}(1+O(\Delta)), \\
& \operatorname{Im}W>0\;\mathrm{or}\;W=0.
\end{align*}

\subsubsection{Wronskian}

We have

\begin{align*}
& \mathrm{Wr}(C_{1,0},C_{4,0})=1, \\
& \mathrm{Wr}(C_{1,0},C_{3,0})=-\frac{\Gamma(\gamma_{3})}{\Gamma(\alpha
_{1})\Gamma(\beta_{1})}=-C_{0}(W).
\end{align*}

Note that any solutions of eq. (\ref{Cps2.6.1.1}) are s.-integrable in the
origin and only one solution ($C_{3,0}$) is s.-integrable on the right end $%
R_{C}$ (for $\operatorname{Im}W>0$), such that there is one solution ($C_{3,0}$)
belonging to $L^{2}(0,R_{C})$ for $\operatorname{Im}W>0$ and the deficiency indexes
of the symmetric operator $\check{h}_{0}$ (see below) are equal to $(1,1)$.

\subsubsection{Symmetric operator $\hat{h}_{0}$}

A symmetric operator $\hat{h}_{0}$ is defined by eq. (\ref{Cps2.4.2.1}) with 
$m=0$.

\subsubsection{Adjoint operator $\hat{h}_{0}^{+}=\hat{h}_{0}^{\ast}$}

It is easy to prove by standard way that the adjoint operator $\hat{h}%
_{0}^{+}$ coincides with the operator $\hat{h}_{0}^{\ast}$,

\begin{equation*}
\hat{h}_{0}^{+}:\left\{ 
\begin{array}{l}
D_{h_{0}^{+}}=D_{\check{h}_{0}}^{\ast}(0,R_{C})=\{\psi_{\ast},\psi_{\ast
}^{\prime}\;\mathrm{are\;a.c.\;in}\mathcal{\;}(0,R_{C}),\;\psi_{\ast},\hat {h%
}_{0}^{+}\psi_{\ast}\in L^{2}(0,R_{C})\} \\ 
\hat{h}_{0}^{+}\psi_{\ast}(\rho)=\check{h}_{0}\psi_{\ast}(\rho),\;\forall
\psi_{\ast}\in D_{h_{0}^{+}}%
\end{array}
\right. . 
\end{equation*}

\subsubsection{Asymptotics}

Because $\check{h}_{0}\psi_{\ast}\in L^{2}(0,R)$, we have%
\begin{equation*}
\check{h}_{0}\psi_{\ast}(\rho)=\eta(\rho),\;\eta\in L^{2}(0,R_{C}), 
\end{equation*}
and we can represent $\psi_{\ast}$ in the form%
\begin{align*}
\psi_{\ast}(\rho) & =c_{1}C_{1,0}(\rho;0)+c_{2}C_{3,0}(\rho;0)+I(\rho), \\
\psi_{\ast}^{\prime}(\rho) & =c_{1}\partial_{\rho}C_{1,0}(\rho
;0)+c_{2}\partial_{\rho}C_{3,0}(\rho;0)+I^{\prime}(\rho),
\end{align*}
where%
\begin{align*}
I(\rho) & =\frac{C_{3,0}(\rho;0)}{C_{0}(0)}\int_{0}^{\rho}C_{1,0}(y;0)%
\eta(y)dy-\frac{C_{1,0}(\rho;0)}{C_{0}(0)}\int_{0}^{\rho}C_{3,0}(y;0)%
\eta(y)dy, \\
I^{\prime}(\rho) & =\frac{\partial_{\rho}C_{3,0}(\rho;0)}{C_{0}(0)}\int
_{0}^{\rho}C_{1,0}(y;0)\eta(y)dy-\frac{\partial_{\rho}C_{1,0}(\rho;0)}{%
C_{0}(0)}\int_{0}^{\rho}C_{3,0}(y;0)\eta(y)dy.
\end{align*}

I) $\rho\rightarrow0$

We obtain with the help of the CB-inequality:%
\begin{equation*}
I(\rho)=O(\rho^{3/2}\ln\rho),\;I^{\prime}(\rho)=O(\rho^{1/2}\ln\rho), 
\end{equation*}
such that we have%
\begin{align*}
& \psi_{\ast}(\rho)=c_{1}C_{1,0\mathrm{as}}(\rho)+c_{2}C_{4,0\mathrm{as}%
}(\rho)+O(\rho^{3/2}\ln\rho), \\
& \psi_{\ast}^{\prime}(\rho)=c_{1}C_{1,0\mathrm{as}}^{\prime}(\rho
)+c_{2}C_{4,0\mathrm{as}}^{\prime}(\rho)+O(\rho^{1/2}\ln\rho).
\end{align*}

II) $\rho\rightarrow R_{C}$

In this case, we prove that $[\psi_{\ast},\chi_{\ast}]^{R_{C}}=0$, $%
\forall\psi_{\ast},\chi_{\ast}\in D_{h_{m}^{+}}$, such that we have%
\begin{equation*}
\Delta_{h_{0}^{+}}(\psi_{\ast})=(\overline{c_{1}}c_{2}-\overline{c_{2}}%
c_{1})=-\frac{i}{2}(.\overline{c_{+}}c_{+}-\overline{c_{-}}c_{-}),\;c_{\pm
}=c_{1}\pm ic_{2}. 
\end{equation*}

\subsubsection{Self-adjoint hamiltonians $\hat{h}_{0\protect\theta}$}

The condition $\Delta_{h_{0}^{+}}(\psi)=0$ gives%
\begin{align*}
& c_{-}=-e^{2i\theta}c_{+},\;|\theta|\leq\pi/2,\;\theta=-\pi/2\sim\theta
=\pi/2,\;\Longrightarrow \\
& \Longrightarrow\;c_{1}\cos\theta=c_{2}\sin\theta,
\end{align*}
or%
\begin{align}
& \psi(\rho)=C\psi_{\theta\mathrm{as}}(\rho)+O(\rho^{3/2}\ln\rho
),\;\psi^{\prime}(\rho)=C\psi_{\theta\mathrm{as}}^{\prime}(\rho)+O(\rho
^{1/2}\ln\rho),  \label{Cps2.6.4.1} \\
& \psi_{\theta\mathrm{as}}(\rho)=C_{1,0\mathrm{as}}(\rho)\sin\theta +C_{4,0%
\mathrm{as}}(\rho)\cos\theta  \notag
\end{align}
We thus have a family of s.a. hamiltohians $\hat{h}_{0\theta}$,%
\begin{equation*}
\hat{h}_{0\theta}:\left\{ 
\begin{array}{l}
D_{h_{0\theta}}=\{\psi\in D_{h_{0}^{+}},\;\psi \;\mathrm{satisfy\;the%
\;boundary\;condition\;(\ref{Cps2.6.4.1})} \\ 
\hat{h}_{0\theta}\psi=\check{h}_{0}\psi,\;\forall\psi\in D_{h_{0\theta}}%
\end{array}
\right. . 
\end{equation*}

\subsubsection{The guiding functional}

As a guiding functional $\Phi_{0\theta}(\xi;W)$ we choose%
\begin{align*}
& \Phi_{0\theta}(\xi;W)=\int_{0}^{R}U_{0\theta}(\rho;W)\xi(\rho)d\rho
,\;\xi\in\mathbb{D}_{\theta}=D_{r}(0,R_{C})\cap D_{h_{0\theta}}. \\
& .U_{0\theta}(\rho;W)=C_{1,0}(\rho;W)\sin\theta+C_{4,0}(\rho;W)\cos\theta,
\end{align*}
$U_{0\theta}(\rho;W)$ is real-entire solution of eq. (\ref{Cps2.6.1.1})
satisfying the boundary condition (\ref{Cps2.6.4.1}).

The guiding functional $\Phi_{0\theta}(\xi;W)$ is simple and the spectrum of 
$\hat{h}_{0\theta}$ is simple.

\subsubsection{Green function $G_{0\protect\theta}(\protect\rho,y;W)$,
spectral function $\protect\sigma_{0\protect\theta}(E)$}

We find the Green function $G_{0\theta}(\rho,y;W)$ as the kernel of the
integral representation 
\begin{equation*}
\psi(\rho)=\int_{0}^{R_{C}}G_{0\theta}(\rho,y;W)\eta(y)dy,\;\eta\in L^{2}(%
\mathbb{R}_{+}), 
\end{equation*}
of unique solution of an equation%
\begin{equation}
(\hat{h}_{0\theta}-W)\psi(\rho)=\eta(\rho),\;\operatorname{Im}W>0, 
\label{Cps2.6.6.1}
\end{equation}
for $\psi\in D_{h_{0\theta}}$. General solution of eq. (\ref{Cps2.6.6.1})
(under condition $\psi\in L^{2}(0,R)$) can be represented in the form%
\begin{align*}
\psi(\rho) & =aC_{3,0}(\rho;W)+\frac{C_{1,0}(\rho;W)}{C_{0}(W)}\eta
_{3}(W)+I(\rho),\;\eta_{3}(W)=\int_{0}^{R_{C}}C_{3,0}(y;W)\eta(y)dy, \\
I(\rho) & =\frac{C_{3,0}(\rho;W)}{C_{0}(W)}\int_{0}^{\rho}C_{1,0}(y;W)%
\eta(y)dy-\frac{C_{1,0}(\rho;W)}{C_{0}(W)}\int_{0}^{\rho}C_{3,0}(y;W)%
\eta(y)dy, \\
I(\rho) & =O\left( \rho^{3/2}\ln\rho\right) ,\;\rho\rightarrow0.
\end{align*}
A condition $\psi\in D_{h_{0\theta}\text{ , }}$(i.e.$\psi$ satisfies the
boundary condition (\ref{Opso2.4.5.1})) gives%
\begin{equation*}
a=-\frac{\cos\theta}{C_{0}^{2}(W)\omega_{\theta}(W)}\eta_{3}(W),\;\omega
_{\theta}(W)=f_{0}(W)\cos\zeta+\sin\zeta, 
\end{equation*}%
\begin{align}
& G_{0\theta}(\rho,y;W)=\pi\Omega_{0\theta}(W)U_{0\theta}(\rho;W)U_{0\theta
}(y;W)+  \notag \\
& \,+\left\{ 
\begin{array}{c}
\tilde{U}_{0\theta}(\rho;W)U_{0\theta}(y;W),\;\rho>y \\ 
U_{0\theta}(\rho;W)\tilde{U}_{0\theta}(y;W),\;\rho<y%
\end{array}
\right. ,  \label{Cps2.6.6.2} \\
& \Omega_{0\theta}(W)=\frac{\tilde{\omega}_{\theta}(W)}{\pi\omega_{\theta
}(W)},\;\tilde{\omega}_{\theta}(W)=f_{0}(W)\sin\theta-\cos\zeta,  \notag \\
& \tilde{U}_{0\theta}(\rho;W)=C_{1,0}^{<}(\rho;W)\cos\theta-C_{4,0}^{<}(%
\rho;W)\sin\theta,  \notag
\end{align}
where we used an equality%
\begin{equation*}
C_{3,0}^{<}(\rho;W)=C_{0}(W)[\tilde{\omega}_{\theta}(W)U_{0\theta}(\rho;W)+%
\omega_{\theta}(W)\tilde{U}_{0\theta}(\rho;W)]. 
\end{equation*}
Note that the functions $U_{0\theta}(\rho;W)$ and $\tilde{U}%
_{0\theta}(\rho;W)$ are real-entire in $W$ and the last term in the r.h.s.
of eq. (\ref{Cps2.6.6.2}) is real for $W=E$. For $\sigma_{0\theta}^{%
\prime}(E)$, we find%
\begin{equation*}
\sigma_{0\theta}^{\prime}(E)=\operatorname{Im}\Omega_{0\theta}(E+i0). 
\end{equation*}

\subsubsection{Spectrum}

\subsubsection{$W=(1+4R_{C}g)/R_{C}^{2}+\tilde{\Delta}$, $\tilde{\Delta}$ $%
\sim0$}

A direct estimation gives 
\begin{equation*}
\Omega_{0\zeta}(W)=\left\{ 
\begin{array}{l}
\Omega_{0\zeta}(W_{0})+O(\sqrt{\tilde{\Delta}}),\;g=g_{0k}\;\mathrm{or}%
\;g\neq g_{0k},\;\zeta\neq\zeta_{0} \\ 
O(1/\sqrt{\tilde{\Delta}}),\;g\neq g_{0k},\;\zeta=\zeta_{0}%
\end{array}
\right. , 
\end{equation*}
$R_{C}g_{0,k}=-N_{0,k}^{2}$, $N_{0,k}=1+2k$,\thinspace$\tan%
\theta_{0}=-f_{0|0}=\left. f_{0}(W)\right\vert _{\tilde{\Delta}=0},\operatorname{Im}%
[\Omega_{0\zeta}(W_{0})]=0.$ This result means that the levels with $%
E=E_{0}\ \ $are absent.

\subsubsection{$\protect\theta=\protect\pi/2$}

First we consider the case $\theta=\pi/2$.

In this case, we have $U_{0\pi/2}(\rho;W)=C_{1,0}(\rho;W)$ and%
\begin{equation*}
\sigma_{0\pi/2}^{\prime}(E)=\frac{1}{\pi}\operatorname{Im}f_{0}(E+i0). 
\end{equation*}
All results for spectrum and spectral function can be obtained from the
corresponding results of subsec. 4.7 by setting there $m=0$.

\paragraph{$w=R_{C}^{2}E>1+4R_{C}g$}

In this case, $f_{0}(E)$ is a finite complex function and we find%
\begin{equation*}
\sigma_{0\pi/2}^{\prime}(E)=\frac{1}{\pi}\operatorname{Im}\mathcal{V}%
_{0}(E)\equiv\varrho_{0\pi/2}^{2}(E)>0. 
\end{equation*}
where $f_{0}(E)=\mathcal{U}_{0}(E)+i\mathcal{V}_{0}(E)$, $\mathcal{U}_{0}(E)=%
\operatorname{Re}f_{0}(E)$, $\mathcal{V}_{0}(E)=\operatorname{Im}f_{0}(E)>0$. The spectrum
of $\hat{h}_{0\pi/2}$\ is simple and continuous, $\mathrm{spec}\hat{h}%
_{0\pi/2}=[(1+4R_{C}g)/R_{C}^{2},\infty)$, and%
\begin{equation*}
\sigma_{0\pi/2}^{\prime}(E)=\left\{ 
\begin{array}{c}
O(\sqrt{\Delta}),\;g\neq g_{0,k} \\ 
O(1/\sqrt{\Delta}),\;g=g_{0,k}%
\end{array}
\right. ,\;\Delta\rightarrow+0, 
\end{equation*}
$\Delta=E-(1+4R_{C}g)/R_{C}^{2}$.

\paragraph{$w=R_{C}^{2}E\leq1+4R_{C}g$, $w<1$}

In this case, we have%
\begin{align*}
& \sigma_{0\pi/2}^{\prime}(E)=\sum_{n=0}^{n_{\max}}Q_{0\pi/2,n}^{2}\delta(E-%
\mathcal{E}_{0,n}),\;Q_{0\pi/2,n}=\frac{2}{R_{C}}\sqrt{\frac {%
R_{C}^{2}g^{2}-N_{0,n}^{4}}{N_{0,n}^{3}}}, \\
& \mathcal{E}_{0,n}=.\frac{1-(R_{C}|g|/N_{0,n}+N_{0,n})^{2}}{R_{C}^{2}}%
,\;R_{C}g<R_{C}g_{0,0}=-N_{1,0}^{2}=-1.
\end{align*}
The spectrum of $\hat{h}_{0\pi/2}$\ is discrete and simple and has the form%
\begin{equation*}
\mathrm{spec}\hat{h}_{0\pi/2}=\{\mathcal{E}_{0,n},\;\mathcal{E}%
_{0,n}<1-4R_{C}|g|,\;n=0,1,...,n_{\max}\}, 
\end{equation*}
$n_{\max}=k$ for $\sqrt{R_{C}|g|}=1+2(k+\delta)$, $0<\delta\leq1$. The
discrete part of the spectrum is absent for $g\geq
g_{0,0}=-N_{0,0}^{2}/R_{C}=-1/R_{C}$.

\paragraph{$w=R_{C}^{2}E\leq1+4R_{C}g$, $w\geq1$}

In this case, we have $\sigma_{0\pi/2}^{\prime}(E)=0$.

Finally, we find.

The spectrum of $\hat{h}_{0\pi/2}$\ is simple, $\mathrm{spec}\hat{h}_{0\pi
/2}=[1+4R_{C}g)/R_{C}^{2},\infty)\cup\{\mathcal{E}_{0,n},\;n=0,1,...n_{\max
}\}$, the discrete part of spectrum is present for $g<g_{0,0}$. The set of
functions 
\begin{equation*}
\left\{ U_{0\pi/2,E}(\rho)=\varrho_{0,\pi/2}(E)C_{1,0}(\rho;E),\;E\geq
(1+4R_{C}g)/R_{C}^{2}; U_{0\pi/2,n}(\rho)=Q_{0\pi/2,n}C_{1,0}(\rho;\mathcal{E%
}_{0,n}),\right.
\end{equation*}
\begin{equation*}
n=0,1,...n_{\max}\} 
\end{equation*}
forms a complete orthogonalized system in $L^{2}(0,R_{C})$.

The same results we obtain for the case $\zeta=-\pi/2$.

\subsubsection{$|\protect\theta|<\protect\pi/2$}

Now we consider the case $|\theta|<\pi/2$.

In this case, we can represent $\sigma_{\theta}^{\prime}(E)$ in the form%
\begin{equation*}
\sigma_{0\theta}^{\prime}(E)=-\frac{1}{\pi\cos^{2}\theta}\operatorname{Im}\frac{1}{%
f_{0\theta}(E+i0)},\;f_{0\theta}(W)=f_{0}(W)+\tan\zeta. 
\end{equation*}

\paragraph{$E>(1+4R_{C}g)/R_{C}^{2}/$}

In this case, we have%
\begin{equation*}
\sigma_{0\theta}^{\prime}(E)=\frac{1}{\pi}\frac{\mathcal{V}_{0}(E)}{[%
\mathcal{U}_{0}(E)\cos\theta+\sin\theta]^{2}+\mathcal{V}_{0}^{2}(E)\cos
^{2}\theta}\equiv\rho_{0\theta}^{2}(E). 
\end{equation*}
The spectrum of $\hat{h}_{0\theta}$ is simple and continuous, $\mathrm{spec}%
\hat{h}_{0\theta}=[(1+4R_{C}g)/R_{C}^{2},\infty)$.

\paragraph{$w=1+4R_{C}g+\Delta$, $\Delta\sim0$}

In this case, we have%
\begin{align*}
& \alpha_{1}=1/2+\left\{ 
\begin{array}{l}
-i\sqrt{R_{C}g}/2+\nu+O(\Delta),\;g>0 \\ 
2\nu,\;g=0 \\ 
\sqrt{R_{C}|g|}/2+\nu+O(\Delta),,\;g<0%
\end{array}
\right. , \\
& \beta_{1}=1/2+\left\{ 
\begin{array}{l}
i\sqrt{R_{C}g}/2+\nu+O(\Delta),\;g>0 \\ 
0,\;g=0 \\ 
-\sqrt{R_{C}|g|}/2+\nu+O(\Delta),,\;g<0%
\end{array}
\right. .
\end{align*}
A direct estimation gives%
\begin{equation*}
\sigma_{0\theta}^{\prime}(E)=\left\{ 
\begin{array}{l}
\left\{ 
\begin{array}{l}
O(\sqrt{\Delta}),\;g=g_{0k}\;\mathrm{or}\;g\neq g_{0k},\;\theta\neq\theta
_{0} \\ 
O(1/\sqrt{\Delta}),\;g\neq g_{0k},\;\theta=\theta_{0}%
\end{array}
\right. ,\;\Delta>0 \\ 
0,\;\Delta<0\;%
\end{array}
\right. , 
\end{equation*}
where $\tan\theta_{0}=-f_{0|0}=-\left. f_{0}(E)\right| _{\Delta=0}$. Note
that the discrete level with $E=(1+4R_{C}g)R_{C}^{2}$ is absent.

\paragraph{$E<(1+4R_{C}g)/R_{C}^{2}$}

In this case, we have $\left. \operatorname{Im}\nu\right| _{W=E}=0$, $\left.
\nu\right| _{W=E}>0$,%
\begin{align*}
\alpha_{1} & =1/2+\nu-i\sqrt{w-1}/4,\;\beta_{1}=\overline{\alpha_{1}}%
,\;w\geq1, \\
\alpha_{1} & =1/2+\nu+\sqrt{1-w}/4,\;\beta_{1}=1/2+\nu-\sqrt{1-w}/4,\;w<1.
\end{align*}

Thus, we have: the function $[f_{0\theta}(E)]^{-1}$ is real except the
points $E_{0n}(\theta)$,%
\begin{equation}
f_{0\theta}(E_{0n}(\theta))=0,   \label{Cps2.6.7.2.3.1}
\end{equation}
such that we obtain%
\begin{align*}
\sigma_{0\theta}^{\prime}(E) & =\sum_{n\in\mathcal{N}_{0}}Q_{0\theta,n}^{2}%
\delta(E-E_{0n}(\theta)),\;Q_{0\theta,n}=\frac{1}{\sqrt{\partial
_{E}f_{0\theta}(E_{on}(\theta))}\cos\theta}, \\
\partial_{E}f_{0}(E) & >0,
\end{align*}
where $\mathcal{N}_{0}$ is a subset of $\mathbb{Z}$ to be described below.
Furthermore, we find 
\begin{equation*}
\partial_{\theta}E_{0n}(\theta)=-[\partial_{E}f_{0}(E_{0n}(\theta))\cos
^{2}\theta]^{-1}<0. 
\end{equation*}

\subparagraph{$g\geq g_{0,0}=-1/R_{C}$}

In this case, the function $f_{0}(E)$ has the properties: $%
f_{0}(E)\rightarrow-\infty$ as $E\rightarrow-\infty$; $f_{0}(E)\rightarrow
f_{0|0}-0=-\tan\theta_{0}-0$ as $E\rightarrow(1+4R_{C}g)R_{C}^{2}-0$;.$%
f_{0}(E)$ increases monotonocally on the interval $(-%
\infty,(1+4R_{C}g)R_{C}^{2})$. Then we find: in the interval $%
(-\infty,(1+4R_{C}g)R_{C}^{2})$, for any fixed $\theta\in(\theta_{0}.\pi/2)$%
, there is one level $E_{0,-1}(\theta)$ which run monotonocally from $-\infty
$ to $(1+4R_{C}g)R_{C}^{2}-0$ as $\theta$ run from $\pi/2-0$ to $\theta_{0}+0
$; there are no discrete levels on the interval $(-%
\infty,(1+4R_{C}g)R_{C}^{2})$ for $\theta\in(-\pi /2,\theta_{0})$. Formally,
eq. (\ref{Cps2.6.7.2.3.1}) has solution $E_{0,-1}(%
\theta_{0})=(1+4R_{C}g)R_{C}^{2}$ for $\theta=\theta_{0}$. However, as was
noted above, such levels are absent We find also%
\begin{equation*}
\mathcal{N}_{0}=\mathcal{N}_{0,-1}=\left\{ 
\begin{array}{l}
\{-1\},\;\theta\in(\theta_{0},\pi/2) \\ 
\varnothing,\;\theta\in(-\pi/2,\theta_{0}]%
\end{array}
\right. . 
\end{equation*}

\subparagraph{$g_{0,k+1}\leq g<g_{0,k+1}$, $\protect\sqrt{R_{C}|q|}=1/2+k+%
\protect\delta$, $0<\protect\delta\leq1$, $k\in\mathbb{Z}_{+}$}

In this case,\ the function $f_{0}(E)$ has the properties: $%
f_{0}(E)\rightarrow-\infty$ as $E\rightarrow-\infty$; $f(\mathcal{E}_{0n}\pm
0)=\mp\infty$ $\ n=0,1,...,n_{\max}=k$; $f_{0}(E)\rightarrow
f_{0|0}(E)-0=-\tan\theta_{0}-0$ as $E\rightarrow(1+4R_{C}g)/R_{C}^{2}-0$. We
find: in each interval $(\mathcal{E}_{0n-1},\mathcal{E}_{0n})$, $n=0,...,k$
(we set $\mathcal{E}_{0,-1}=-\infty$), for any fixed $\theta\in(-\pi/2,\pi/2)
$, there is one level $E_{0n}(\theta)$ which run monotonocally from $%
\mathcal{E}_{n-1}+0$ to $\mathcal{E}_{n}-0$ as $\zeta$ run from $\pi/2-0$ to
-$\pi/2+0$; in the interval $(\mathcal{E}_{k},(1+4R_{C}g)R_{C}^{2})$, for
any fixed $\theta\in(\theta_{0},\pi/2)$, there is one level $E_{0k+1}(\theta)
$ which run monotonocally from $\mathcal{E}_{k}+0$ to $(1+4R_{C}g)R_{C}^{2}-0
$ as $\zeta$ run from $\pi/2-0$ to $\theta_{0}+0$; thare are no levels in
the interval $(\mathcal{E}_{k},(1+4R_{C}g)/R_{C}^{2})$ for $%
\theta\in(-\pi/2,\theta_{0})$. Formally, eq. (\ref{Cps2.5.7.2.3.1}) has
solution $E_{0,k+1}(\theta _{0})=(1+4R_{C}g)/R_{C}^{2}$ for $%
\theta=\theta_{0}$. However, as was noted above, such levels are absent.. We
find also%
\begin{equation*}
\mathcal{N}_{0}=\mathcal{N}_{0,k}(\theta)=\left\{ 
\begin{array}{l}
\{0,1,...,k\},\;\theta\in(-\pi/2,\theta_{0}] \\ 
\{0,1,...,k+1\},\;\zeta\in(\theta_{0},\pi/2)%
\end{array}
\right. . 
\end{equation*}

Finally, we obtain. The spectrum of $\hat{h}_{0\theta}$ is simple and $%
\mathrm{spec}\hat{h}_{0\theta}=[(1+4R_{C}g)/R_{C}^{2},\infty)\cup
\{E_{0n}(\theta),\;n\in\mathcal{N}_{0}\}$. The set of functions 
\begin{equation*}
\left\{ U_{0\theta,E}(\rho)=\varrho_{0\theta}(E)U_{0\theta}(\rho
;E),\;E\geq(1+4R_{C}g)/R_{C}^{2};\;U_{0\theta,n}(\rho)=Q_{0\theta,n}U_{0%
\theta}(\rho;E_{0\theta}(\zeta),),\;n\in\mathcal{N}_{0}\right\} 
\end{equation*}
forms a complete orthogonalized system in $L^{2}(0,R_{C})$.

\section{Conclusions}

As we found, two dimensional oscillator and coulomb problems on pseudoshpere
are described by the same equations in terms of the variables $\alpha$ and $%
\beta$ This means that each point of the spectra of one of these theories
corresponds a point of the spectra of the other theory, i.e. there is
one-to-one correspondence between points of the the planes $E_{O}, \lambda$ 
and $E_{C}, g$.

\section{Acknowledgement}I.Tyutin  thanks RFBR Grant 11-01-00830 for partial support.

\end{document}